\documentclass[12pt]{report}

\usepackage{graphicx,epsfig,tabularx}
\usepackage{amsmath,amssymb,url}
\usepackage{psfrag}

\newcommand{\mZero}{\ensuremath{\mathrm{m}_0}}
\newcommand{\mOneHalf}{\ensuremath{\mathrm{m}_{1/2}}}

\newcommand{\mW}{\ensuremath{\mathrm{m}_{\mathrm{W}}}}

\newcommand{\pbinv}{\ensuremath{\mathrm{pb}^{-1}}}
\newcommand{\fbinv}{\ensuremath{\mathrm{fb}^{-1}}}
\newcommand{\ppbar}{\ensuremath{\mathrm{p}\bar{\mathrm{p}}}}

\newcommand{\mevcc}{\ensuremath{\mathrm{MeV}/c^2}}
\newcommand{\mevc}{\ensuremath{\mathrm{MeV}/c}}
\newcommand{\gevcc}{\ensuremath{\mathrm{GeV}/c^2}}
\newcommand{\gevc}{\ensuremath{\mathrm{GeV}/c}}
\newcommand{\tevcc}{\ensuremath{\mathrm{TeV}/c^2}}
\newcommand{\tevc}{\ensuremath{\mathrm{TeV}/c}}
\newcommand{\wpm}{\ensuremath{\mathrm{W}^\pm}}
\newcommand{\hpm}{\ensuremath{\mathrm{H}^\pm}}
\newcommand{\epem}{\ensuremath{\mathrm{e}^+\mathrm{e}^-}}
\newcommand{\tanb}{\ensuremath{\tan{\beta}}}
\newcommand{\mh}{\ensuremath{\mathrm{m}_{\mathrm{h}}}}
\newcommand{\mH}{\ensuremath{\mathrm{m}_{\mathrm{H}}}}
\newcommand{\mA}{\ensuremath{\mathrm{m}_{\mathrm{A}}}}

\newcommand{\sel}{\ensuremath{\tilde{\mathrm{e}}}}

\newcommand{\rpar}{R-parit\'{e}}

\newcommand{\Mone}{\ensuremath{\mathrm{M}_1}}
\newcommand{\Mtwo}{\ensuremath{\mathrm{M}_2}}
\newcommand{\Mthree}{\ensuremath{\mathrm{M}_3}}
\newcommand{\mZ}{\ensuremath{\mathrm{m}_{\mathrm{Z}}}}
\newcommand{\thw}{\ensuremath{\theta_{\mathrm{W}}}}

\newcommand{\swtwo}{\ensuremath{\sin^2{\thw}}}

\newcommand{\cwtwo}{\ensuremath{\cos^2{\thw}}}

\newcommand{\LEFT}{\ensuremath{\mathrm{L}}}
\newcommand{\RIGHT}{\ensuremath{\mathrm{R}}}

\begin{document}
\begin{titlepage}
\pagestyle{empty}

\topmargin-1cm
\noindent ORSAY\hfill LAL 09-218\\
n$^\circ$ d'ordre : \hfill D\'ecembre 2009\\
\vspace{1cm}
\begin{center}
\vspace{1.5cm}
M\'emoire d'Habilitation \`a Diriger des Recherches\\
\vspace{3cm}
\begin{bf}
\begin{large}
Measuring Supersymmetry \\
\end{large}
\end{bf}
\vspace{2cm}
Dirk ZERWAS\\
\vspace{3.5cm}
Habilitation soutenue le 11 D\'ecembre 2009 devant la commission d'examen\\
\vspace{1.5cm}
\begin{tabular}{llll}
MME.     & S.   & DAWSON   & rapporteur \\
M.       & K.   & DESCH    & rapporteur\\
         & D.   & FOURNIER & \\
         & H.   & OBERLACK & rapporteur \\
         & J.   & ORLOFF   & \\
         & G.   & WORMSER  & \\
\end{tabular}
\end{center}

\cleardoublepage

\section*{Abstract}
\pagestyle{empty}
Supersymmetry is an attractive extension of the standard model of 
particle physics. It associates to every bosonic degree of freedom
a fermionic one and vice versa. 
Supersymmetry unifies the coupling constants of the 
electromagnetic, weak and strong forces at a high scale 
and provides a candidate for the 
elusive dark matter. Supersymmetry could be discovered at the LHC, the
proton--proton collider at CERN which has started operations in 2008. \
The LHC 
is foreseen to have a center--of--mass energy of 14~TeV, opening
up a new mass range to be explored to search for supersymmetric
particles with the ATLAS and CMS experiments.
The development and production of electronics for these detectors
has been a challenge, e.g. for
the readout board for the electromagnetic calorimeter. 
Reconstructing the physics events with the best precision, in particular
the reconstruction and identification of electrons and photons in the large 
QCD background has been prepared in extensive test beam studies 
and Monte Carlo simulations.
If the Higgs boson and supersymmetry are discovered, the properties of the
(s)particles such as the masses, branching ratios must be measured precisely, 
either at the LHC or at a future \epem{} linear collider. 
The SFitter project aims to determine the underlying theoretical model 
parameters from the correlated experimental measurements including 
theoretical errors. The methods are applied to the extraction
of the fundamental parameters of supersymmetry as well the 
measurement of the Higgs boson couplings at the LHC.
The extrapolation of the supersymmetric
parameters from the weak scale to the Grand Unification Scale
could provide the basis towards the inclusion of gravity.

\cleardoublepage

\topmargin 15cm
\vspace{13cm}
\noindent 
{\it Therefore, since brevity is the soul of wit,\\ 
And tediousness the limbs and outward flourishes, \\
I will be brief.}\\
\begin{flushright}Hamlet, Act 2 Scene 2, William Shakespeare\end{flushright}

\cleardoublepage

\end{titlepage}

\pagenumbering{arabic}
\tableofcontents
\cleardoublepage

\chapter{Introduction}

The Standard Model of elementary particle physics~\cite{Glashow:1961tr,Salam:1968rm,Weinberg:1967tq} describes matter
and its interactions with an unprecedented degree of precision.
Matter is built of fermions (quarks and leptons) and their
interactions are mediated via bosons.
The photon ($\gamma$) is responsible for the electromagnetic 
interactions, the charged vector bosons (\wpm) and the Z boson
are the carriers of the weak force. The gluons mediate the strong interaction. 
Masses are generated via the Higgs mechanism, leading to 
an additional neutral scalar particle, the Higgs 
boson (H)~\cite{Higgs:1964ia,Higgs:1966ev,Higgs:1964pj,Englert:1964et,Guralnik:1964eu}.

Supersymmetry is the next logical step after gauge theories~\cite{Wess:2009cy}.
In supersymmetric theories a fermionic degree of freedom is 
associated to every bosonic degree of freedom and 
vice-versa~\cite{Golfand:1971iw,Volkov:1973ix,Wess:1974tw,Fayet:1974jb,Martin:1997ns} 
(and references therein).
Ultra-violet completions of the Standard Model lead to quantum corrections 
on the Higgs boson mass of the order of the new physics scale, larger
than the Higgs boson itself, whereas in supersymmetry 
these large corrections cancel. Supersymmetry can solve this 
hierarchy problem in a natural
way. In the Standard Model
the coupling constants of the strong, weak and electromagnetic
interactions, extrapolated via renormalization 
group equations (RGEs), do not meet in one point. Supersymmetry
modifies this behavior to unify the constants at about 10$^{16}$~GeV
(grand unification scale, GUT scale)~\cite{Dawson:1979zq,Einhorn:1981sx,Amaldi:1991cn}.
Additionally 
supersymmetry is attractive as it provides a candidate for 
the elusive dark matter.  

To construct supersymmetric models such as 
the minimal supersymmetric extension of the Standard Model (MSSM)
or minimal Supergravity (mSUGRA), the Higgs sector 
has to be extended to give masses to up and down type quarks
The two Higgs doublets are necessary: to avoid anomalies and
large flavor changing neutral currents (FCNC).
In the Standard Model the Higgs field and its complex conjugate field are used to give
masses to up and down type quarks, while in a supersymmetric theory
the complex conjugate field cannot be used, necessitating the second doublet.
The Higgs sector of the MSSM contains five Higgs bosons: h, H, A and \hpm{}.
The lightest Higgs boson must be light with a mass of less than 140~\gevcc{}.
Additionally the sfermions, the partners of the fermions, are predicted. The supersymmetric 
partners of the neutral gauge 
and Higgs bosons mix to form the neutralinos. The charginos are mixtures of
the supersymmetric partners of the charged Higgs and gauge bosons.

\begin{figure}[htb]
\begin{minipage}[htb]{0.48\textwidth}
\centering
\includegraphics[width=\textwidth]{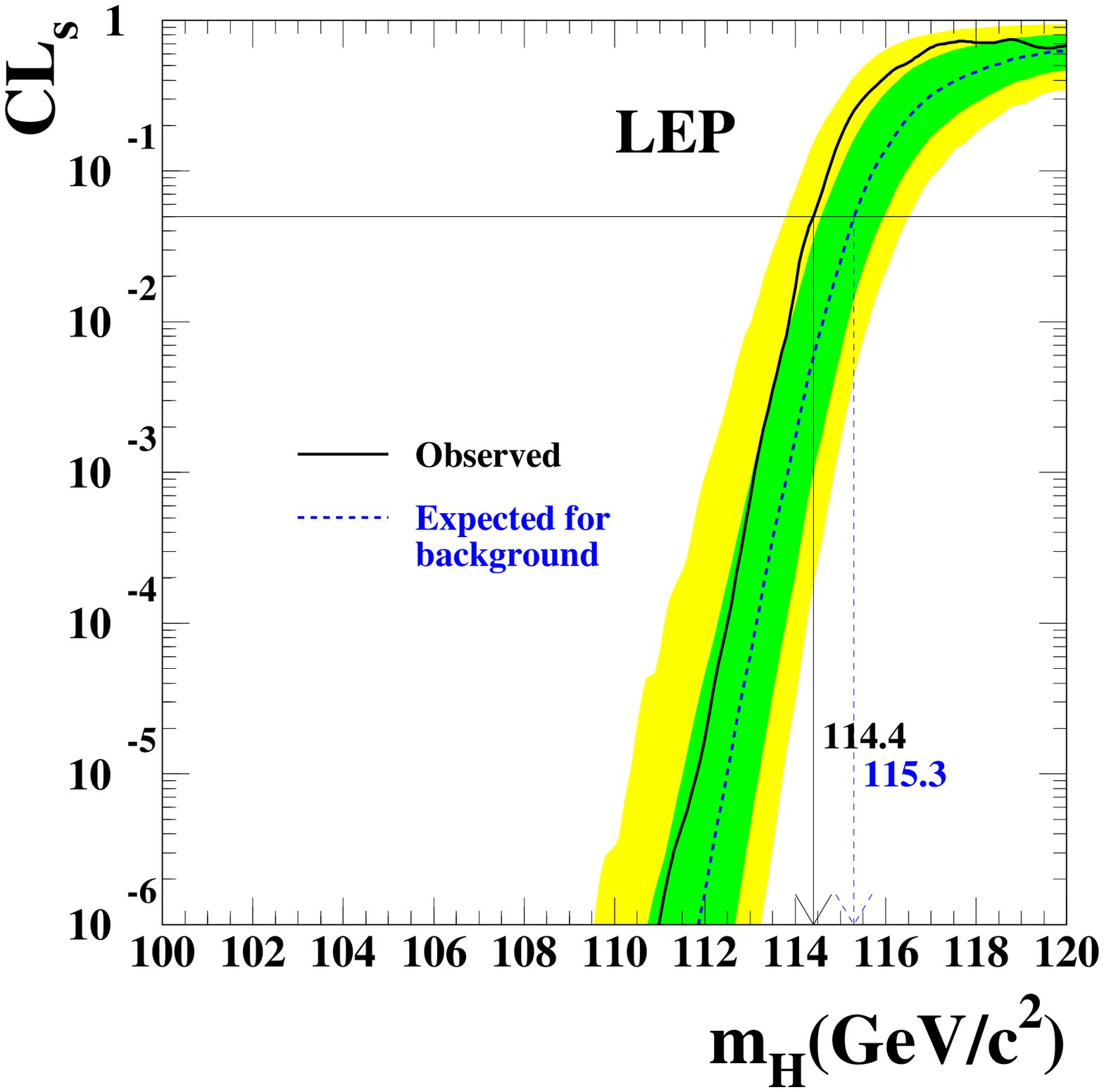}
\vspace{-1.cm}
\caption{The expected and observed lower limit on the Standard Model Higgs
boson mass obtained by the LEP collaborations is shown~\cite{Barate:2003sz}.} 
\label{fig:HiggsLEPLimit}
\end{minipage}
\hspace{0.3cm}
\begin{minipage}[htb]{0.48\textwidth}
\vspace{2cm}
\centering
\includegraphics[width=\textwidth]{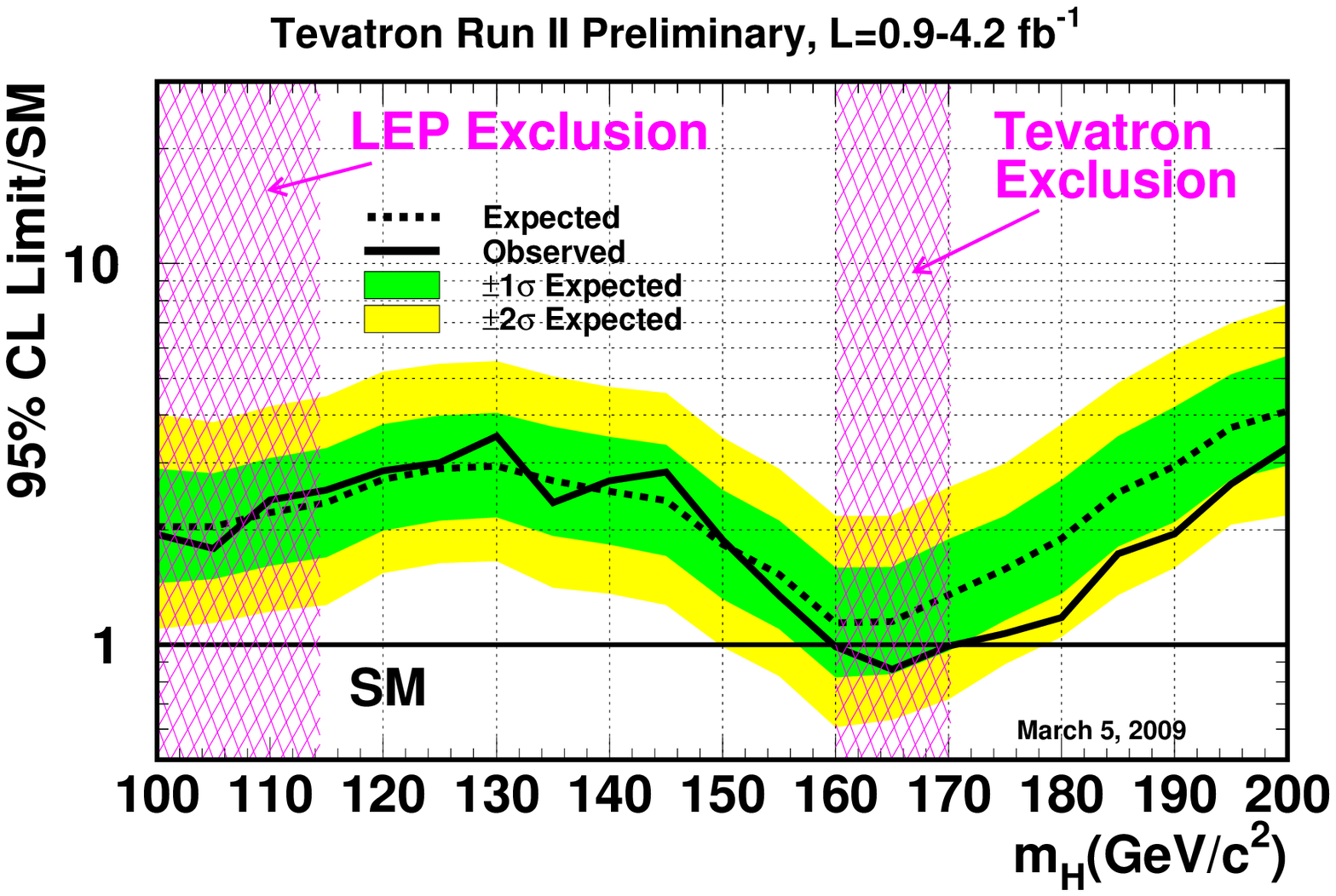}
\vspace{-1cm}
\caption{The expected and observed lower limit on the Standard Model Higgs
boson mass obtained by the CDF and Do collaborations is shown~\cite{Bernardi:2008ee}.} 
\label{fig:HiggsTeVLimit}
\end{minipage}
\end{figure}

The most precise measurements of parameters of 
the Standard Model are obtained from \epem{} colliders such as LEP and SLC,
as well as the proton--antiproton (\ppbar{}) collider TeVatron 
at Fermilab.
ALEPH, DELPHI, L3 and OPAL took data from 1989 to 2000 at LEP, operating 
at center--of--mass energies from 91~GeV to 209~GeV. The SLD detector 
operated at SLAC's \epem{} SLC at a center--of--mass energy of about 91~GeV.
CDF and D0 are 
taking data at the TeVatron, operating at a center--of--mass energy of 1.96~TeV.
In 2009 the proton--proton (pp) collider LHC at CERN will effectively
start operations. The center--of-mass energy is foreseen to be 7~TeV which will 
be increased to 14~TeV in the coming years. ATLAS and CMS
are the two multi-purpose detectors located at the LHC. They are 
optimized for the search of the Higgs boson(s)
and new physics. The \epem{} Linear Collider project (ILC) 
is being prepared for a center--of--mass energy from 500~GeV to 1~TeV. 
A decision on whether the ILC will be built is expected after the analysis 
of the first LHC data.
R\&D is being performed for a linear collider (CLIC) which may be capable
of running at 3~TeV.

The search for the Standard Model Higgs boson at LEP has led
to a limit of 114.4~\gevcc{}~\cite{Barate:2003sz} as shown in 
Figure~\ref{fig:HiggsLEPLimit}, slightly smaller than the expected limit 
of 115.3~\gevcc{}. Time will tell whether the events observed
at LEP around this mass will turn out to have been the first indication of the
Higgs boson or not. 

The TeVatron is particularly sensitive to the Standard Model Higgs boson for 
masses of the order 150-180~\gevcc{}. In this mass region the decay of the
Higgs boson to a pair of \wpm{} bosons is dominant. The combination of several channels 
from both CDF and D0 has lead to the first new limit beyond the LEP bound:
a Standard Model Higgs boson with a mass of 160 to 170~\gevcc{} is excluded at 95\% confidence
level~\cite{Bernardi:2008ee,Bernardi:2009} as shown in Figure~\ref{fig:HiggsTeVLimit}.

\begin{figure}[htb]
\begin{minipage}[htb]{0.48\textwidth}
\vspace{-1.5cm}
\includegraphics[width=\textwidth]{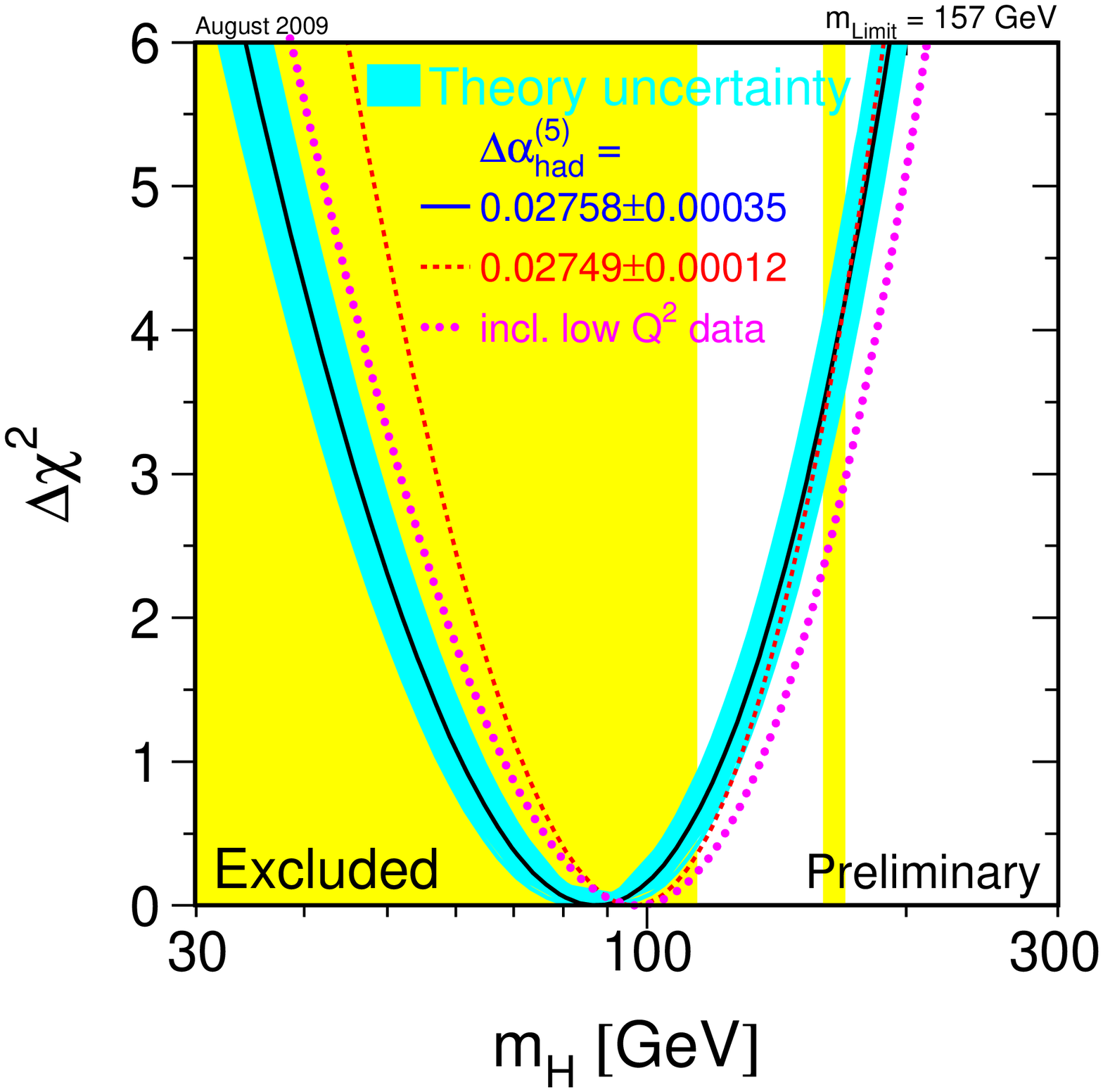}
\vspace{-1.0cm}
\caption{The result of the electroweak fit is shown as function of the Higgs boson mass. 
The yellow shaded area shows the region 
excluded by the direct search.}
\label{fig:HiggsElectroweakFit}
\end{minipage}
\hspace{0.3cm}
\begin{minipage}[htb]{0.48\textwidth}
\includegraphics[width=\textwidth]{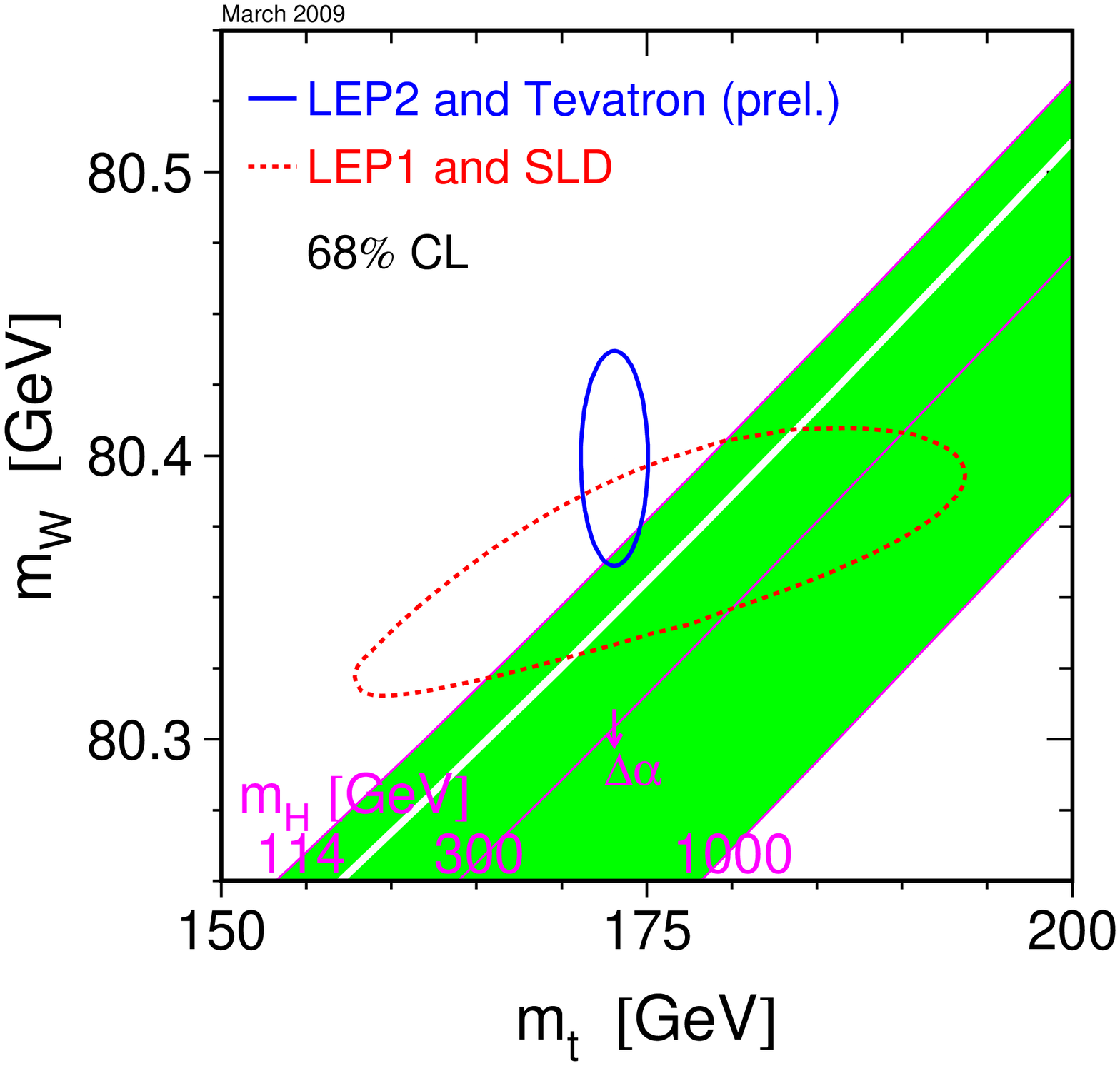}
\vspace{-1cm}
\caption{The measurement of the \wpm{} boson mass as function of the top quark mass is shown
(solid line). The dotted line shows the indirect constraints on the \wpm{} boson mass and top quark
mass. The straight lines show different Higgs mass hypotheses (update of~\cite{Alcaraz:2007ri}).} 
\label{fig:mTopmW}
\end{minipage}
\end{figure}

The precise measurement of the electroweak parameters at LEP and SLC are discussed in 
detail in~\cite{Alcaraz:2007ri}. The most precise individual measurements of the 
\wpm{} boson mass are from CDF~\cite{Aaltonen:2007ps} and D0~\cite{Abazov:2009cp}
which have reached a precision of 48~\mevcc{} and 43~\mevcc{} respectively. The combination
of all \wpm{} mass measurements leads to $80.399\pm0.023\ \gevcc$. 
The top quark mass has now been measured with a precision better than percent
as shown in~\cite{Varnes:2008tc}. 
The predictions of the electroweak observables depend via quantum corrections
on the mass of the Higgs boson, especially the \wpm{} boson and top quark masses. 
Their sensitivity is illustrated in Figure~\ref{fig:mTopmW} where the solid 
line denotes the experimental measurements of the \wpm{} boson and top quark masses
and the straight lines different Higgs boson mass hypotheses. 
Combining all measurements in a fit the Higgs 
boson mass can be determined to be $84^{+34}_{-26}\gevcc$. This leads to 
an upper limit of 157~\gevcc{} on the Higgs boson mass at 95\% confidence level.
Thus the data prefer a light Higgs boson mass.

\begin{figure}[htb]
\begin{minipage}[htb]{0.48\textwidth}
\centering
\includegraphics[width=\textwidth]{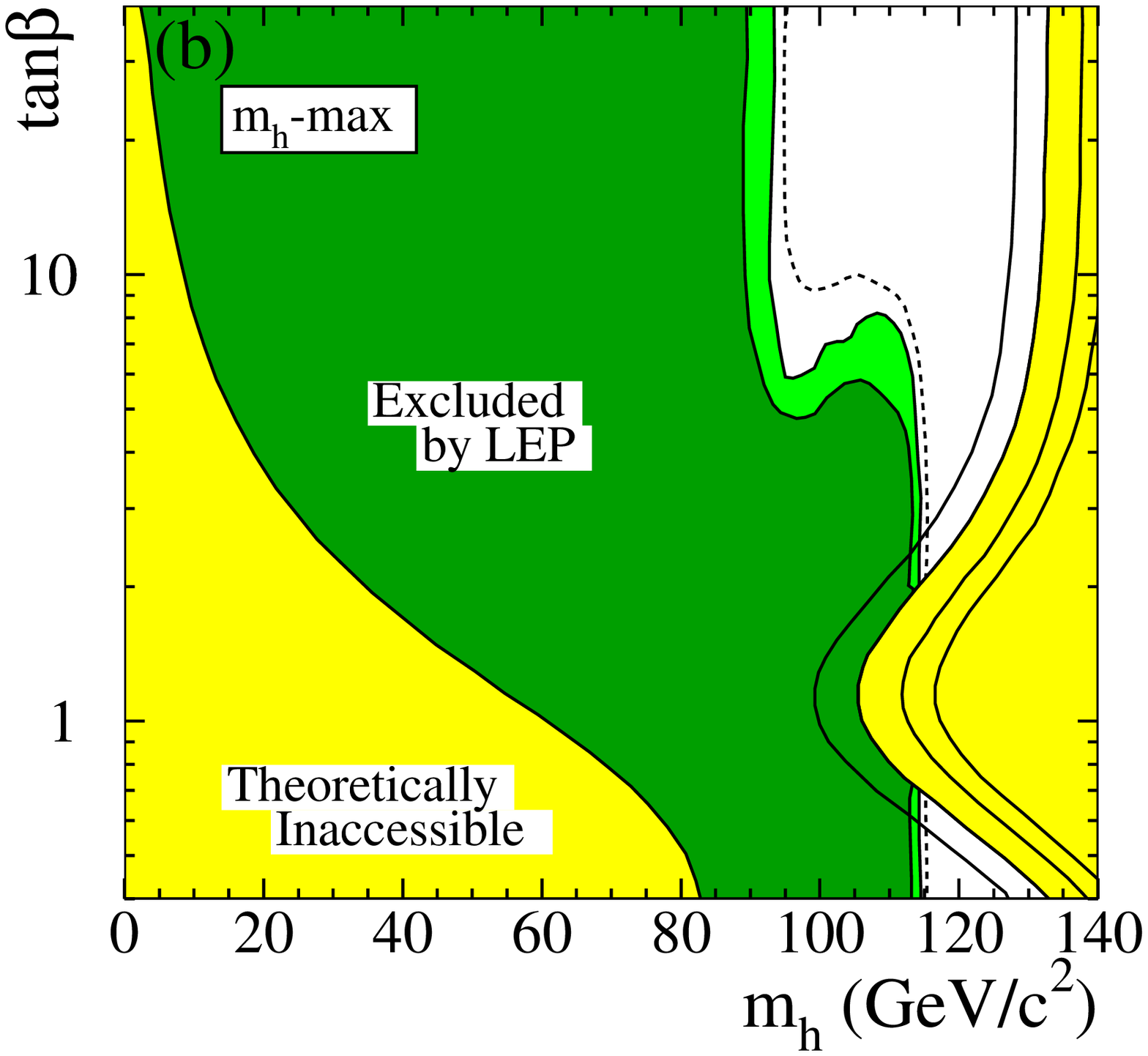}
\vspace{-0.2cm}
\caption{The region excluded by LEP and by theory in the plane $\tan\beta$ versus
mass of the lightest Higgs boson~\cite{Schael:2006cr}.}
\label{fig:LEPMSSMHiggs}
\end{minipage}
\hspace{0.3cm}
\begin{minipage}[htb]{0.48\textwidth}
\centering
\includegraphics[width=\textwidth]{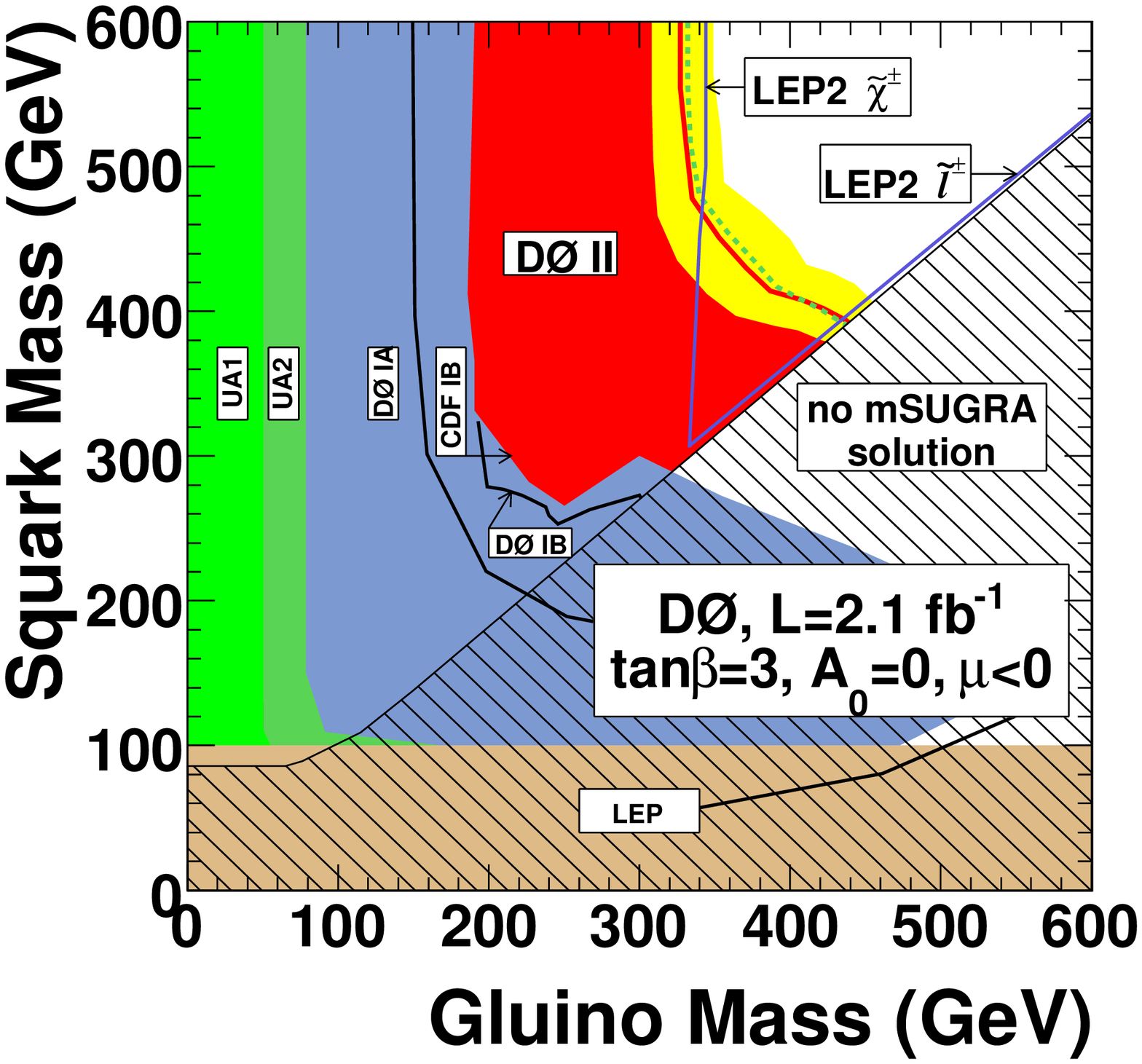}
\vspace{-0.7cm}
\caption{The regions excluded by the TeVatron and LEP 
in the plane of the squark mass versus the gluino mass are shown~\cite{Grivaz:2008tf}.} 
\label{fig:TeVatronSquarks}
\end{minipage}
\end{figure}

In addition to the Standard Model production mode, albeit 
with modified couplings, the associated production of Higgs bosons, such as
the h produced in association with the A,
can also be used in the search for the MSSM Higgs bosons.
Limits on neutral MSSM Higgs bosons have been determined to be be 92.8~\gevcc{} for the h and
93.4~\gevcc{} for the A boson~\cite{Schael:2006cr}. The searches for the neutral MSSM
Higgs bosons can also be interpreted as limit on the ratio of the vacuum expectation 
values (vevs) of the two Higgs doublets $\tan\beta$. $\tan\beta$ between 0.7
and~2 are excluded as shown in Figure~\ref{fig:LEPMSSMHiggs}. 
Some variation of the excluded region is possible as it depends on the top quark mass and 
the parameters of the supersymmetric model.

As the cross section for the pair production of scalars is proportional to 
$\beta^3$, where $\beta=\sqrt(1-4\mathrm{m}_{\hpm}^2/s)$, the mass limit
on the charged Higgs bosons from the combination of the LEP experiments has not reached the kinematic
limit. Charged Higgs boson masses of less than 78.6~\gevcc{} are excluded 
at 95\% confidence level~\cite{:2001xy}.

LEP has set mass limits
on selectrons and muons (cross section $\sim\beta^3$) at 95~\gevcc{} and slightly weaker limits 
on the stau slepton of 87~\gevcc{}~\cite{LEPSUSY:sleptons}. 
Charginos with masses of up to 103~\gevcc{} (cross section $\sim\beta$), i.e. essentially up
to the kinematic limit, have been excluded~\cite{LEPSUSY:charginos}. 
Combining all searches a lower limit on the mass of the lightest
neutralino of 50~\gevcc{} has been obtained in mSUGRA~\cite{LEPSUSY:neutralinos}. 

The TeVatron is on the forefront of the search for supersymmetric colored
particles, the squarks and gluinos. In the absence of a signal, lower limits 
on the mass of the sparticles have been set. These depend on the specific scenario 
chosen. As an example, for a given typical choice of parameters, 
squarks of mass less than 392~\gevcc{} have been excluded and a limit of 327~\gevcc{}
on the mass of the gluino has been obtained.
For equal squark and gluino masses a lower 
limit of 408~\gevcc{} has been determined~\cite{Grivaz:2008tf} (and references therein)
as shown in Figure~\ref{fig:TeVatronSquarks}. 

Several aspects of the search for supersymmetry 
and its measurement will be discussed in the following. 
In chapter~\ref{chap:SUSYPheno}, the phenomenology of the
most important supersymmetric models will be reviewed.
Chapter~\ref{chap:exp} contains a brief discussion 
of experimental aspects of the search for supersymmetry. In particular
a part of the Front End Electronics of the ATLAS electromagnetic 
calorimeter as well as the performance of the reconstruction of electrons
and photons are described in more detail, based on Refs.~\cite{Aad:2009wy,Zerwas:2009zz,Buchanan:2008zza}.
In chapter~\ref{chap:determination} the determination of supersymmetric
parameters from experimental measurements as well 
as the determination of the Higgs boson couplings at the LHC 
will be discussed. The results summarized in this chapter are 
explained in more detail in Refs.~\cite{Lafaye:2004cn,Lafaye:2007vs,Lafaye:2009vr}.
In the last chapter the results will be summarized and an
outlook for the future will be given. 
\cleardoublepage

\chapter{Phenomenology of Supersymmetry}
\label{chap:SUSYPheno}

To construct a supersymmetric extension of the Standard Model, such as
the MSSM and mSUGRA,
an extended Higgs sector is necessary in order to give masses to up and 
down type quarks as well as 
the addition of partners of the standard model particle. 
Further extensions, such as the MSSM with an additional singlet (NMSSM), have an
even richer spectrum of Higgs bosons and possibly new signatures with respect 
to the MSSM~\cite{Ellwanger:2009wj}. In this chapter the discussion is restricted to the MSSM and mSUGRA.
The parameters of the MSSM are defined at the electroweak scale, whereas 
the parameters of mSUGRA are defined at the GUT scale.
R--parity will be conserved in the following, therefore supersymmetric 
particles are produced in pairs, (cascade-) decaying to the lightest 
supersymmetric particle (LSP). The LSP is stable, neutral and weakly
interacting, leading to the characteristic signature for supersymmetry at a collider,
missing energy.

The slope of the gauge couplings' running is modified at a scale of about 1~TeV
by Supersymmetry, leading to a unification of the couplings
at about 10$^{16}$~GeV. As this scale is close to the Planck scale 
(10$^{18}$~GeV), 
Supersymmetry could be a telescope to physics at this scale.
A side effect of this unification at a higher scale compared to the Standard Model
is that the proton can have a longer lifetime.

The Standard Model accounts for only 4\% of the energy of the universe. 24\% are 
due to dark matter. The LSP of supersymmetry is prime candidate for dark matter
as it is neutral and weakly interacting. 

From an experimental point of view it is intriguing to note that the Standard Model
electroweak fits favor a light Higgs boson with a mass less than 157~\gevcc{}.
Supersymmetry, as realized in the MSSM, 
predicts a lightest Higgs boson which should have a mass less 
than about 140~\gevcc{}. 

Ultra-violet completions of the Standard Model lead to 
quantum corrections on the mass of the Higgs boson of the order of this new physics scale. 
This unnatural behavior is corrected in supersymmetry 
where an equal number of fermions and bosons leads to a cancellation of the corrections.
In order to preserve the natural cancellation
in the Higgs sector the breaking of Supersymmetry, i.e. lifting the mass
degeneracy of standard and supersymmetric particles, cannot be arbitrarily large, but must be at most of the
order of~TeV.

These arguments motivate the search for Supersymmetry.
In the following sections a few basic considerations of supersymmetry and
the phenomenology of two models, the MSSM and mSUGRA, will be discussed.

\section{Supersymmetry}

Supersymmetry was presented first in~\cite{Golfand:1971iw,Volkov:1973ix,Wess:1974tw}. 
The construction of viable models was pioneered by Fayet and Iliopoulos~\cite{Fayet:1974jb}
\par
At the Planck scale $\mathrm{M}_{\mathrm{P}}=2.4\cdot 10^{18}$~\gevcc{}, 
gravitational effects are important.
The corrections to the Higgs boson mass in the Standard Model
would have to be fine-tuned
to one part in $10^{16}$ to give a Higgs mass 
of the order of the electroweak scale if the Standard Model were to be valid up to this scale.
As supersymmetric models contain two scalars for every fermion, the quadratic
divergences cancel and only the logarithmic parts remain. 
The fermion masses, including radiative corrections,
are logarithmically divergent.
Including additionally soft supersymmetry breaking,
qualitatively the corrections to the Higgs boson mass then take on the 
following form
($\mathrm{m_{soft}}$ is the mass splitting between the fermion
and the scalars)~\cite{Martin:1997ns}:
\begin{equation}
\Delta\mH^2=\mathrm{m_{soft}^2}
\left[\ \frac{\lambda}{16\pi^2}\ln (\Lambda/\mathrm{m_{soft}})
+\dots\ \right]
\end{equation}
If $\lambda\sim 1$ and $\Lambda$ the Planck scale, for a soft supersymmetry
breaking mass of 1~\tevcc{}, the correction to the Higgs boson mass will
be about 500~\gevcc{}. Since the correction increases about linearly with
$\mathrm{m_{soft}}$, therefore the masses of at least some of the 
supersymmetric particles must be less than about 1~\tevcc{}. 
\par
Supersymmetry is a spacetime symmetry. 
The Haag-Lopuzanski-Sohnius~\cite{Haag:1974qh} extension of the 
Coleman-Mandula~\cite{Coleman:1967ad} theorem
restricts the forms of such symmetries in interacting 
quantum field theories.
In practice for extensions of the Standard Model, the generators
$Q$, $Q^\dagger$ have to satisfy an algebra of the form~\cite{Martin:1997ns}:
\begin{equation}
\begin{array}{l}
\{Q,Q^\dagger\}=\sigma_\mu P^\mu\\[10pt]
\{Q,Q\}=\{Q^\dagger,Q^\dagger\}=0\\[10pt]
\left[P^\mu,Q\right]=\left[P^\mu,Q^\dagger\right]=0 \\ 
\end{array}
\end{equation}
where $Q$ generates the transformation of a fermion to a boson and vice
versa, and $P^\mu$ is the momentum generator of spacetime 
translations. As in~\cite{Martin:1997ns} the spinor indices have been 
omitted.
\par
Supermultiplets are introduced in which a boson and 
a fermion must be present.
Scalars and fermions are arranged in a matter (or chiral) multiplet,
vector bosons and fermions are arranged in a gauge (or vector)
multiplet. 
\par
Supersymmetry breaking terms, which do not reintroduce
quadratic divergences, are added to the Lagrangian
by hand~\cite{Girardello:1981wz} (in the notation of~\cite{Martin:1997ns}):
\begin{equation}
{\cal L}=-\frac{1}{2}(M_\lambda\lambda^a\lambda^a+c.c.)-
(m^2)^i_j\phi^{j\ast}\phi_i
-\left(\frac{1}{2}b^{ij}\phi_i\phi_j+\frac{1}{6}a^{ijk}\phi_i\phi_j\phi_k
+c.c.\right)
\end{equation}
There is a gaugino mass term ($M_\lambda$) for each gauge group, scalar
mass terms ($(m^2)^i_j$, $b^{ij}$) and a trilinear coupling $a^{ijk}$.
These terms are added {\it ad hoc} to 
parametrize the current ignorance of the mechanism of supersymmetry breaking.
\par
If supersymmetry is gauged
locally, gravity can be incorporated, leading to supergravity.
The graviton, a massless spin-2 particle, 
must have a supersymmetric 
partner (gravitino, spin-$\frac{3}{2}$).
The gravitino 
can help solving the problem of generating the
soft supersymmetry breaking masses flavor-blind where
a hidden sector is introduced which communicates with the
visible sector via gravitational interactions
(gravitino condensation in the hidden sector).
The hidden sector is connected to the gaugino and scalar
soft supersymmetry breaking terms. Additional terms ensure
that different 
soft breaking masses can be generated for each superparticle.

\section{The Minimal Supersymmetric Extension of the Standard Model}

Each Standard Model particle (or rather degree of freedom) receives
a new particle as partner. The following section follows closely 
the notation of~\cite{Martin:1997ns,Haber:1984rc}.
The parameters of the MSSM are defined at the electroweak scale.
\par
\begin{table}[ht]
\begin{center}
\begin{tabular}{|c|c|c|}
\cline{1-3}
spin-0          & spin-1/2               & spin-1  \\
\cline{1-3}
$\tilde{\mathrm{q}}_{\mathrm{R}}$, $\tilde{\mathrm{q}}_{\mathrm{L}}$ 
& q         &         \\
           &     $\tilde{\mathrm{g}}$               &  g     \\
$\tilde{\ell}_{\mathrm{R}}$, $\tilde{\ell}_{\mathrm{L}}$ & $\ell$ &         \\
           &$\tilde{\gamma}$               & $\gamma$ \\
H        & $\tilde{\mathrm{H}}$, $\tilde{\mathrm{Z}}$      & Z \\
h A    & $\tilde{\mathrm{h}}$                   &        \\
\hpm    & $\tilde{\mathrm{H}}^\pm$, $\tilde{\mathrm{W}}^\pm $  & \wpm\ \\
\cline{1-3}
\end{tabular}
\caption{\label{tab:part} The (s)particle content of the MSSM} 
\end{center}
\end{table}
Two Higgs doublets of opposite 
hypercharge are used to avoid triangle anomalies and FCNCs. 
Different Higgs fields
give masses to the up-- and down--type particles. 
The complete list of (s)particles and their superpartners
is shown in Table~\ref{tab:part}.
\par
\begin{equation}
\begin{array}{lcl}
\mathrm{V}&=&(m_1^2+|\mu|^2)H_1^{i\ast}H_1^i
+(m_2^2+|\mu|^2)H_2^{i\ast}H_2^i
-m_{12}^2(\epsilon_{ij}H^i_1H^j_2+h.c.)\\[15pt]
&&+\frac{\displaystyle 1}{\displaystyle 8}(g^2+{g'}^2)
[H_1^{i\ast}H_1^i-H^{j\ast}_2H^j_2]^2
+\frac{\displaystyle 1}{\displaystyle 2}g^2|H^{i\ast}_1H^i_2|^2\\
\end{array}
\end{equation}
Up-type fermion masses are proportional to $v_1$, the vacuum expectation
value of the first doublet, and 
down-type fermion masses to $v_2$, the vacuum expectation value 
of the second doublet.
The parameter $\mu$ is the supersymmetric Higgs mass parameter.
The parameters governing the Higgs sector are 
usually taken to be the mass of the lightest neutral CP-even Higgs boson (h)
or CP--odd Higgs boson (A),
\tanb{} and \mW{} from the electroweak sector.
\par
At tree level at least one Higgs boson should be lighter or as light as the
Z boson.
However, this stringent bound for the neutral Higgs bosons is 
loosened substantially by radiative
corrections, which are driven mostly by the heavy top quark mass and
the top squark masses:
\begin{equation}
\Delta\mh^2=\frac{3g^2}{8\pi^2}\frac{\mathrm{m_t^4}}{\mW^2}
\ln\frac{\mathrm{m_{\tilde{t}_1}m_{\tilde{t}_2}}}{\mathrm{m_t^2}}
\end{equation}
The only strong 
prediction remaining after including these corrections
is that the lightest neutral Higgs boson (h) must have a mass less
than about 140~GeV.
\par
In the leptonic sector each charged
lepton receives two scalar partners, called $\LEFT$-- and
$\RIGHT$--sleptons ($\sel_{\LEFT}$, $\sel_{\RIGHT}$),
where ``$\LEFT$'' and ``$\RIGHT$'' refer to the chirality state
of the corresponding lepton. 
For the neutrino there is one sneutrino ($\tilde{\nu}_\LEFT$).
In the gauge sector
the partners of the neutral gauge and Higgs bosons mix to form the 
neutralinos: $\chi^0_1$,$\chi^0_2$,$\chi^0_3$,$\chi^0_4$. 
In turn the partners of the charged gauge and Higgs bosons mix to form
the chargino mass eigenstates $\chi^\pm_1$ and $\chi^\pm_2$.
The gluon, as a massless spin-1 particle
is grouped with the gluino, its fermionic partner.
\par
\rpar{}~\cite{Farrar:1978xj} is used to avoid 
certain yukawa terms ($\lambda_{ijk}$, $\lambda'_{ijk}$, $\lambda''_{ijk}$)
which are allowed in a general supersymmetric Lagrangian,
but can lead to a rapid proton decay inconsistent with 
experimental limits. The conservation of this multiplicative 
quantum number in the following studies has as consequence that 
supersymmetric particles must be produced in pairs. They decay,
possibly via cascade decays, until the lightest supersymmetric particle
is reached, which is stable.  
\par
The lightest neutralino is a good candidate for the lightest supersymmetric
particle.
It is color-neutral, not charged and interacts only
weakly with matter. 
The lightest neutralino is a good candidate to 
solve the dark matter problem of the universe.
\par
Certain relationships among supersymmetric particles hold:
\begin{equation}
\mathrm{m}_{\tilde{\ell}_\LEFT}^2= \mathrm{m}_{\tilde{\nu}}^2-\cwtwo\mZ^2\cos
2\beta
\end{equation}
Consequently the left-handed slepton must always be heavier than the
sneutrino, if \tanb{} is greater than unity. Similar relationships
hold in the squark sector.
\par
The mass matrix for down-type squarks and charged sleptons 
can be written in the following way in the basis of left-handed and
right-handed sfermion:
\begin{equation}
\left(\begin{array}{cc}
\mathrm{m_{\tilde{f}_\LEFT}^2}& 
\mathrm{m_f}(\mathrm{A_{\tilde{f}}}-\mu\tan\beta)\\[10pt]
\mathrm{m_f}(\mathrm{A_{\tilde{f}}}-\mu\tan\beta)&\mathrm{m_{\tilde{f}_\RIGHT}^2}\\
\end{array}\right)
\end{equation}
the off-diagonal matrix elements are proportional to the trilinear coupling
$\mathrm{A_{\tilde{f}}}$ and to the fermion's mass. 
Since the mixing angle is governed by the fermion mass, it is expected that
in the first two generations the weak eigenstates are equal to the 
mass eigenstates, so the trilinear mass terms for the first two generations
are usually neglected in the scalar sector.
For the up-type
squarks $\tanb$ must be replaced by $\cot\beta$. 
\par
The MSSM is governed by about 120~parameters. Not allowing
CP violation, requiring the trilinear terms to be real, 
requiring the soft supersymmetry breaking parameters in the scalar sector to be
generation-diagonal to avoid additional sources of flavor 
changing neutral currents, reduces the number of parameters substantially. 
Additionally, the trilinear mass terms of the first two generations are neglected
as their impact on the mass mixing is essentially undetectable experimentally.

Another experimental requirement on the soft mass breaking terms in the
first and second generation squark terms is derived from the inability to flavor
tag, i.e., differentiate between u--, d--, s-- and c--quarks. 
Adding everything together, a phenomenological MSSM is constructed. This 
pMSSM is governed by the following parameters in the supersymmetric sector
and will be called MSSM in the following for simplicity:
the Higgsino mass term ($\mu$), 
the mass of the CP--odd Higgs boson mass ($\mA$), 
the ratio of the vacuum expectation values ($\tan\beta$),
the gaugino mass parameters (\Mone{}, \Mtwo{}, \Mthree{}),
6~soft breaking parameters for sleptons 
($\mathrm{M_{\tilde{e}_\LEFT}}$, $\mathrm{M_{\tilde{e}_\RIGHT}}$, 
$\mathrm{M_{\tilde{\mu}_\LEFT}}$, $\mathrm{M_{\tilde{\mu}_\RIGHT}}$,
$\mathrm{M_{\tilde{\tau}_\LEFT}}$, $\mathrm{M_{\tilde{\tau}_\RIGHT}}$),
2~soft breaking parameters for the first and second generation squarks 
($\mathrm{M_{\tilde{q}_\LEFT}}$, $\mathrm{M_{\tilde{q}_\RIGHT}}$),
3~soft breaking parameters for the third generation squarks 
($\mathrm{M_{\tilde{t}_\RIGHT}}$, 
$\mathrm{M_{\tilde{b}_\RIGHT}}$, $\mathrm{M_{\tilde{q}_{3\LEFT}}}$)
and 
3~trilinear parameters in the scalar fermion
sector 
($\mathrm{A_{\tau}}$, $\mathrm{A_{t}}$, $\mathrm{A_{b}}$).

In addition to the supersymmetric parameters, for a precise prediction of supersymmetric observables,
the following Standard Model parameters are necessary:
the electromagnetic coupling constant at the Z--mass ($\alpha(\mZ)$)
the Fermi constant,
the QCD coupling constant at the Z--mass ($\alpha_S$),
the Z--boson pole mass,
mass of the bottom quark in the MSbar scheme,
the pole mass of the top quark and 
the pole mass of the tau lepton.
Thus in total $20+7$ parameters govern the MSSM.
\par
The electroweak scale where the supersymmetric parameters
are defined needs to be specified. It is sometimes taken 
to be the geometric mean of the stop quark masses. This scale would 
then, in some scenarios, depend on the mass of a sparticle which has not 
yet been measured.
Therefore the proposal of the SPA project~\cite{AguilarSaavedra:2005pw} is followed
by fixing the scale to 1~\tevcc{}.

\section{Minimal Supergravity}

While the MSSM is a general model depending 
on few assumptions, for practical purposes it is useful
to define also a restricted model, motivated by supergravity: mSUGRA.
The mSUGRA parameters are mostly defined at the GUT scale. The parameters
are then evolved via renormalization group equations (RGEs) to the electroweak scale at 1~TeV. 
Every mSUGRA parameter set has a corresponding MSSM equivalent, however 
not every MSSM parameter set has an mSUGRA equivalent. It is therefore interesting to study
both approaches. 
\par
\begin{figure}[htb]
\begin{center}
\epsfig{file=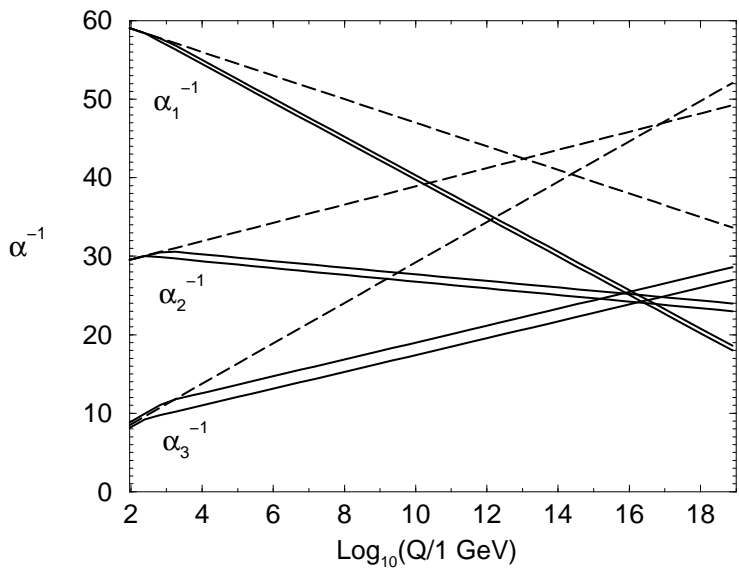,width=10cm,
bbllx=0pt,bblly=0pt,bburx=560pt,bbury=475pt}
\end{center}
\vspace{-1cm}
\caption{\label{fig:gut} Running of $\alpha_i, i=1,3$ in the Standard Model
(dashed lines)
and in the MSSM (full lines).}
\end{figure}
The Standard Model couplings of the 
three gauge groups SU(3), SU(2) and U(1), calculated according to the 
RGEs as a function of the mass scale,
do not intersect in one point, as shown in Figure~\ref{fig:gut}
($\alpha^{-1}_a=g_a^2/(4\pi)$). 
In the MSSM~\cite{Amaldi:1991cn} supersymmetric particles introduce an intermediate 
scale, modifying the slope after passing the threshold.
In Figure~\ref{fig:gut} (taken from~\cite{Martin:1997ns}),
the three couplings intersect at at the GUT scale ($M_U\approx 10^{16}$~GeV).
Inspired by flavor neutral gravity, it is assumed that the gaugino mass parameters are also
equal at the GUT scale. As a consequence
\Mone{}, \Mtwo{} and \Mthree{} are reduced to one
parameter at the GUT scale: \mOneHalf{}. The following
relationship holds at any scale:
\begin{equation}
\frac{\Mone}{g_1^2}= \frac{\Mtwo}{g_2^2} = \frac{\Mthree}{g_3^2} = \frac{\mOneHalf{}}{g_\mathrm{U}^2}
\end{equation}
$g_U$ is the coupling at the unification scale, $g_3^2$ is related 
to the strong coupling constant, and the couplings $g_1$, $g_2$
are related to the usual electroweak constants by $g_2=g$ and
$g_1=\sqrt{5/3}g'$
\par
Additionally one assumes that 
the soft supersymmetry breaking terms for the sfermions are also
universal at the GUT scale ($\mathrm{m_0}$).
This, however, does not imply that all sfermions have equal
mass at the electroweak scale. Radiative corrections for left-handed
sfermions are different from right-handed sfermions, since their 
couplings to gauge bosons are different.
The solutions of the renormalization group equations for the
masses of the right-handed slepton, the left-handed slepton
and the sneutrino at the electroweak scale take on the following form:
\begin{equation}
\begin{array}{lcl}
\mathrm{m}_{\tilde{\ell}_\RIGHT}^2&=&\mZero^2+0.22\Mtwo^2-
\swtwo\mZ^2\cos 2\beta+\mathrm{m_\ell^2}\\[10pt]
\mathrm{m}_{\tilde{\ell}_\LEFT}^2&=&\mZero^2+0.75\Mtwo^2-0.5(1-2\swtwo)
\mZ^2\cos 2\beta+\mathrm{m_\ell^2}\\[10pt]
\mathrm{m}_{\tilde{\nu}}^2&=&\mZero^2+0.75\Mtwo^2+0.5\mZ^2\cos 2\beta\\
\end{array}
\end{equation}
\par
A universal trilinear coupling is assumed
at the GUT scale and 
one requires that the electroweak symmetry breaking is driven by
supersymmetry~\cite{Ibanez:1992rk}. This
determines the supersymmetric Higgs
parameter $\mu$ up to its sign.
\par
Thus in the supersymmetric sector the following parameters remain:
the ratio of the Higgs vacuum expectation values defined at the electroweak scale ($\tan\beta$),
the sign of $\mu$ in the Higgs sector, the universal gaugino 
breaking parameter ($\mOneHalf$), the universal sfermion breaking parameter ($\mZero$)
and the trilinear term ($\mathrm{A_0}$). The last three
parameters are defined at the GUT scale. Additionally, as before for the MSSM, the 
Standard Model parameters have to be specified.

\section{Benchmark points}

\begin{table}[htb]
\begin{center}
\begin{tabular}{lcccc}
\hline
            & SPS1a & SPS1a$'$ & SU3  & LM1 \\
\hline
$m_0$       & 100   & 70     & 100  & 60  \\
$m_{1/2}$   & 250   & 250    & 300  & 250 \\
$\tan\beta$ & 10    &  10    & 6    & 10 \\
$A_0$       & -100  & -300   & -300 &  0 \\
\hline
\end{tabular}
\caption{Summary of mSUGRA parameter sets with similar collider 
phenomenology. $\mu$ is positive for all cases.}
\end{center}
\label{tab:msugraPoints}
\end{table}
Given the number of parameters, 
supersymmetry can lead to a wealth of different signatures at colliders. 
To enable comparisons between experiments and/or colliders,
sets of parameters are defined to represent 
typical signatures.  Most commonly known is the point Snowmass Points and Slopes 
(SPS)~1a~\cite{Allanach:2002nj} which has been
studied in detail in the LHC-ILC report~\cite{Weiglein:2004hn}. Points with similar phenomenology
have been studied in ATLAS (SU3) and CMS (LM1~\cite{Ball:2007zza}). A
summary of the points is shown in Table~\ref{tab:msugraPoints}.
Common to all points are relatively light supersymmetric particles, thus
an early discovery at the LHC would be possible. 

\begin{figure}[htb]
\begin{minipage}[htb]{0.48\textwidth}
\vspace{0cm}
\centering
\includegraphics[width=\textwidth,bbllx=0,bblly=206,bburx=611,bbury=727]{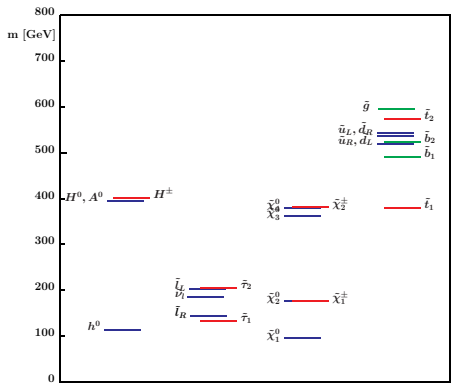}
\caption{The particle spectrum of the SPS1a point is shown.} 
\label{fig:SPS1a}
\end{minipage}
\hspace{0.3cm}
\begin{minipage}[htb]{0.48\textwidth}
\vspace{0cm}
\centering
\end{minipage}
\end{figure}
In the following SPS1a
will be studied as an example. The resulting 
spectrum of (s)particles is shown in Figure~\ref{fig:SPS1a}.
The parameters of this point 
lead to first and second generation squarks with masses of the order of 500~\gevcc{}.
The gluino has a mass of about 600~\gevcc{}. The lightest right--handed sleptons
have a mass of~150~\gevcc{}, their left--handed counter parts have masses of 200~\gevcc{}.
The lightest Higgs boson mass is at the edge of the LEP bound.
There is a group of light neutralinos and chargino at about 200~\gevcc{} and below as well
as a group of heavy chargino and neutralinos. The right--handed squarks 
decay predominantly directly to their standard model partner and the lightest neutralino. 
The left--handed squarks have a long decay chain via the second lightest neutralino and 
a right--handed slepton open. As the lightest stau (with a left--handed component) 
is lighter than the second--lightest neutralino and the lightest chargino, the chargino
decays with a branching ratio of almost 100\% to stau and neutrino. The neutralino
decay to stau and tau has a branching ratio of~90\%.

\cleardoublepage

\chapter{Colliders and Experiments}
\label{chap:exp}

The Large Hadron Collider (LHC) at CERN has started operations on September 10, 2008. 
The LHC will collide proton beams with a center--of--mass energy of up to
14~TeV. Two multi-purpose detectors, ATLAS and CMS, are located on the LHC ring. 
The ILC, an \epem{} collider with
a center--of--mass energy of 500~GeV, extendable to 1~TeV is awaiting approval
pending the results of the LHC. The ILC is expected to start operations eight 
years after approval.

In this chapter the LHC is briefly described. The ATLAS detector is discussed with emphasis
on its performance, especially for electron and photon reconstruction. 
A hardware aspect of the readout of the Liquid Argon calorimeters
is analyzed in more detail. The ILC and an associated detector concept is discussed last.

\section{LHC}

The LHC is described in detail in Ref.~\cite{Evans:2008zzb}. The nominal mode of 
operation of the LHC is to provide a 
luminosity of $10^{34}\mathrm{cm}^{-2}\mathrm{s}^{-1}$
at a center--of--mass energy of 14~TeV colliding proton beams.
The bunch crossing frequency will be 40~MHz. 
The beams circulate in a tunnel with a circumference of 26.7~km which is connected 
to the CERN accelerator complex via two 2.5~km long transfer lines.

The luminosity at a collision point is described by the following formula:
\begin{equation}
L=\frac{N_b^2 n_b f_{\mathrm{rev}}\gamma}{4\pi\epsilon\beta^\ast}F
\end{equation}
where $\gamma$ is the relativistic gamma factor (for protons at 7~TeV about 7000),
$\epsilon$ the normalized transverse beam emittance (3.75$\mu$m), $\beta^\ast$ the beta function 
at the interaction point (0.5~m), $f_{\mathrm{rev}}$ the revolution frequency, $N_b$ is the number 
of particles per bunch (maximum $1.1\cdot 10^{11}$) 
and $n_b$ the number of bunches per beam, nominally 2808. Note that 
$N_b$ enters as a square while $n_b$ only linearly as the luminosity is defined 
per interaction point. Bunches are separated in the beam pipe by 7.5~m. 
To minimize beam collisions in the 130~m common beam pipe in the interaction regions
(corresponding to about 17 unwanted collisions for each bunch of a beam), 
a small crossing of 150--200~$\mu$rad will be used. 
$F$ is the geometrical reduction factor due to the non--zero cross angle. 
The size of a bunch is expected to be about 15~$\mu$m in the direction transverse to the beam axis
and about 7~mm in the direction of the beam axis (z--direction).

The 
LHC tunnel consists of eight straight sections and eight arcs with a smaller radius, the 
LHC magnets are built to reach a strength of up to 9~T with 8.33~T being the nominal 
operating field for a beam momentum of 7~\tevc{}. 
A total of 1232 superconducting 
dipoles have been built in industry. Adding in quadrupole and other beam equipment, the LHC
consists of about 4000~elements. 

The dipoles are operated at a temperature of 1.9~K. This low temperature is achieved in a two stage 
process, first by cooling down to about 90~K with liquid Nitrogen and then cooling down to the operating
temperature with super-fluid liquid Helium. The LHC has been divided into 
eight sectors. The cool-down of one sector to operating temperature 
takes about two months. The energy that will be stored in the beams will be around 362~MJ 
(60~kg of TNT) and an additional 600~MJ is stored in the magnet system. A fast system is able to
extract the beam if necessary and dilute it before reaching an absorber.

\begin{figure}[htb]
\begin{center}
\includegraphics[width=10cm]{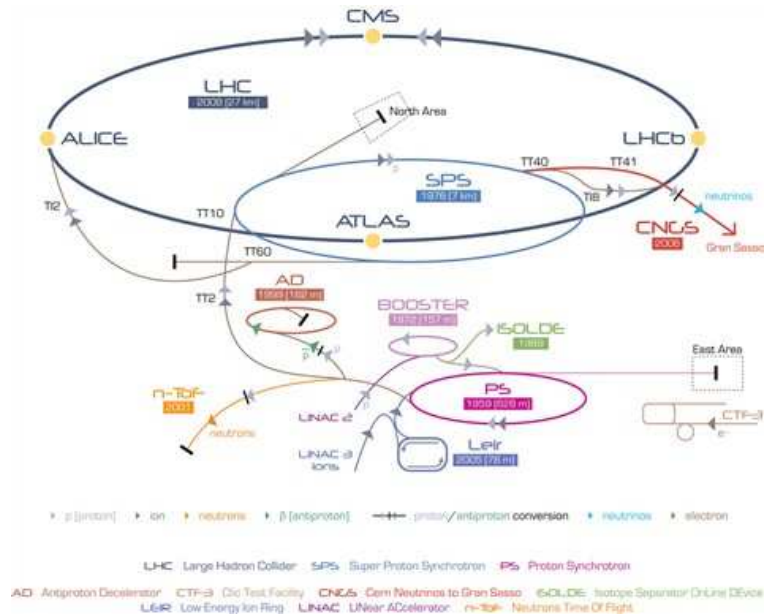}
\end{center}
\caption{\label{fig:LHCinjection} The CERN accelerator complex from the Linear accelerator to the LHC.}
\end{figure}

The injection sequence of the beams is shown in Figure~\ref{fig:LHCinjection}.
The protons are first accelerated to 50~\mevc{} in a linear accelerator. In the Proton Synchrotron
Booster (circular) they reach 1.4~\gevc{} before being transferred to the Proton Synchrotron (PS).
In the PS the protons are accelerated to 25~\gevc{} and injected into the Super Proton Synchrotron
(SPS).
The protons with an energy of 450~GeV are then transferred from the SPS to the LHC. The beam is captured 
and accelerated by a superconducting RF system. In contrast to electron beams, the synchrotron 
radiation loss for protons is small $\sim(\mathrm{m}_\mathrm{e}/\mathrm{m}_\mathrm{p})^4$.
The energy gain per turn during the ramp to 7~TeV is 485~keV.
The expected beam lifetime (until the luminosity has decreased to $1/e$)
is expected to be about 15~h. The average filling time should on average be about 7~h.

The integrated luminosity per year is expected to be 10~\fbinv{} running at $10^{33}\mathrm{m}^{-2}\mathrm{s}^{-1}$.
For the nominal luminosity of $10^{34}\mathrm{m}^{-2}\mathrm{s}^{-1}$, an integrated
luminosity of 100~\fbinv{} is expected.

Beams were injected in both rings on September 10, 2008 and circulated many times. 
The RF--system has captured the beams leading to a nearly infinite theoretical lifetime.
However, during the commissioning of a sector to the highest beam energy, a superconducting 
soldering joint turned normal conducting with a resistance of a few n$\Omega$. The small 
resistance was large enough to heat the super-fluid liquid helium turning it into its gaseous 
state. The volume expansion was so quick that the safety valves did not follow quickly enough.
The shock wave displaced several magnets. About 50 elements of the machine have been 
repaired or replaced by spares. Additional safeguards are being
put in place.

After the incident in 2008, repairs of the cryogenic as well as the magnet system are 
underway. The current startup plan aims to restart the machine in November 2009 with a maximum
center--of--mass energy of 7~TeV. The ramp up to the nominal center--of--mass energy of 14~TeV
is foreseen for later.

\section{The ATLAS Detector}

\begin{figure}[htb]
\includegraphics[width=\textwidth]{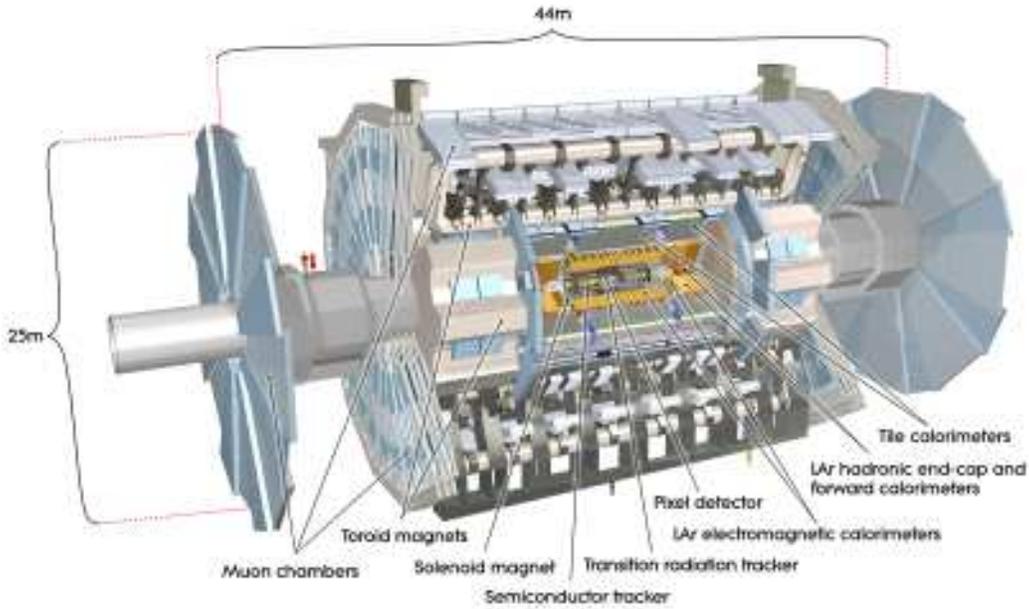}
\caption{\label{fig:ATLAS} View of the ATLAS detector.}
\end{figure}

The ATLAS detector~\cite{:2008zzm}, as shown in Figure~\ref{fig:ATLAS}, is a typical $4\pi$ detector. 
Its innermost parts consist of the 
tracking devices in a 2~T solenoidal magnetic field covering the pseudo-rapidity 
($\eta=-\log\tan\theta/2$, $\theta$ is the polar angle with respect to the beam axis)
of $|\eta|<2.5$. Electromagnetic and hadronic calorimeters
follow further away from the interaction point covering $\eta$ down to 4.9. 
The muon detectors are further out in toroidal magnetic field covering $|\eta|$ up to 2.4. 

At the LHC the total inelastic cross section for proton--proton interactions is 80~mb. At nominal
luminosity every event will be accompanied by an average of 23 minimum bias events. The LHC 
will collide protons every 25~ns in the center of ATLAS. The high cross section necessitates 
a fast readout of the detector as well as radiation hard or tolerant electronics for components
on the detector. 

The tracking detectors are immersed in a solenoidal field of 2~T. It is expected that 1000 particles
will come from the collision point with a frequency of 40~MHz within a pseudo-rapidity of 2.5, thus
the track density will be large.
The innermost tracking device is the Pixel detector, which is segmented in R--$\phi$ and z. Each track 
crosses three pixel layers, the innermost is at a radius of 5~cm. 
The intrinsic accuracy in the barrel is 10~$\mu$m in R--$\phi$
and 115~$\mu$m in z from 80.4~million readout channels.

The Pixel detector is followed by the SCT, consisting of silicon strips with a slightly coarser 
granularity. The intrinsic accuracy in the barrel is 17~$\mu$m in R--$\phi$ and 580~$\mu$m in z.
There are 6.3~million readout channels for the SCT.

The TRT provides more points for tracking than the Pixel and SCT, however each point has less 
precision. 
On average 36~points are provided for tracking with R--$\phi$ information
with an intrinsic accuracy of 130$\mu$m per straw up to $|\eta|<2$. Yet 
no information is available in the z~direction. The TRT has 351000 channels.
In addition to its tracking capabiities the TRT also provides particle 
identification. The TRT measures the transition radiation which is expected
to yield a higher signal for electrons than for pions. This information
is used for the identification of electrons described later in this Chapter.

The resolution for the tracking detectors has been determined in a realistic simulation 
of the ATLAS detector. Using the parametrization $\sigma_X = \sigma_X(\inf)(1\oplus p_X/p_T)$,
where $\sigma_X(\inf)$ is the asymptotic resolution at infinite momentum 
and $p_X$ is a constant for which the intrinsic and multiple--scattering terms
are equal. The resolution in the 
barrel in simulation correspond to $\sigma_X(\inf)=0.34$~TeV$^{-1}$ and $p_X=44$~GeV.
For a transverse momentum of 100~\gevc{}, the relative transverse momentum resolution
$p_T\cdot\sigma(1/p_T)$ is about 0.04.

\begin{figure}[htb]
\includegraphics[width=\textwidth]{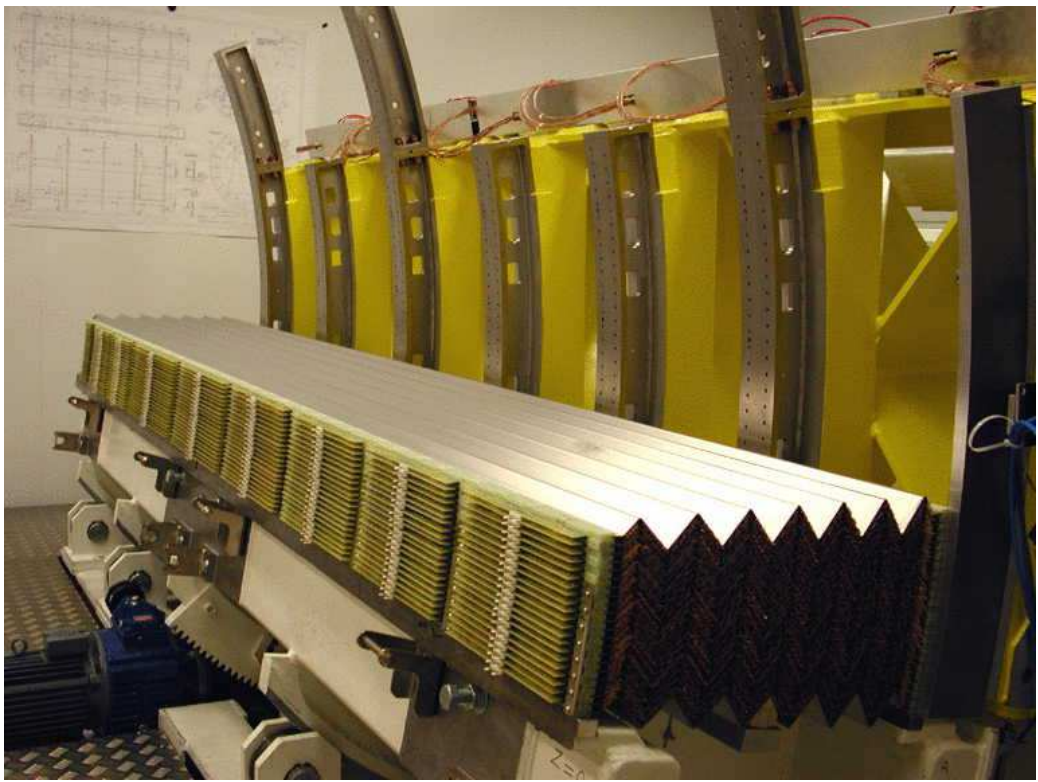}
\caption{\label{fig:Calo} View of a module of the accordion calorimeter of the ATLAS barrel during
assembly.}
\end{figure}
The electromagnetic calorimeter of ATLAS is a sampling calorimeter. The passive material is 
lead (and stainless steel for the mechanical stability). The active material is Liquid Argon,
operating at a temperature of about 90~K. The particularity of the design is the accordion 
structure as shown in Figure~\ref{fig:Calo} 
ensuring a coverage in $\phi$ without dead zones. The depth of the calorimeter 
varies roughly, as function of $\eta$, from 25~$X_0$ to 30~$X_0$. 
An electrical field of 2~kV is applied in the 2~mm Liquid Argon gap.

The calorimeter has three segments in depth providing $\eta$ coverage up to 3.2 in addition
to a presampler (PS) covering $|\eta|<1.7$. 
The first segment, the strips, typically has a cell size of $\Delta\eta = 0.003$,
coarser in $\phi$ with a size of 0.1. The fine granularity allows to separate efficiently photons from
$\pi^0$. The main part of an electromagnetic shower is deposited in middle section where the typical 
readout cell size in $\eta$ and $\phi$ direction is $0.025\times0.025$. The thin back section
has the same cell size in $\phi$ as the middle and is coarser in $\eta$ with a cell size of 0.05.
The expected energy resolution for electrons and photons is 10\%/$\sqrt{E}$ and a global constant
term of $0.7$~\% must be achieved. The linearity is required to be of the order of a
per mil or better to measure precisely the absolute masses of particles.


The hadronic and forward calorimeters in the endcap are also based on Liquid Argon technology.
The absorber of the hadronic calorimeter is made of copper. The calorimeter has four segments in depth. 
Their size in $\Delta\eta\times\Delta\phi$ is $0.1\times0.1$ in the region of $|\eta|<2.5$ and $0.1\times0.2$
at larger $\eta$ for 
a total of 5632 readout channels.
The forward calorimeter (FCAL) consists of three high--density modules. The first one, for 
electromagnetic measurements, is made of copper, the other two are made of tungsten. Concentric rods 
and tubes are installed parallel to the beam axis. The gap between rod and tube, filled with Liquid Argon
as the sensitive material can be as small as 0.25~mm. This ensures a fast signal in the forward region
where the pileup effects are expected to be stronger.

The tile calorimeter is situated behind the electromagnetic calorimeter in the barrel. 
Scintillating tiles provide the signal of the sampling calorimeter while steel is used as absorber.
Three layers are provided in depth. The readout cells are of size $0.1\times0.1$ in the first two layers
and $0.1\times0.2$ in the last layer. Measured in interaction lengths, the tile calorimeter is almost 10$\lambda$
deep at $\eta=0$.

The muon system provides coverage for $|\eta|<2.7$. The muons are deflected in large air core toroids.
1800 hall sensor monitor the magnetic field. Monitored drift tubes are used over most of the $\eta$ range to provide
information for the track reconstruction, cathode strip chambers are used in the endcap due to the higher occupancy. 
Resistive plate chambers (barrel) and thin gap chambers (endcap) 
are used for triggering.

ATLAS uses a three stage trigger system to reduce the rate from 40~MHz in level--1 to 75~kHz and further down
to 200~Hz in the two high level trigger systems using the full detector information. 

\subsection{Electromagnetic Test beam}

A vigorous test beam program has been pursued by ATLAS. Components of all sub-detectors have been
tested. A pre-series module of the electromagnetic calorimeter was tested 
in the test beam with electron beams of fixed energy from 10~GeV to 245~GeV.  
Four barrel modules, corresponding
to 12.5\% of the barrel calorimeter 
and three endcap modules (19\% of the endcap calorimeters) which are now installed in ATLAS have also been 
tested. 

The main results of this test beam campaign of nearly five years (1998--2002) have been reported in 
Refs.~\cite{Aharrouche:2008zz,Aharrouche:2007nk,Aharrouche:2006nf,Aubert:2005dh,Colas:2005jn,Aubert:2002mw,Aubert:2002dm}.
The uniformity, i.e., the variation of the calorimeter response for a fixed energy of 245~GeV, 
in a barrel module of effective size $\Delta\eta\times\Delta\phi=1.2\times0.2$ is better than 0.5\%. 
The uniformity of the calorimeter is especially important for the measurement of thin resonances, e.g.,
the Higgs boson decaying to two photons, 
or sharp endpoints (the lepton--lepton invariant mass in a supersymmetric decay chain). 
If the intrinsic width is smaller or about equal to the uniformity, the 
width of the distribution, e.g., of the invariant photon--photon mass, will be dominated by 
the detector effects. The result obtained is in excellent agreement with the requirements
set forth before the construction of ATLAS started: uniformity in a region of 
$0.2\times 0.4$ better than 0.5\%. The number shows the excellent control of the long and difficult
production process of all components of the electromagnetic calorimeter.

In addition to the uniformity the energy resolution with its sampling term and local constant term 
contributes to the width of the measured invariant masses and edges. In the test beam 
the energy resolution was measured to be (at a fixed $\eta$ position) better than 10\%$/\sqrt{E}$ 
and the local constant term to be better than 0.3\%. The performance expected for
the electromagnetic calorimeter of ATLAS has been achieved.

While uniformity and energy resolution determine the width, the energy scale (or central value) needs to be 
known precisely. Therefore the linearity of the calorimeter is very important. 
The linearity was measured to be of the order of 2~per mil in a range from 20~GeV to 180~GeV in a special test beam
setup. 

The test beam has been tremendously important to develop the methods for reconstruction of electrons 
and photons. However the results obtained with the modules now installed in the ATLAS detector cannot
be transferred directly from the test beam to ATLAS, because the material distribution in the test beam is 
quite different from the one in situ in ATLAS. The strategy chosen was to develop the methods 
on the test beam, optimize the test beam Monte Carlo, heavily used for the linearity study, and apply the 
methods to the simulation of the full ATLAS detector. The Monte Carlo provides the (indirect) link
between the test beam and the (future) ATLAS data.

In order to have a reliable measurement from the electromagnetic calorimeter it is also 
necessary to monitor every aspect (stability of the electronics system etc). 
The dependence of the energy response as function of the liquid Argon temperature
has been measured to be 2\%/K for 
a fixed beam energy of 245~GeV. The temperature stability in the test beam 
cryostat was shown to be better than 7~mK.
The Front End electronics, in particular the Front End boards
(FEBs) where the signal is amplified, formed and digitized, were tested and 
their design was validated.

\subsection{The Front End Board of the LAr Calorimeters}

The Front End Electronics of ATLAS consist of several components. Here, as an example
the Front End Board (FEB), described in detail in Ref.~\cite{Buchanan:2008zza} is discussed. 
The FEB is located at large radius on the cryostat in the transition region
between barrel and endcap. 
It amplifies the signal of the Liquid Argon detectors of ATLAS, with the 
exception of the HEC, where the amplifiers are located inside the cryostat.

After preamplification the FEB forms the triangular ionization signal 
into a bipolar shaper where the peak of the signal
is reached typically after 50~ns. This corresponds to only 2 bunch crossings, whereas the ionization
signal has a duration of approximately 450~ns. Additionally the bipolar shaping allows
to treat the minimum bias/pileup events, which are typically at low transverse momentum, as an additional
noise contribution to the resolution of the electromagnetic calorimeter.
The shaper also serves as a first step summing
device for the trigger. The analog signal is then sampled at 40~MHz and saved in a 
switch capacitor array (SCA) for about 1.5~$\mu$s until the first level trigger
decision is received. Upon a trigger, the signal is digitized and sent to the Read Out Device (ROD)
via an optical link.

The FEB treats 128~readout channels of the calorimeter. 1627 FEBs were produced and configured
in 19 different flavors. The production, assembly and test was shared among three labs: 
NEVIS (University of Columbia),
Brookhaven National Laboratory (BNL) and LAL. While NEVIS was responsible for the production
and the first digital tests, LAL and BNL each configured, tested and assembled 50\% of the FEBs.
The operation lasted for almost a year. Three engineers, three technicians and a physicist were
involved in the development, maintenance, improvement and functioning of the test setup up 
at LAL alone.
This allocation does not cover the 10 years of R\&D of the FEB components in radiation 
tolerant technology (see~\cite{Buchanan:2008zza} for more details).

\begin{figure}[htb]
\begin{minipage}[htb]{0.48\textwidth}
\includegraphics[width=\textwidth]{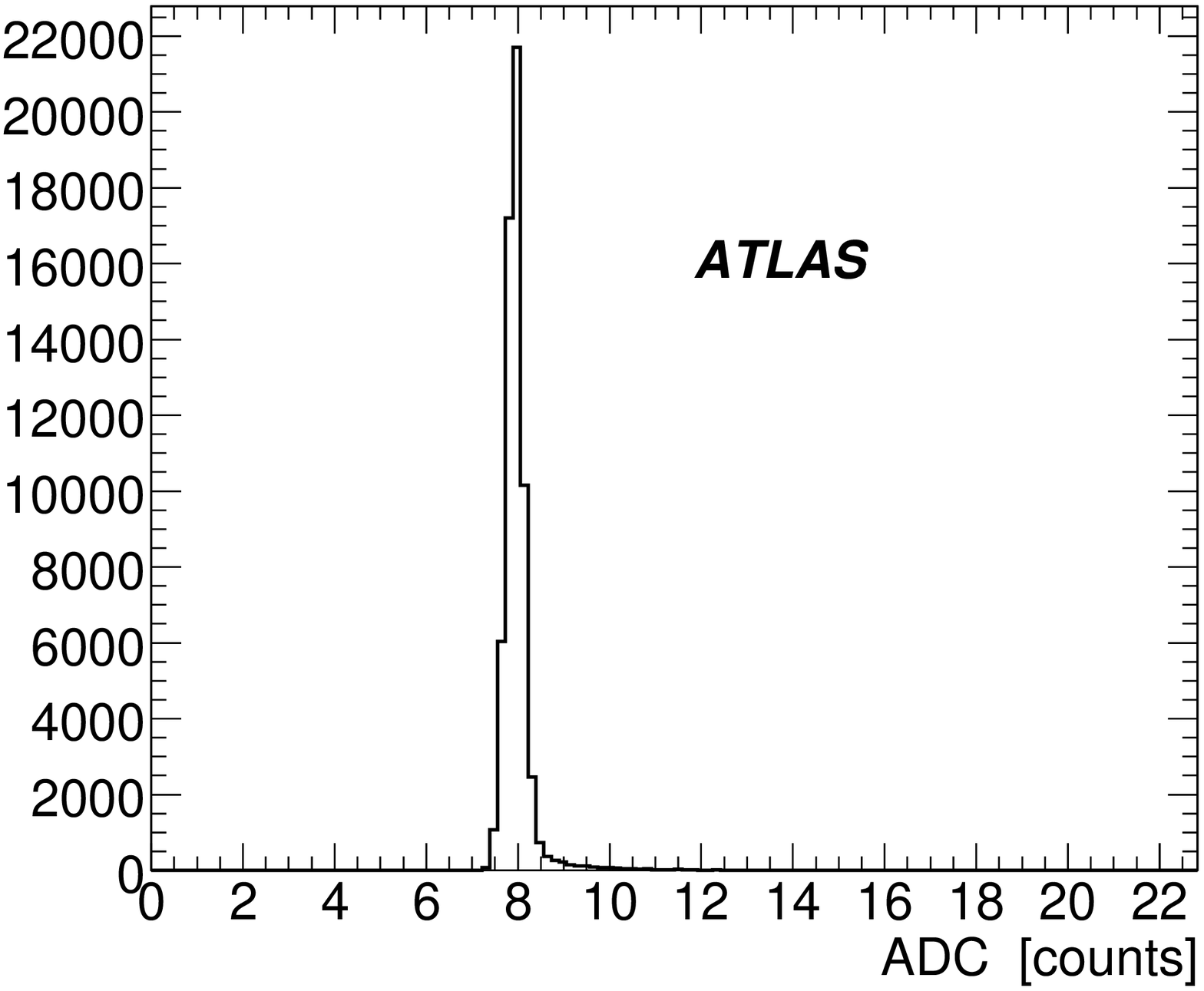}
\vspace{-1.2cm}
\caption{Distribution of the noise RMS in ADC counts of the High Gain for a flavor of the FEB (from~\cite{Buchanan:2008zza}).} 
\label{fig:FEBnoise}
\end{minipage}
\hspace{0.3cm}
\begin{minipage}[htb]{0.48\textwidth}
\vspace{1.2cm}
\includegraphics[width=\textwidth]{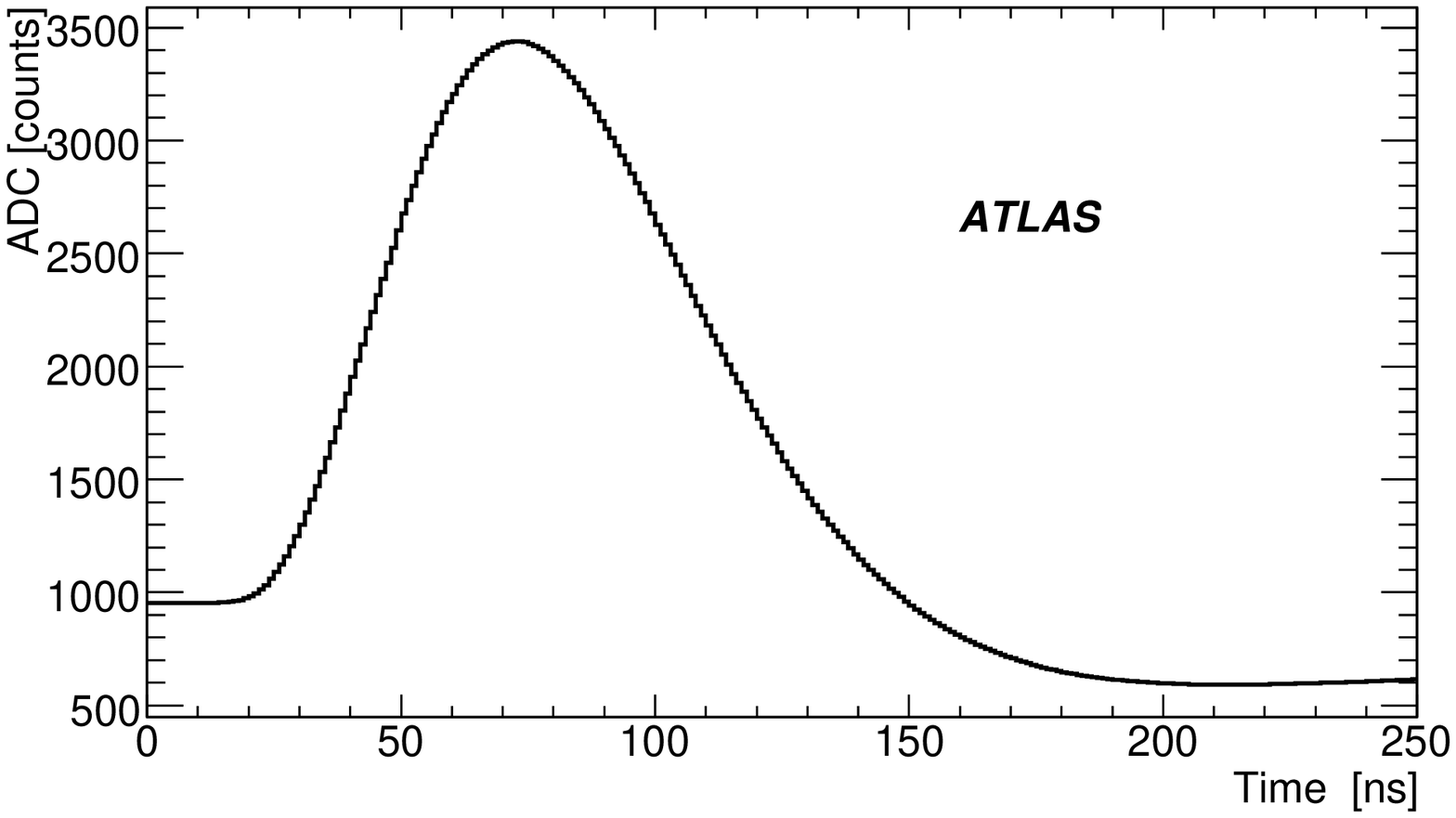}
\vspace{-1.1cm}
\caption{Reconstructed signal of the FEB in medium gain as function of the time (from~\cite{Buchanan:2008zza}).} 
\label{fig:FEBshape}
\end{minipage}
\end{figure}

At LAL, the FEBs were tested one by one in specially developed test stand.
The shaper output is in three gains. First the pedestals and the
electronic noise were measured. The distribution of the 
noise RMS of the high gain channel is shown in Figure~\ref{fig:FEBnoise}
for all FEBs of a flavor. This test allowed to detect dead channels.

To test the signal integrity, a calibration pulse was sent through the electronics chain
and the signal was recorded in steps of 1~ns with interleaved events as shown in  
in Figure~\ref{fig:FEBshape}. The analysis of the signal form yielded in particular the time 
of the signal peak. The peak was required to fall into an acceptance window 7~ns long. The minimum
and maximum of the time window were fixed to be the same for all FEBs of the same preamplifier flavor.
A wrong time constant of the shaper, usually observed as a signal
reaching its maximal amplitude before the minimum time required, was detected this way.

\begin{figure}[htb]
\begin{minipage}[htb]{0.48\textwidth}
\includegraphics[width=\textwidth]{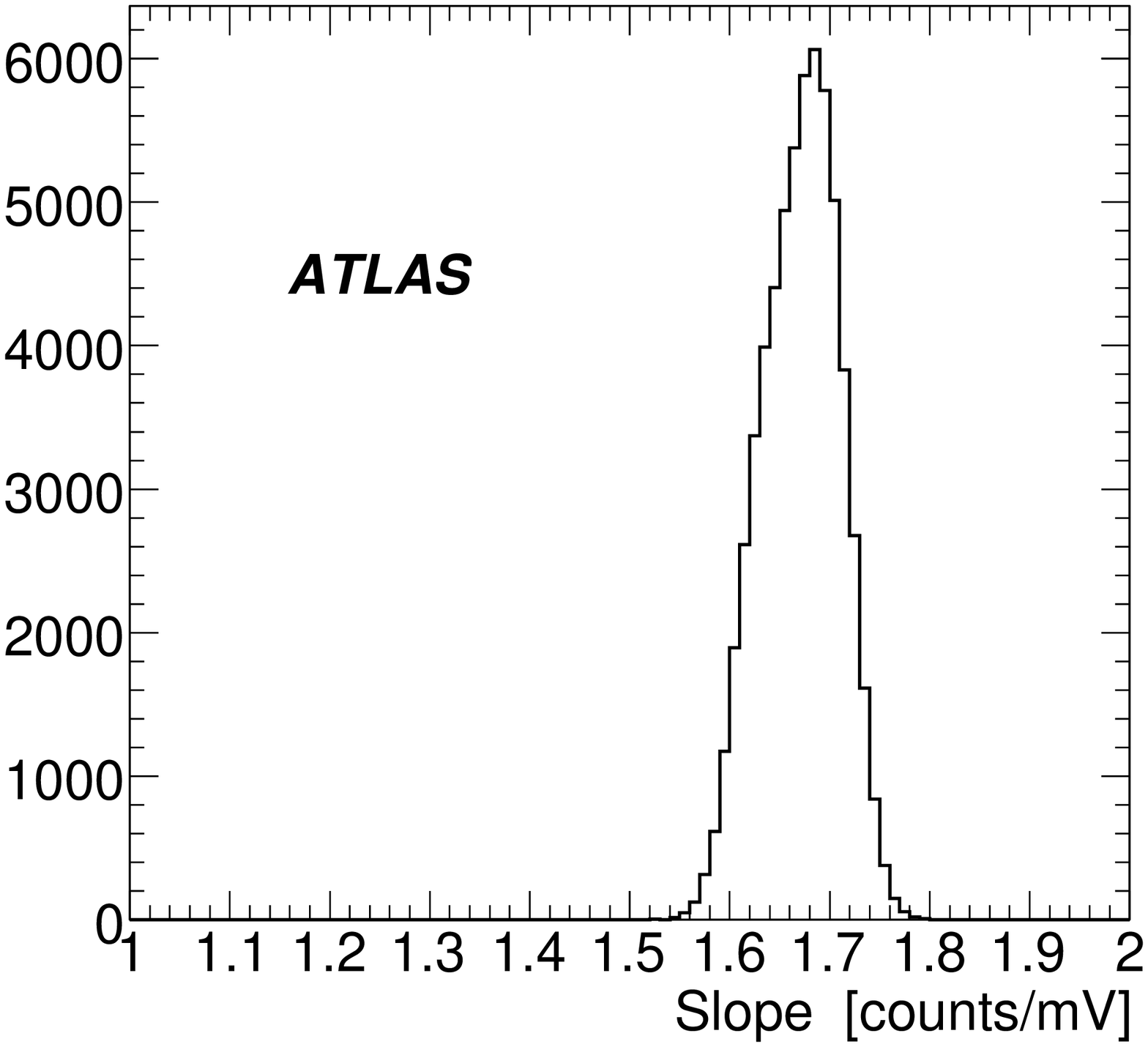}
\vspace{-1.2cm}
\caption{Distribution of the Low Gain slope from a linear fit for a FEB flavor (from~\cite{Buchanan:2008zza}).}
\label{fig:FEBslope}
\end{minipage}
\hspace{0.3cm}
\begin{minipage}[htb]{0.48\textwidth}
\vspace{2.5cm}
\includegraphics[width=\textwidth]{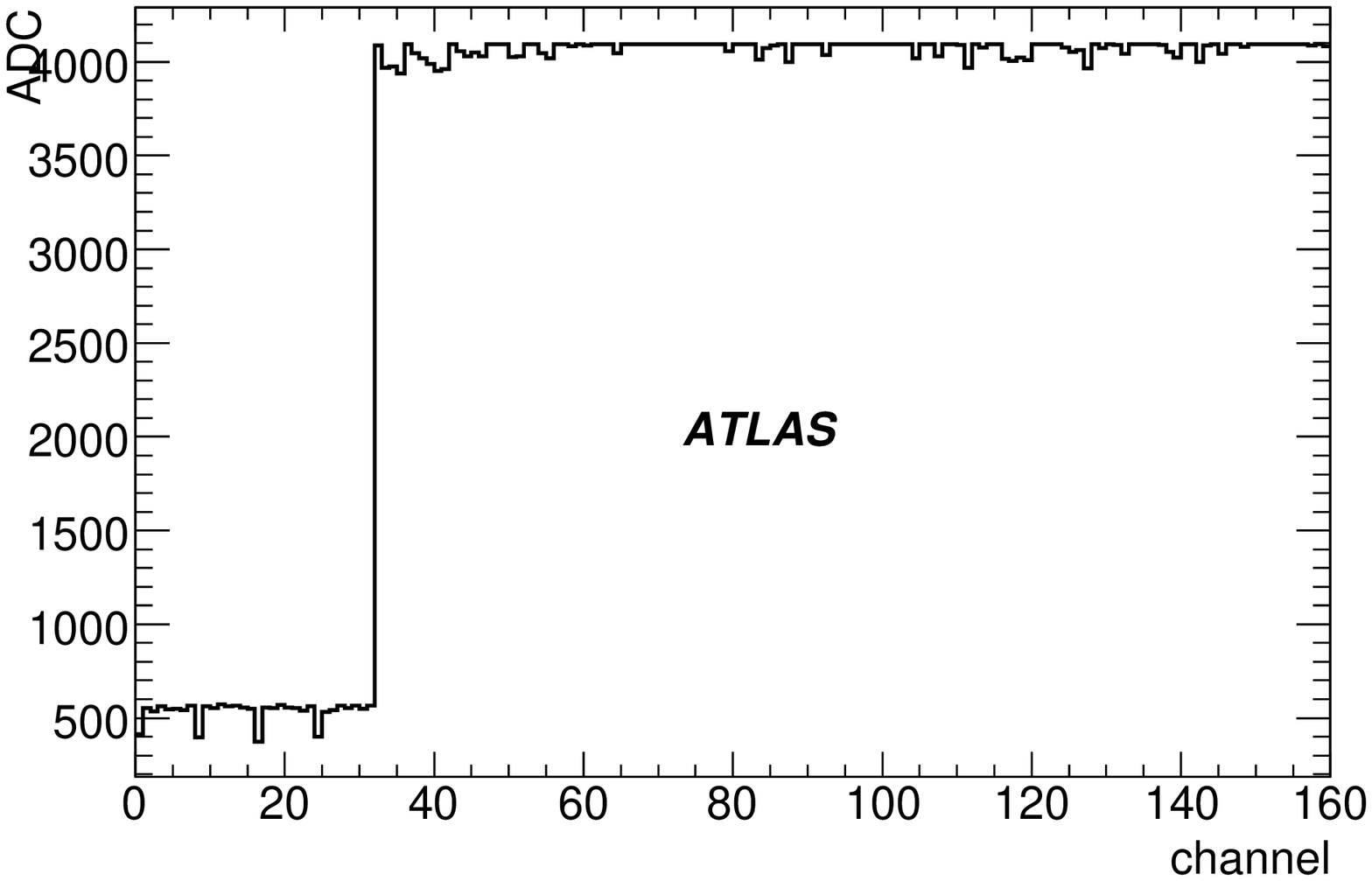}
\vspace{-0.9cm}
\caption{Test of the FEB trigger output. No signal is expected for the first 32 entries and for the next 128 
entries a large signal should be observed (from~\cite{Buchanan:2008zza}).}
\label{Fig:FEBTrigger} 
\end{minipage}
\end{figure}

The distribution of the measured linear part of the gain is shown in Figure~\ref{fig:FEBslope}.
The slope is sensitive for the detection of  
dead channels, accelerated shaper timing and channels with linearity problems.
In normal running in ATLAS only one of the three gains is read out. The gain choice
is made by a dedicated chip (GSEL) on the FEB. The integrity of the choice, corresponding
in practice to the setting of two thresholds in the medium, was verified.

The FEB provides the signals for the trigger level--1. In order to avoid excessive 
triggering due to noisy channels, individual calorimeter readout cells can be turned off.
Figure~\ref{Fig:FEBTrigger} shows the test for one FEB as example. This test also 
checks the correct signal routing of the trigger signals on the FEB.
Additionally the gains as well as the shapes of the trigger outputs were measured and compared 
to references as well.

On the 786 boards configured and tested at LAL with roughly 75000 shapers, SCAs and Preamplifiers,
801 repairs had to be performed. Part of the repairs (faulty chip replacement) was performed 
in industry, another part in house.

The FEBs are cooled with water circulating in plates which are mounted individually on each
FEB. To test that the system has no leaks, 
an air pressure test at 3~bar was performed. The water system in the experiment is an under-pressure
system which ensures that, in case of a leak, air would enter the system, but no water 
would flow out.
The FEBs were shipped to CERN after a further functional
verification. 

Since the completion of this operations the test bench has been 
kept functional. It was used again in 2008/2009 for a small production
of about 40~additional boards. After more than a year of operation on the detector several 
boards were malfunctioning due to corrosion. The corrosion was traced to an error in the cleaning 
procedure after chip replacement at the subcontractor. In addition the peaking time
of the signal was unstable over time. The FEBs were removed from the detector, cleaned 
if necessary, or replaced. The  
peaking time of the shaper was stabilized by cutting pin connections 
between the shaper and FEB board ($1627$ boards times $32$ circuits per board).

The production, assembly and test of the FEBs was a difficult endeavor. It was a successful
operation as the time delay given by the collaboration was respected. A large number 
of problems on the boards were detected and repaired before installation.

\section{Performance of the ATLAS Detector}

The first physics results expected from collision data will be essential to 
prepare the physics calibration and measure the performance in situ.
In order to evaluate the current understanding of these issues, physics studies and a test
of the analysis model, ATLAS has performed a Computing System Commissioning
(CSC) evaluation~\cite{Aad:2009wy}. 

For the CSC studies more than 25 million events were simulated, digitized und reconstructed.
The event samples comprised single particles, essentially for calibration and identification,
as well as complex physics events at the nominal center-of-mass energy of 14~TeV. 
The best known description of the detector, including the 
experience gained from the test beam studies of all sub-detectors, was used in the simulation.

\begin{figure}[htb]
\begin{minipage}[htb]{0.48\textwidth}
\centering
\includegraphics[width=\textwidth]{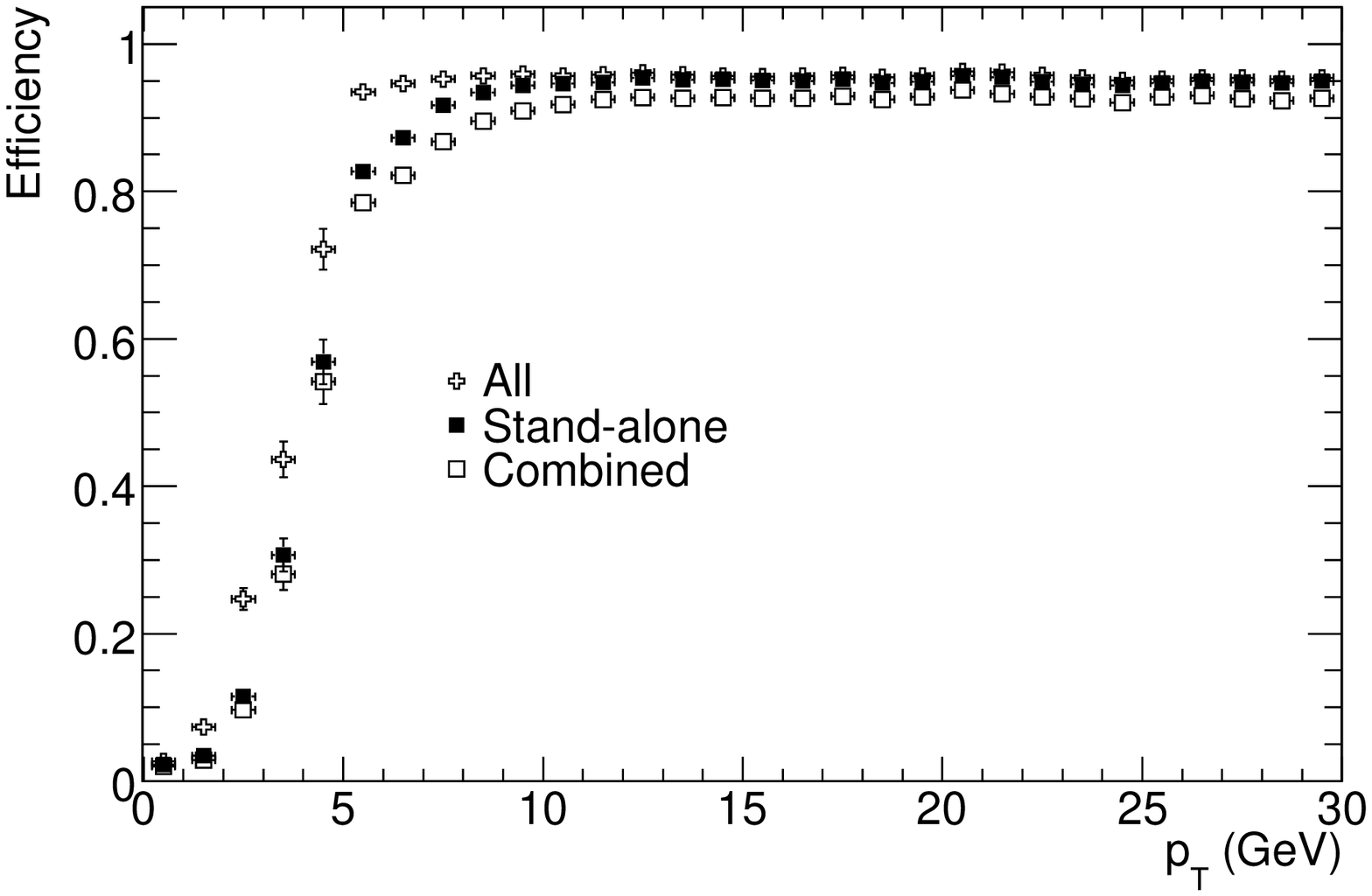}
\caption{(The efficiency for reconstructing muons is shown as function of the transverse momentum for  
the muon spectrometer reconstruction, combined reconstruction and combined with the use of segment tags
(taken from~\cite{:2008zzm}).} 
\label{fig:muonPerf}
\end{minipage}
\hspace{0.3cm}
\begin{minipage}[htb]{0.48\textwidth}
\centering
\includegraphics[width=\textwidth]{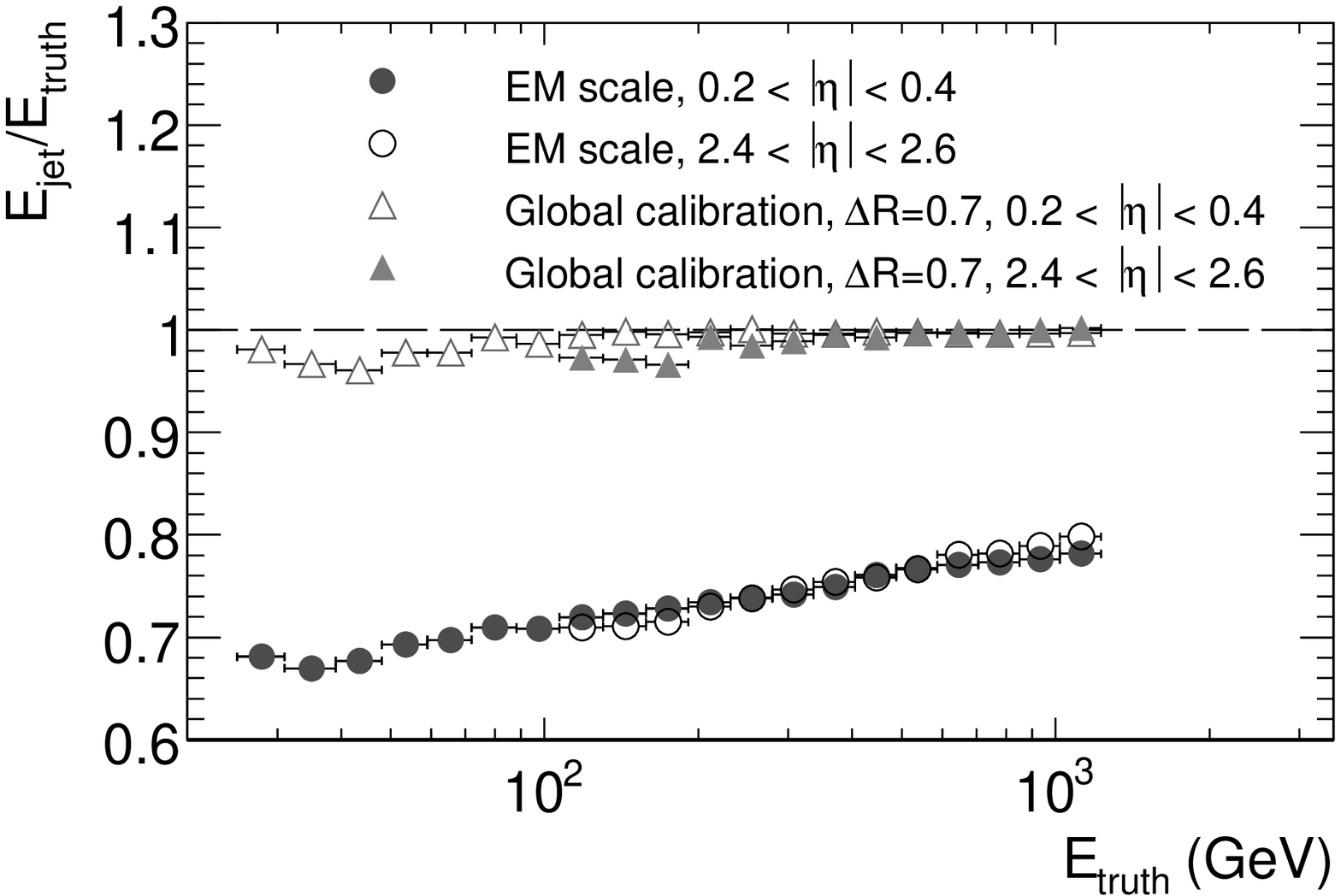}
\caption{The signal linearity of jets reconstructed with $\mathrm{R_{cone}=0.7}$, 
in two different regions of $|\eta|$ is shown as a function of $\mathrm{E_{truth}}$ for 
jets at the electromagnetic energy scale and fully calibrated jets (taken from~\cite{:2008zzm}).} 
\label{fig:jetPerf}
\end{minipage}
\end{figure}

Muons are reconstructed with an efficiency of more than 95\% for tracks with transverse momenta
of more than 5~\gevc{}. Good coverage in $\eta$ is obtained in combination with a low fake
rate as shown in Figure~\ref{fig:muonPerf}~(Left). Muons can be reconstructed in the muon 
spectrometer alone as well as in the inner detector. The combination of the two significantly
reduces the fake rate by an order of magnitude. In order not to rely on Monte Carlo for the 
determination of the efficiency, the production and decay of the Z boson will be used.
With an integrated efficiency of  100~\pbinv{}, using the well known Z boson mass,
a statistical precision of 0.1\% and a systematic error of 1\% can be obtained. 
With only half as much integrated luminosity 
the muon energy scale will be known with an accuracy of $\pm0.5$~GeV for transverse
momenta of 50~\gevc{}.

Jet reconstruction is performed in several stages. First clusters are reconstructed 
in the calorimeters and calibrated locally~\cite{Aad:2009wy}. 
Jets are then reconstructed with a jet algorithm, most frequently 
the cone and the kt algorithms though others are also under study.
Noise, pileup and the underlying event are corrected for as the last step.
The jet energy scale, as shown in Figure~\ref{fig:jetPerf}, is correct 
at the percent level. Photon plus jet events and Z boson plus jets will
be used to carry out the in situ calibration. Due to the large production
cross sections, only 10~\pbinv{} will be necessary to reach a percent precision
for low (less than 80~GeV) transverse energies.

\begin{figure}[htb]
\begin{minipage}[htb]{0.48\textwidth}
\centering
\includegraphics[width=\textwidth]{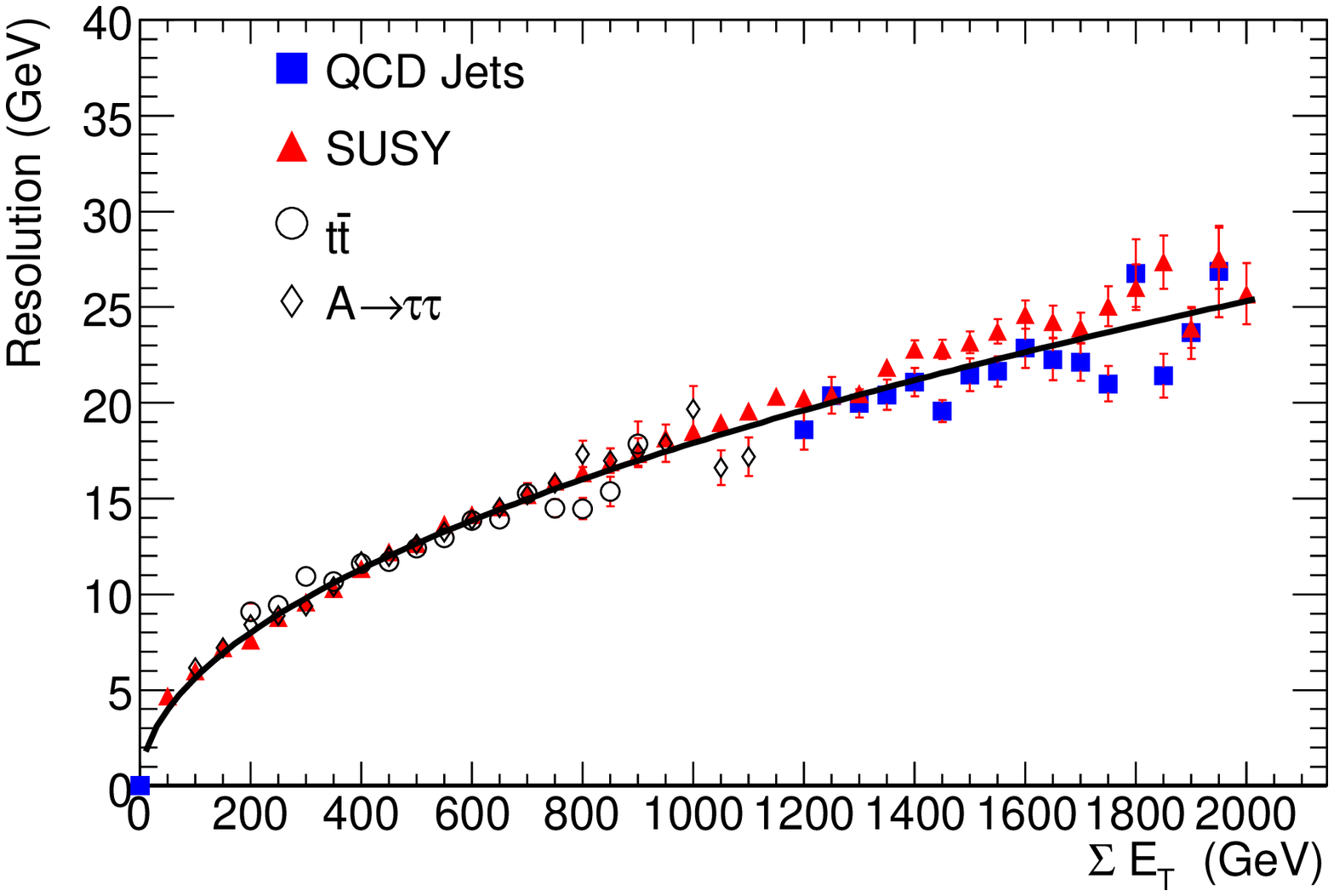}
\caption{The resolution of the $E_T^{miss}$ vector is shown for QCD di--jet events and
Higgs bosons (A) including a parametrization $0.57\sqrt{\Sigma \mathrm{E_T}}$. 
(taken from~\cite{:2008zzm}).} 
\label{fig:ETmissPerf}
\end{minipage}
\hspace{0.3cm}
\begin{minipage}[htb]{0.48\textwidth}
\centering
\includegraphics[width=\textwidth]{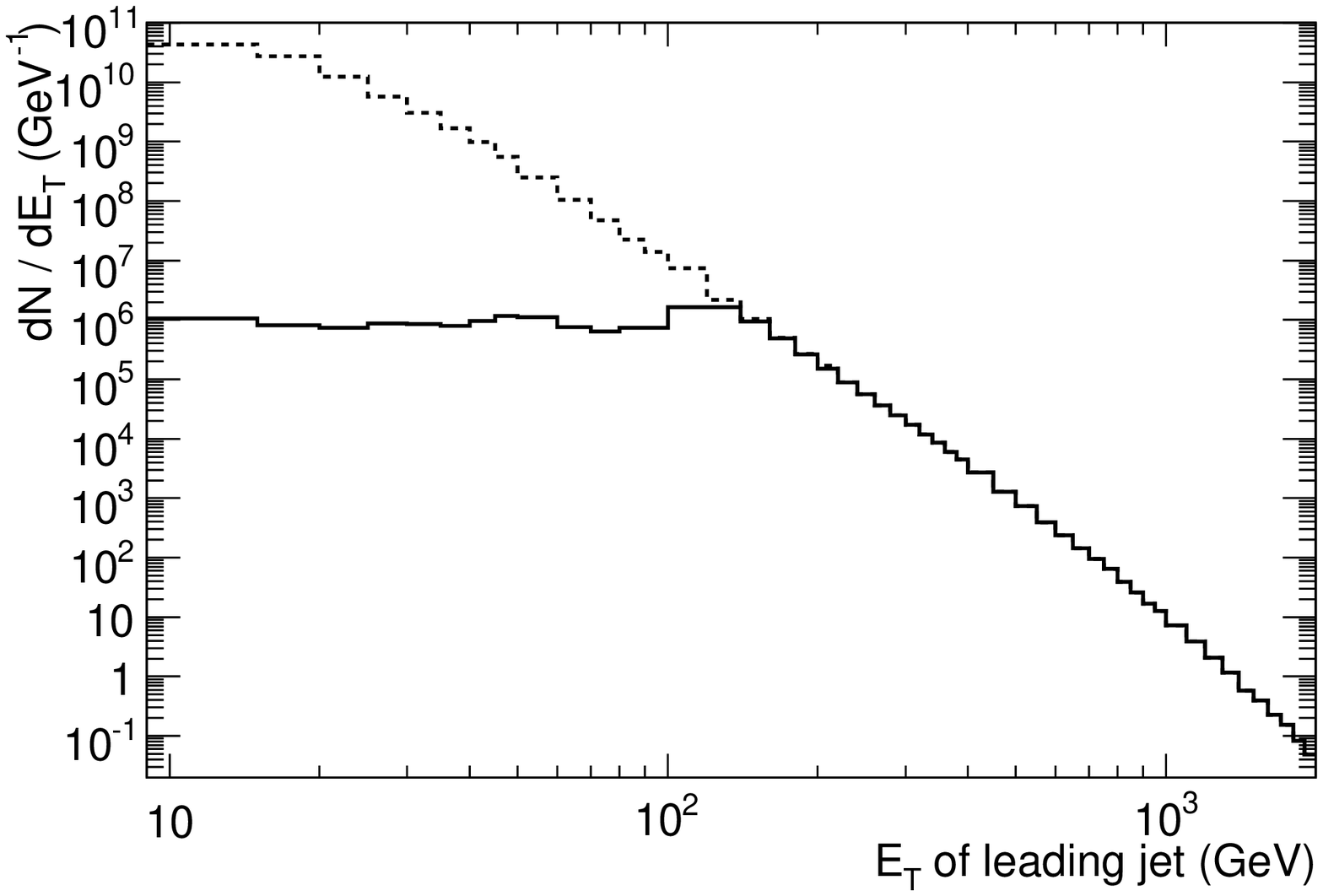}
\caption{The expected differential spectrum for single jets is shown after the level-1 trigger and 
pre-scale factors as well as without (taken from~\cite{:2008zzm}).} 
\label{fig:TriggerPerf}
\end{minipage}
\end{figure}

In the $E_T^{miss}$ reconstruction the energy deposits in the readout cells of the calorimeters
are refined by the identification of muons, electrons and jets. Not using this knowledge would lead 
to an offset of about 30\% with respect to the true missing energy. The linearity is about 5\% and 
the $E_T^{miss}$ resolution is expected to be
0.57$\cdot\sqrt{\Sigma\mathrm{E_T}}$,
as shown in Figure~\ref{fig:ETmissPerf}. The scale will be verified for low $E_T^{miss}$ 
with the Z boson decaying to taus. 

The hadronically decaying taus are either reconstructed 
by a tracking based or a calorimeter based algorithm with an efficiency of 30\% for a jet rejection
of 10$^3$. Clean samples of taus will come from the production
of the $W$ and $Z$ bosons. The visible mass reconstructed from 
a leptonically and a hadronically decaying tau will peak at about 54~\gevcc{}. This can 
then be used for the calibration of the hadronic $\tau$ decay.

For b-tagging a working point at 60\% efficiency is defined, where 
the rejection of light jets is about~30 for the simplest algorithm. 
Using the longitudinal and transverse impact parameter significance,
the rejection is increased by a factor two. Reconstruction
of the secondary vertex leads to a light jet rejection of 150.
The b--tagging performance will be calibrated in situ with top quark
pair production to a precision of 5--10\% and with muon plus jets at low 
momentum.

Trigger menus have been developed for different values of the 
instantaneous luminosity, expected to be several orders of magnitude 
below the nominal 10$^{34}\mathrm{cm}^{-2}\mathrm{s}^{-1}$ at startup.
The total rates are within budget and the turn--on curves are reasonably sharp.
Differential jet production will be measured over many orders of magnitude, as shown
in~Figure~\ref{fig:TriggerPerf}.

\subsection{Electron and Photon Reconstruction}

The reconstruction, calibration and identification of electrons and photons 
are an example of the transfer of test beam knowledge to ATLAS simulation and 
reconstruction. 

In the standard algorithm for isolated electrons and photons,
clusters are reconstructed from calibrated
readout cells. The cluster algorithm uses a fixed size, independent of the energy
of the cluster. A track is then matched to the cluster geometrically and uses
the ratio of calorimetric energy to tracker momentum. To 
qualify the candidate as electron candidate the track is required
not to be associated to the reconstructed conversions. Conversions are reconstructed
independently in the tracking system alone.
Clusters with conversion tracks
and/or clusters without associated tracks are classified as photon candidates.

For the calibration the difficulty lies in the simultaneous optimization of resolution
and linearity (energy scale). 
Two methods of electron and photon calibration have been developed to address this 
problem. The first one is a weighting technique which uses the presampler and 
the energy deposited in the calorimeter segments to derive from Monte Carlo
the correction factors to be applied. These factors correct for the energy loss upstream
of the calorimeter. The material in front of the calorimeter (inner detector, services,
mechanical structures) is several radiation lengths, leading to bremsstrahlung and 
photon conversions. The second method uses the Monte Carlo to correlate the energy lost 
with the shower depth and derive a correction (dead matter calibration).
The application of these methods leads to a linearity of the order of per mil
up to several hundred GeV. 
In addition to this longitudinal weighting, local corrections are also applied to correct
the cluster position as well as to correct for the local energy variations. 
For details see Ref.~\cite{Aad:2009wy}.

As the cross--sections for QCD jet production are several orders of magnitude larger than,
for example, the production of the $Z$ boson decaying to electrons, a high
jet rejection is necessary while maintaining high efficiency. The electron and photon
identification is based on the use of the shower shapes, calculated from the order 100 
readout cells building an electromagnetic cluster, from tracking information and 
from track-matching
information. In particular on the tracking side the TRT transition radiation measurement
is used to differentiate between charged pions and electrons. On the calorimeter side
the fine granularity of the first calorimeter segment (strips) allows to reduce
the pion (decaying to two photons) by a factor three.

\begin{figure}[htb]
\begin{minipage}[htb]{0.48\textwidth}
\includegraphics[width=\textwidth]{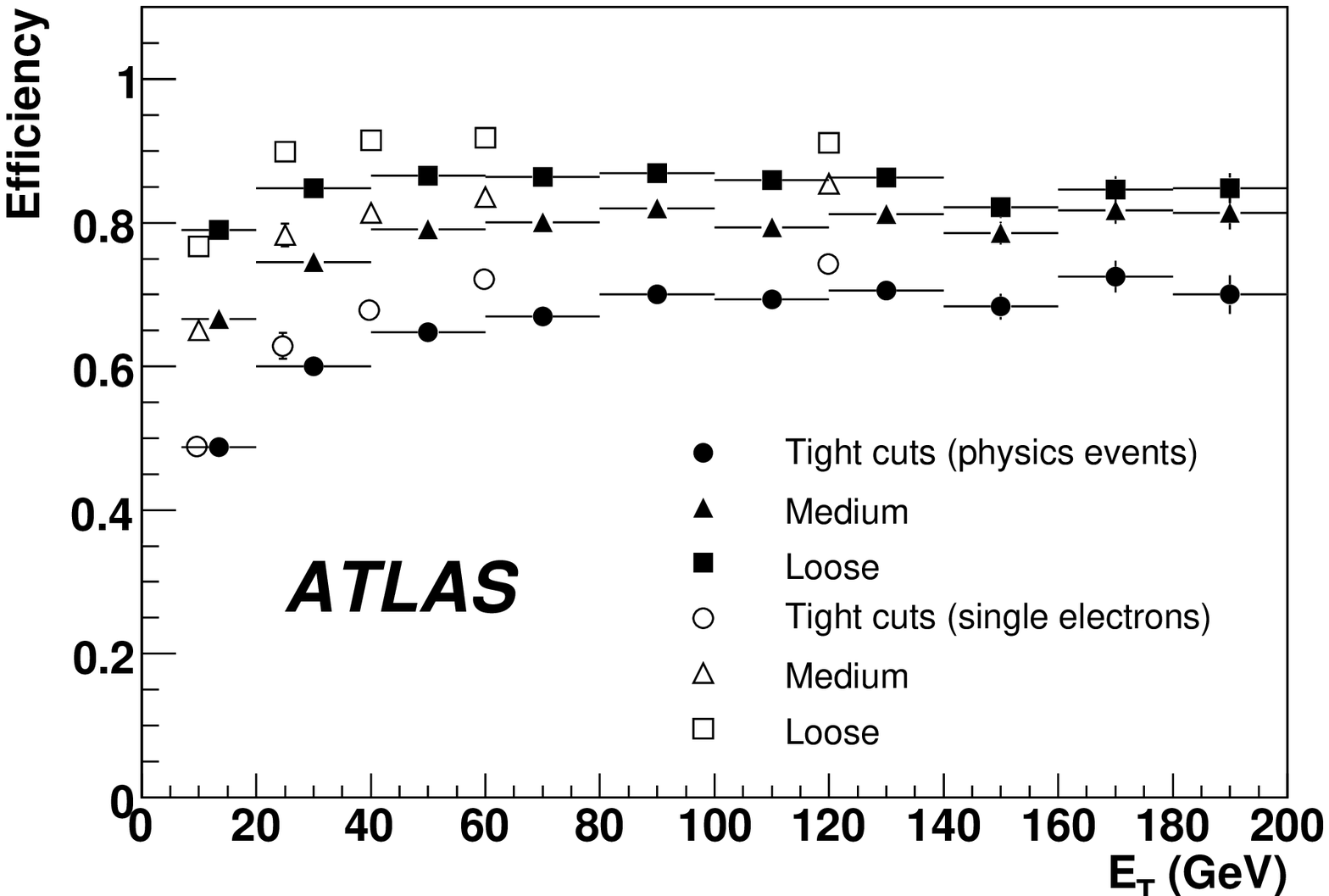}
\caption{The electron efficiency is shown as function of the transverse energy (taken from~\cite{Aad:2009wy})
for supersymmetric events as well as single particles.} 
\label{fig:ElectronID}
\end{minipage}
\hspace{0.3cm}
\begin{minipage}[htb]{0.48\textwidth}
\vspace{0.3cm}
\includegraphics[width=\textwidth]{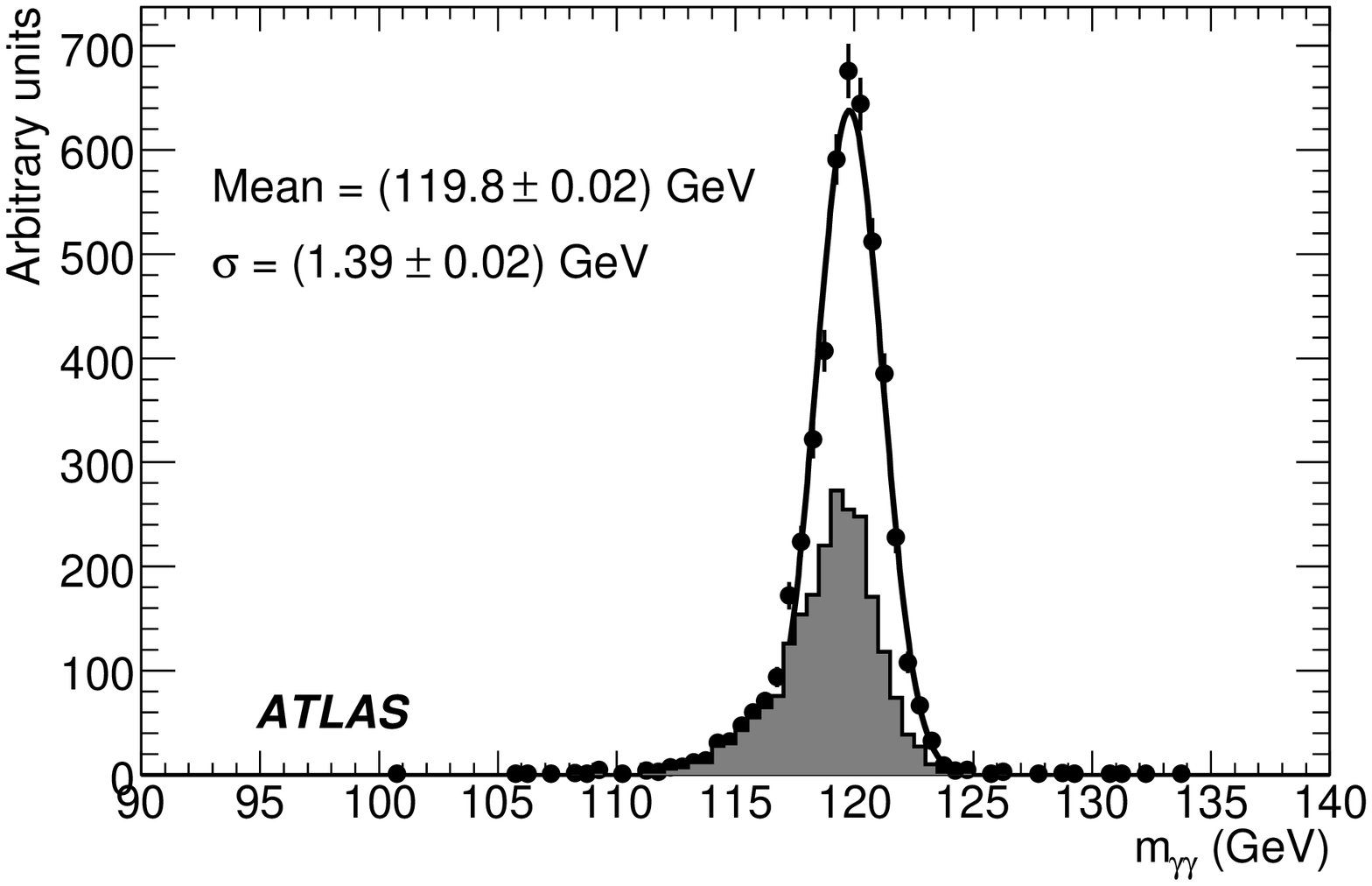}
\vspace{-0.7cm}
\caption{The reconstructed Higgs mass is shown. The shaded histogram are the events
where at least one photon has converted (taken from~\cite{Aad:2009wy}).} 
\label{fig:Hgamgam}
\end{minipage}
\end{figure}

The differential
jet cross sections drop sharply as function of the transverse energy. 
The cross section for some discovery channels of new physics might be large.
On the other hand, sensitivity to rare new physics processes with electrons and photons 
must be retained. These different needs led ATLAS to define three electron identification
qualities, one (tight) with the highest possible jet rejection of 10$^5$, a second one
(medium) with a rejection of 2200 (with higher efficiency) and third one with a jet rejection of 570. 
The corresponding efficiencies
are shown in Figure~\ref{fig:ElectronID} as function of the transverse energy.
For photons a jet rejection of 8000 has been obtained for an efficiency of 85\%.

The Z boson data will be essential to determine the electron reconstruction and identification 
efficiency in situ. One electron is well identified (tag) and the other electron
is used to study the efficiency (probe). Good agreement is obtained between the tag\&probe
method and Monte Carlo truth. A statistical error on the efficiency of $\pm 0.1\%$ and a systematic
error of $\pm 1.5\%$ can be obtained with an integrated luminosity of 100~\pbinv{}.

The Monte Carlo can only help to determine the best possible reconstruction for the linearity.
The absolute scale must be determined by using the process Z$\rightarrow ee$. The ATLAS 
electromagnetic calorimeter has a uniformity of 0.5\% in regions of 
$\Delta\eta\mathrm{x}\Delta\phi=0.2\mathrm{x}0.4$, as shown in the test beam studies.
These regions are inter--calibrated to obtain 
a long range constant term of about 0.4\% with
an integrated luminosity of 100~\pbinv{}. 
Thus the goal of a global constant term of 0.7\% is within reach.

As an illustration of the performance of the electron and photon reconstruction 
the reconstructed invariant mass of the photons from the decay of the Higgs boson
to two photons is shown in Figure~\ref{fig:Hgamgam}. The energy scale is correct 
to about 500~MeV. The shaded area are events where at least one of the photons has converted.

In addition to the standard algorithm described here, a second algorithm, more appropriate
for the reconstruction of non isolated electrons, e.g., in the vicinity of jets, has also 
been developed by ATLAS. As the first source of electrons at the LHC will be the 
copious production of electrons from $J/\psi/\Upsilon$ decays with
rather low transverse energy, it will be useful to use
the two algorithms to study the overlap and possible improvements.

The lessons learned from test beam measurements have been transferred successfully
to the reconstruction of physics events simulation with a detailed description of  
the ATLAS detector. 
The performance of the reconstruction, e.g., electrons and photons, is in agreement 
with expectations providing a good basis for future studies and most importantly 
the collision data from 2009 on.

\begin{boldmath}
\section{\epem{} Linear Collider}
\end{boldmath}

\begin{figure}[p]
\begin{center}
\includegraphics[width=8.5cm]{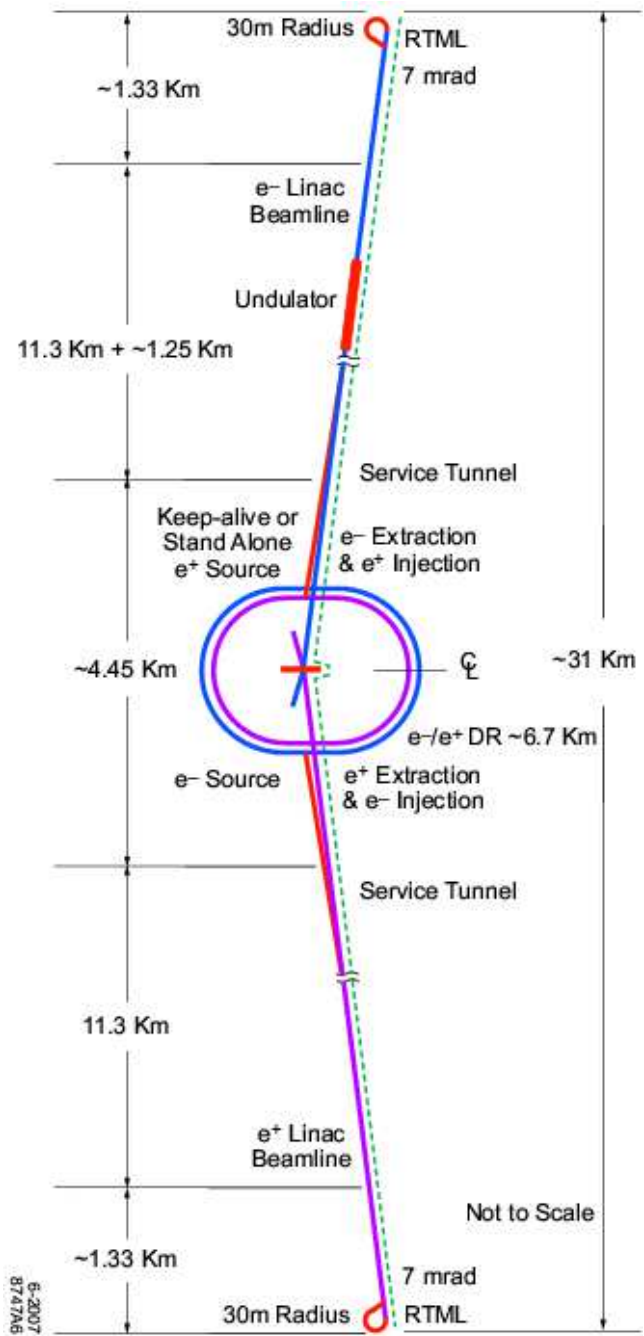}
\end{center}
\caption{Layout of the ILC.}  
\label{fig:ILC}
\end{figure}

Two designs for a linear collider are being pursued currently: ILC and CLIC. 
The ILC is an \epem{} collider design with a center--of--mass energy of 500~GeV upgradable to 
1~TeV~\cite{Brau:2007zza}. While the nominal center--of--mass energy is lower than that of the LHC, all the 
energy is available in the collisions in contrast to the LHC. The properties
of an \epem{} collider complement the LHC studies. Therefore the ILC is considered to be 
the next project for high energy physics, depending on the results obtained by the LHC. 

The layout of the ILC is shown in Figure~\ref{fig:ILC}. The tunnel length is about 31~km. Electrons 
and positrons are accelerated in tunnels of 11~km length. The beam delivery system is about 4.5~km long.
The crossing angle is planned to be 14~mrad. Two detectors will be installed in a push--pull system, 
alternating their data--taking periods. 

The main technological challenge of the ILC is the acceleration of the electrons, whereas at the LHC 
the challenge is the magnet system. The ILC design calls 
for the use of superconducting cavities, operating at a frequency of 1.3~GHz, with an average gradient
of 31.5~MeV/m. In addition to this challenge, it is planned to have an electron polarization of more than
80\%. The positrons are produced by conversion of photons (passing the electron beam through an undulator)
into \epem{} pairs. The current design of the machine has a positron polarization of 30\%, which
in a upgrade could be increased to 60\%. The polarisation must be measured with a precision
of 0.1\% via the compton effect.

The center--of--mass energy can be varied from 200~GeV to 500~GeV. Short periods at 91~GeV 
(Z--resonance) for calibration will also be possible. The precision of the knowledge of the beam energy
will feed directly into the absolute uncertainty of the measurement of any particle mass. 
It is expected to be able to measure the beam energy to 200~ppm by using two different methods. Upstream
the deflection of the beam through a dipole field will be measured. Downstream 
the synchrotron radiation of the beam passing through a string of dipole magnets will be measured.
The beam has a pulse rate of 5~Hz with a pulse length of about 1~ms. In each pulse there will 
be 2625~bunches with $2\cdot 10^{10}$ particles per bunch.
The design luminosity of the ILC is $10^{34}\mathrm{m}^{-2}\mathrm{s}^{-1}$.
The RMS of the beam size at the interaction point is expected to be about 6~nm (vertical) 
and about 640~nm (horizontal). In addition to initial state radiation, 2.4\% (RMS) of the energy will be 
lost due to beamstrahlung. The downstream beam energy measurement will measure the beam energy spread as 
well as the beam energy distribution after beamstrahlung has occurred.

While the preparation of the ILC is well advanced, R\&D is also 
performed for a collider further in the future .
A second generation \epem{} collider called CLIC (\cite{Battaglia:2004mw} and references therein)
is being developed. The CLIC concept
is based on a novel scheme decelerating one beam to accelerate another one. 
CLIC has the potential to reach a center--of--mass energy of 3~TeV.

\begin{figure}[htb]
\begin{center}
\includegraphics[width=\textwidth]{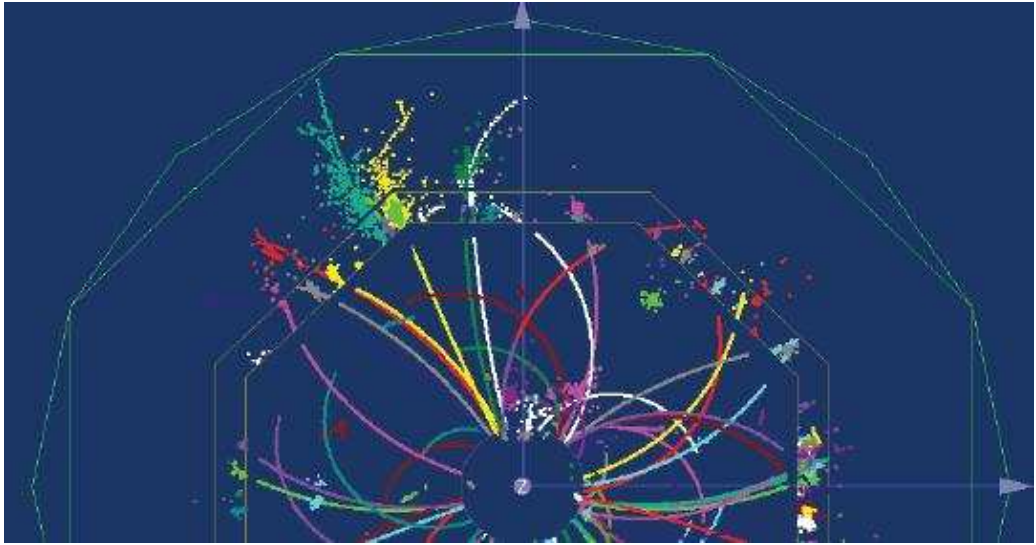}
\end{center}
\caption{Event display of a simulated event in a detector for the ILC. The fine 
granularity of the calorimeter used in the particle flow is clearly seen.}  
\label{fig:Calice}
\end{figure}
Currently there are three detector concepts under study, competing for the two
ILC detector spots. The two main ones are 
ILD (International Linear Detector)~\cite{ILDLOI} and SiD (Silicon Detector)~\cite{SiDLOI}.
They differ in the technology chosen for the tracking sub-detectors. 
SiD has silicon based tracking with a strong magnetic field 
of 5~T, thus choosing to build a compact detector.
ILD has chosen a larger volume design with a lower magnetic field of 4~T and 
with a large gaseous TPC in addition to the silicon vertex
detector. The goal for the tracking system is to reach a precision
of $\sigma(\mathrm{1/p_T}) = 10^{-4}\mathrm{GeV}^{-1}$ above 10~\gevc{}, an order 
of magnitude more precise than the precision achieved by the LEP detectors.

The calorimeters chosen by both collaborations are similar: the electromagnetic calorimeter
is a very high granularity detector. For this reason
it is sometimes called tracker--calorimeter, with 100~million
readout channels (the ATLAS electromagnetic calorimeter has about 150000 readout channels). 
The energy resolution for electrons is about 15\%/$\sqrt{\mathrm{E}}$ for electrons,
which is less ambitious than LHC experiments (e.g. ATLAS 10\% or CMS 2\%). However
the main physics processes expected at the ILC are multi--jet processes where the individual
performance is less important than the combined reconstruction of all sub--detectors
in the particle flow concept. A jet resolution using Particle Flow of 30\%/$\sqrt{\mathrm{E}}$ is expected, improving
on the 50\% expected in ATLAS. An event display showing the fine granularity of the electromagnetic
calorimeter is shown in Figure~\ref{fig:Calice}.

\cleardoublepage

\chapter{Determination of Supersymmetric Parameters and Higgs Couplings}
\label{chap:determination}

The determination of the fundamental parameters, be it supersymmetric parameters 
or the Higgs couplings, critically depends on detailed
experimental simulations of measurements and errors at the LHC and at
the ILC. While the techniques described here are general, the well--understood parameter point
SPS1a~\cite{Allanach:2002nj} is used as a specific 
example. This point has a favorable phenomenology for both
LHC and ILC. 

SPS1a was invented as a parameter set for the study of the LHC and ILC  
discovery and measurement potential. 
The Standard Model electroweak 
measurements, the b--physics precision observables, $(g-2)_\mu$~\cite{Davier:2009zi}
with the addition of the WMAP~\cite{Spergel:2006hy} measurement of the relic
density already allows to delimit interesting regions of parameter space
without the direct observation of supersymmetric particles. 
In particular, as shown in~\cite{Buchmueller:2007zk,Bechtle:2009ty}, 
the supersymmetric
mSUGRA fit of these observables results in a preferred value for the lightest Higgs boson mass
of $\mathrm{m_h}=113.3\gevcc$, pushing the Higgs boson mass 
closer to the limit of direct searches at LEP of 
114.4~\gevcc{}~\cite{Barate:2003sz}. The effect of using the LEP exclusion
is about 3~\gevcc{}.
The best--fit point is close, albeit with rather large errors, to the definition of 
SPS1a~\cite{Buchmueller:2008qe}. This is a further motivation to study the phenomenology of this parameter set 
in detail. Using a different technique, the MSSM parameters have 
been studied using the precision observable as shown in Ref.~\cite{AbdusSalam:2009qd}.

Discovering supersymmetry, and in particular SPS1a, at the LHC is considered to be 
feasible for masses up to 2.5~\tevcc{} (squarks and gluinos). 
In general, supersymmetry provides signatures with several high-pT jets
and large missing energy. The search for the lightest Higgs boson around the 
LEP limit is considered to be more difficult. It necessitates more statistics
(around 30~\fbinv{}) than for supersymmetry, where 1~\fbinv{} are sufficient to be
sensitive to masses of squarks of several hundred~\gevcc{}. The complexity
in the Higgs sector is due to the necessity to search for rare, but clean
final states ($\gamma\gamma$) in large QCD/Standard Model backgrounds.
Once the new particles are discovered, detailed measurements can be performed.

To reconstruct the fundamental parameters, 
precise theoretical predictions are necessary.
An overview of the tools is given in Ref.~\cite{Allanach:2008zn}.
The production rates for the Higgs boson are taken from~\cite{Spira:1995mt}.
The most important in the supersymmetric sector are
SuSpect, SOFTSUSY and 
SPheno~\cite{Djouadi:2002ze,Allanach:2001kg,Porod:2003um}
for the calculation of the sparticle and Higgs masses. 
SUSY-Hit provides the branching ratios via  
SDecay for the decay of supersymmetric particles and 
HDecay for the decays of the Higgs 
bosons~\cite{Muhlleitner:2003vg,Djouadi:2006bz,Djouadi:1997yw}. 
\epem{} cross sections are calculated by PYTHIA and 
SPheno~\cite{Sjostrand:2006za,Porod:2003um}, while
proton--proton cross sections, first calculated in~\cite{Dawson:1983fw}, 
are provided by Prospino2.0 at NLO~\cite{Beenakker:1996ch,Beenakker:1997ut,Beenakker:1999xh,Plehn:1998nh}. 

The pioneering work on extracting the fundamental parameters
was performed in~\cite{Blair:2002pg,Allanach:2004ud}, followed by
Fittino~\cite{Bechtle:2005vt} and  
SFitter~\cite{Lafaye:2007vs}.
The package Super-Bayes originated
in the study of the dark matter aspect of supersymmetry~\cite{de Austri:2006pe}.
Several other projects are also underway.

In this chapter, first the signatures of the Standard Model Higgs boson sector and 
the supersymmetric signals will be discussed. 
Then the most favorable case will be studied: mSUGRA including
all supersymmetric signals, followed by the study of the weak scale model, the 
MSSM. As a final 
study the question of the determination of the Higgs couplings 
as well as the sensitivity of these to new physics will be addressed.
For the latter study the question will be: if new physics signals
are not observed at the LHC, will the precision of the measurement
of the Higgs boson couplings be sufficient to be sensitive to new physics?

\section{Low Mass Higgs Boson Signatures}

\begin{figure}[htb]
\begin{minipage}[htb]{0.48\textwidth}
\includegraphics[width=\textwidth]{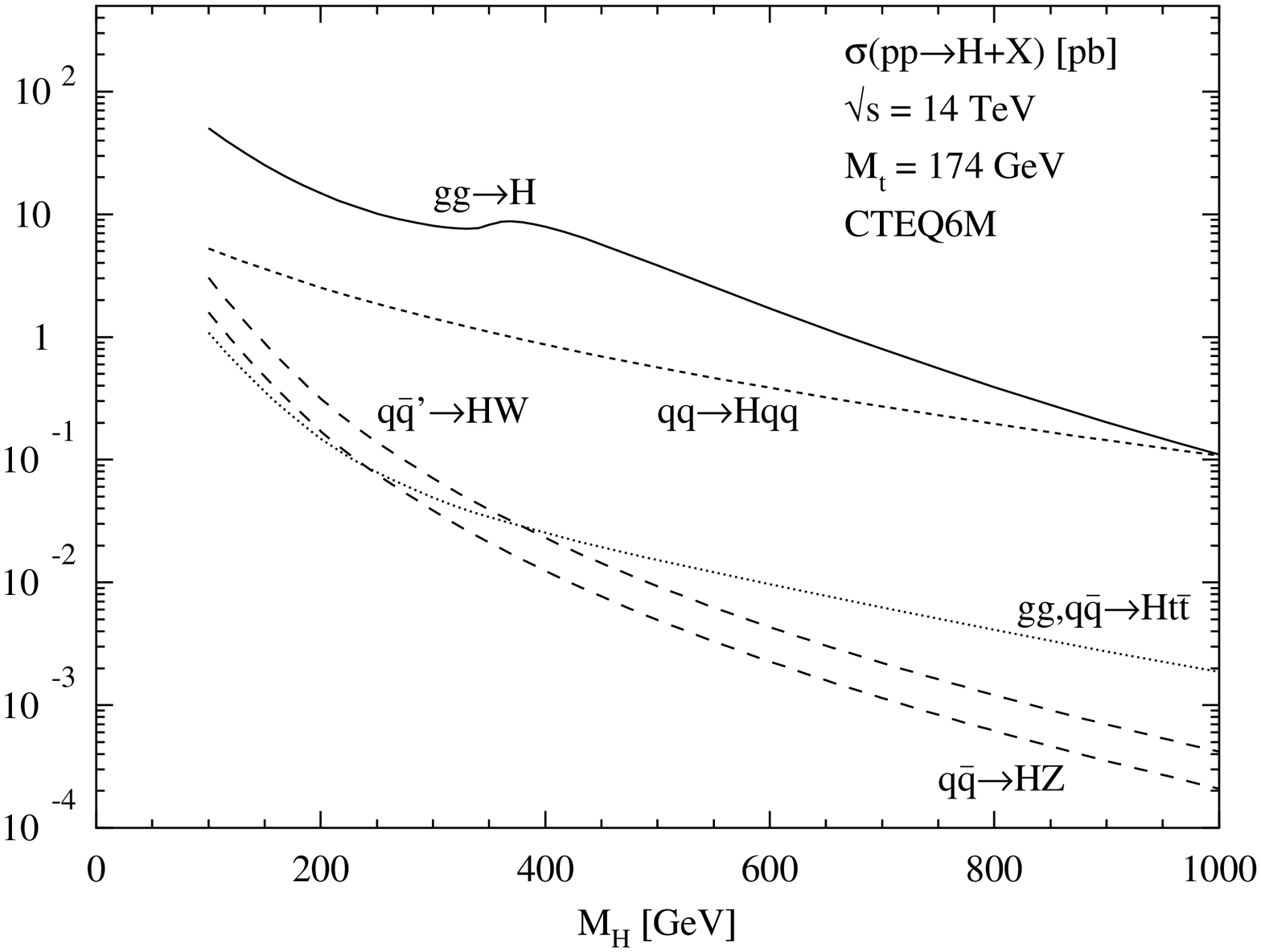}
\vspace{-1.0cm}
\caption{The production cross section for a Standard Model Higgs boson at the LHC is shown 
as function of its mass for the main production mechanisms.}
\label{fig:HiggsProd}
\end{minipage}
\hspace{0.3cm}
\begin{minipage}[htb]{0.48\textwidth}
\vspace{-0.8cm}
\includegraphics[width=0.7\textwidth,angle=-90]{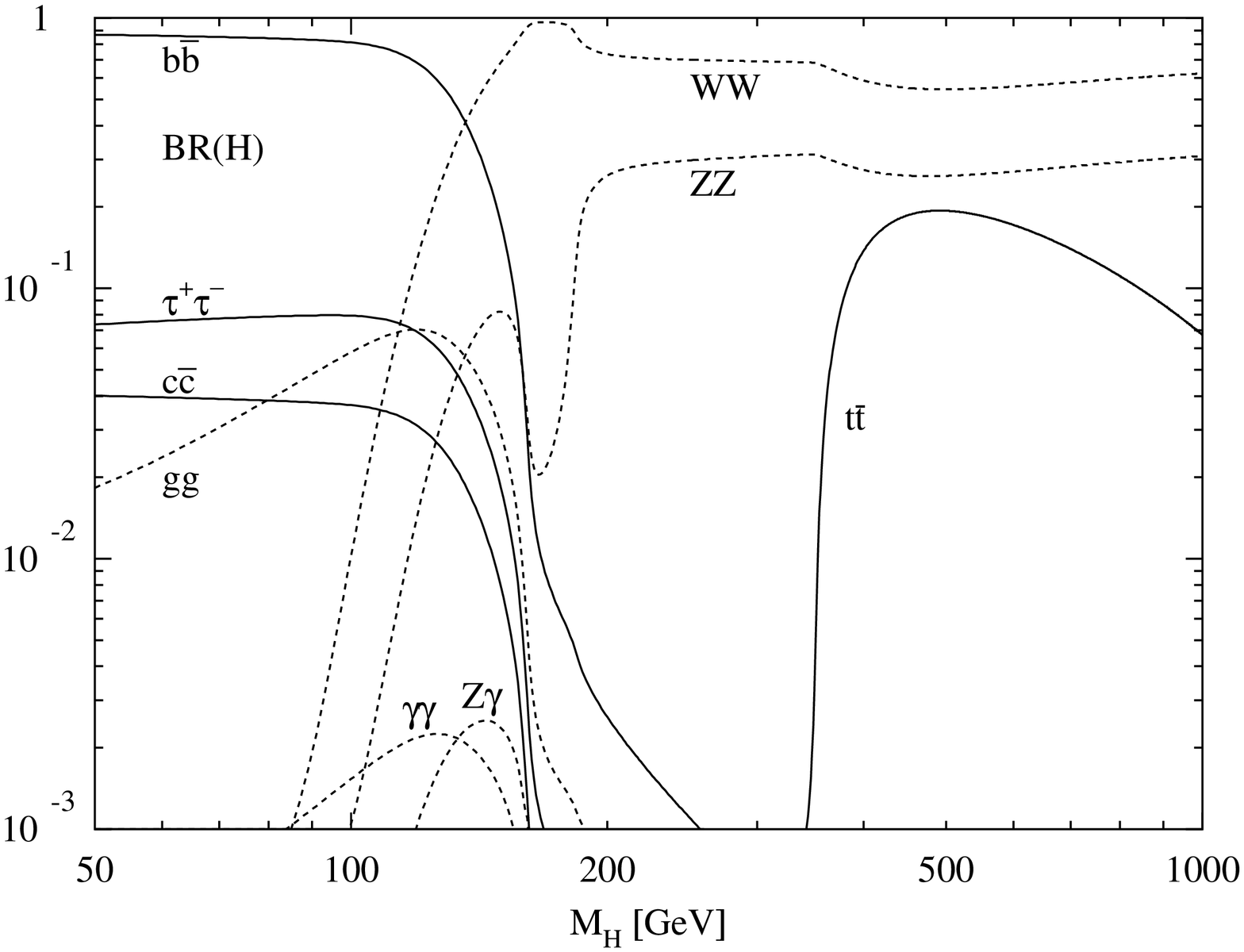}
\caption{The branching ratios of the Standard Model Higgs boson are shown as
function of its mass.}
\label{fig:HiggsBR}
\end{minipage}
\end{figure}
\begin{table}[htb]
\begin{small} \begin{center}
\begin{tabular}{l|l||r|r@{ }l|r||r|r}
 production & decay & 
 $S+B$ & 
 \multicolumn{2}{|r|}{$B$ }& 
 $S$ & 
 $\Delta S^\mathrm{(exp)}$ &  
 $\Delta S^\mathrm{(theo)}$ \\ \hline
 $gg \to H$ & $ZZ$ & 
  13.4 & 6.6 & ($\times$ 5) & 6.8 & 3.9 & 0.8 \\
 $qqH$ & $ZZ$ & 
  1.0  & 0.2 & ($\times$ 5) & 0.8 & 1.0 & 0.1 \\
 $gg \to H$ & $WW$ & 
  1019.5 & 882.8 & ($\times$ 1) & 136.7 & 63.4 & 18.2 \\
 $qqH$ & $WW$ & 
  59.4 & 37.5 & ($\times$ 1) & 21.9 & 10.2 & 1.7 \\
 $t\bar{t}H$ & $WW (3 \ell)$ & 
  23.9 & 21.2 & ($\times$ 1) & 2.7 & 6.8 & 0.4 \\
 $t\bar{t}H$ & $WW (2 \ell)$ & 
  24.0 & 19.6 & ($\times$ 1) & 4.4 & 6.7 & 0.6 \\
 inclusive & $\gamma\gamma$ & 
  12205.0 & 11820.0 & ($\times$ 10) & 385.0 & 164.9 & 44.5 \\
 $qqH$ & $\gamma\gamma$ & 
  38.7 & 26.7 & ($\times$ 10) & 12.0 & 6.5 & 0.9 \\
 $t\bar{t}H$ & $\gamma\gamma$ & 
  2.1 & 0.4 & ($\times$ 10) & 1.7 & 1.5 & 0.2 \\
 $WH$ & $\gamma\gamma$ & 
  2.4 & 0.4 & ($\times$ 10) & 2.0 & 1.6 & 0.1 \\
 $ZH$ & $\gamma\gamma$ & 
  1.1 & 0.7 & ($\times$ 10) & 0.4 & 1.1 & 0.1 \\
 $qqH$ & $\tau\tau (2 \ell)$ & 
  26.3 & 10.2 & ($\times$ 2) & 16.1 & 5.8 & 1.2 \\
 $qqH$ & $\tau\tau (1 \ell)$ & 
  29.6 & 11.6 & ($\times$ 2) & 18.0 & 6.6 & 1.3 \\
 $t\bar{t}H$ & $b\bar{b}$ & 
  244.5 & 219.0 & ($\times$ 1) & 25.5 & 31.2 & 3.6 \\
 $WH/ZH$ & $b\bar{b}$ & 
 228.6 & 180.0 & ($\times$ 1) & 48.6 & 20.7 & 4.0 
\end{tabular}
\end{center} \end{small} \vspace*{0mm}
\caption[]{Signatures used for the analysis for a Higgs mass of
  120~\gevcc{}. The Standard Model event numbers for $30~\fbinv$ include
  cuts~\cite{duehrssennote}. The factor after the background rates
  describes how many events are used to extrapolate into the signal
  region. The last two columns give the one-sigma experimental and
  theory error bars on the signal (taken from~\cite{Lafaye:2009vr}).}
\label{tab:HiggsChannels}
\end{table}
The Higgs boson can be produced in gluon fusion~\cite{Georgi:1977gs}, 
weak boson fusion and top pair production. 
The expected cross sections are shown in Figure~\ref{fig:HiggsProd}.
While tremendous progress on the understanding of the Monte Carlo modeling of the background
processes has been made, it is essential for most of the analyzes to determine the background
from the data, e.g., via side-band background subtraction. This means that at the LHC 
channels with the capability of the reconstruction of the Higgs mass (with the best possible precision)
are favored over counting analyses.

The gluon fusion process at the LHC is known with an uncertainty of 
13\%~\cite{Anastasiou:2005pd,Anastasiou:2003ds,Catani:2007vq,Catani:2003zt,Bozzi:2005wk}. 
NLO corrections~\cite{Spira:1995rr,Spira:1997dg} as well as NNLO calculations
with an effective $ggH$ 
coupling~\cite{Harlander:2002wh,Anastasiou:2002yz,Anastasiou:2008tj,Ravindran:2003um}
lead to a cross section of 37~pb for a 120~\gevcc{} Higgs boson.
For the weak boson fusion process, the cross section is about 
4.5~fb~\cite{Rainwater:1998kj,Plehn:1999xi,Rainwater:1999sd,Kauer:2000hi}. 
This production mode relies on a central jet veto between the forward tagging
jets~\cite{Barger:1994zq,Rainwater:1996ud,DelDuca:2006hk}. 
For this signal, NLO rates as well as distributions have been 
calculated~\cite{Han:1992hr,Figy:2003nv,Ciccolini:2007jr,Ciccolini:2007ec,Arnold:2008rz}
and an error of 
7\%~\cite{Han:1992hr,Figy:2003nv,Ciccolini:2007jr,Ciccolini:2007ec,Arnold:2008rz} is used.

The production channels of gluon fusion and weak boson fusion are not
independent: Higgs boson production with two jets has
contributions from both gluon fusion and weak boson fusion. With
appropriate cuts, one channel can be enhanced over the
other~\cite{DelDuca:2001eu,Nikitenko:2007it}. The error on this
classification is covered by the error estimate. 

A cross section of 450~pb is expected for the channel ttH, 
the radiation of a Higgs boson from a
top quark. NLO rate calculations have been performed~\cite{Beenakker:2001rj,Beenakker:2002nc,Dawson:2002tg}. 
A theory error of around 13\% is associated to this channel.

For the associated production of the Higgs boson and a weak vector boson (Z or W),
a cross section of 2.2~pb is expected. 
For this channel a theory error of 7\% is assumed.

The analysis will be constrained to the single production of Higgs bosons. 
Pair production is essentially sensitive to the self--coupling of the
Higgs and extremely difficult at the 
LHC~\cite{Dicus:1987ic,Glover:1987nx,Plehn:1996wb,Djouadi:1999rca,Dawson:1998py,Baur:2002qd,Baur:2002rb,Dahlhoff:2005sz,Baur:2003gp}.

For a Higgs boson of mass 120~\gevcc{} the highest branching ratio is to a pair of b--quarks.
The branching ratio is roughly 90\% as shown in Figure~\ref{fig:HiggsBR}. 
However, discovering the Higgs boson in this decay channel is 
particularly difficult, because 
reconstructing a Higgs boson invariant
mass peak from two b--jets above the QCD continuum is hopeless.
Therefore the 
associated production in combination with a pair of top quarks have been studied.
The combinatorial background
is more complex than previously estimated, making it difficult to establish
this signal at the 5$\sigma$ level~\cite{Ball:2007zza,Aad:2009wy} for 
30~\fbinv{}. 
This development is taken into account by reducing the signal by 50\% 
with respect to the analysis in Ref.~\cite{duehrssennote}.

\begin{table}[htb]
\begin{tabular}{l|r}
luminosity measurement & 5 \% \\\hline
detector efficiency & 2 \% \\\hline
lepton reconstruction efficiency & 2 \% \\\hline
photon reconstruction efficiency & 2 \% \\\hline
WBF tag-jets / jet-veto efficiency & 5 \% \\\hline
$b$-tagging efficiency & 3 \% \\\hline
$\tau$-tagging efficiency (hadronic decay) & 3 \% \\\hline
lepton isolation efficiency ($H \rightarrow 4\ell$) & 3 \% 
\end{tabular}
\begin{tabular}{l|r|l}
 & 
 $\Delta B^\mathrm{(syst)}$ &  
 correl \\ \hline
 $H \to ZZ$ & 
  $1 \%$ & yes \\
 $H \to WW$ & 
  $5 \%$ & no \\
 $H \to \gamma\gamma$ & 
  $0.1 \%$ & yes \\
 $H \to \tau\tau$ & 
  $5 \%$ & yes \\
 $H \to b\bar{b}$ & 
  $10 \%$ & no \\
\end{tabular}
\caption{Experimental systematic errors used in the analysis. Left: systematic
  errors applying to both signal and background. Reconstruction and
  tagging efficiencies are defined per particle, e.g., $H \to
  \gamma\gamma $ has a $4 \%$ error on the photon
  reconstruction. Right: systematic background errors, either fully
  correlated or independent between channels.  Tables are the same as~\cite{duehrssennote,Lafaye:2009vr}.}
\label{tab:HiggsSystError}
\end{table}
The decay of the Higgs boson to two photons has the advantage of a precise 
mass reconstruction. Thus in spite of the low branching ratio of order of $10^{-3}$,
the excellent mass resolution of the calorimeters
at the LHC allows 
to separate the signal from the background of QCD and prompt photons. 
As the natural width of the Higgs boson at 120~\gevcc{} is
smaller than the experimental resolution (about 2~\gevcc{}), 
the uniformity of the calorimeters
is essential for the discovery. For the measurement of the Higgs boson 
mass, once sufficient statistics are accumulated, a precise knowledge of the
photon energy scale in the calorimeter is essential.
The preparation of this measurement through test beam studies gives 
additional confidence in this channel.
The channel can be measured separately
in the gluon fusion, weak boson fusion and top pair production. The expected
precision of the mass measurement is $\mathcal{O}(200\mevcc)$. 

The associated production of ZH or WH was considered
to be hopeless initially (in contrast to the TeVatron where it is the main source for the signal 
search). In a theoretical study the Higgs boson is required 
to be strongly boosted with an inherent loss of efficiency due to this 
requirement.
The idea is then to use the subjet structure of the Higgs boson decay
to gain additional leverage~\cite{Butterworth:2008iy} more than compensating the loss
of efficiency. 
This promising new channel is also included in the analysis.
The results of the theoretical analysis have been confirmed recently by an ATLAS 
study at the level of 10\%~\cite{ATLAS:2009hw}.

The branching ratio for Higgs boson to a pair of Z bosons is smaller 
at 120~\gevcc{} than at much higher masses. It is considered to
be the golden channel with the Z bosons decaying to electrons or muons.
Some information can still be obtained from this channel, 
either produced in gluon fusion or in weak boson fusion.

The decay of the Higgs boson 
to two leptonically (electrons and muons) decaying \wpm{} bosons is also used in the analysis.
This counting analysis, due to the neutrinos in the \wpm{} decay, 
does not allow the reconstruction of the Higgs boson mass. 
The background is rejected using the angular correlation of the leptons.

The decay of the Higgs boson to a pair of $\tau$--leptons can be 
used for the weak boson fusion process. Here the Higgs boson is boosted
(in contrast to the gluon fusion process) which allows the use of the
collinear approximation to reconstruct the Higgs boson mass. 

\begin{table}[htb]
\begin{tabular}{l|r}
$\sigma$ (gluon fusion)                   & 13 \% \\\hline
$\sigma$ (weak boson fusion)              & 7 \% \\\hline
$\sigma$ ($VH$-associated)                & 7 \% \\\hline
$\sigma$ ($t\bar{t}$-associated)          & 13 \% 
\end{tabular}
\hspace*{20ex}
\begin{tabular}{l|r}
$\text{BR}(H \rightarrow ZZ)$             & 1 \% \\\hline
$\text{BR}(H \rightarrow WW)$             & 1 \% \\\hline
$\text{BR}(H \rightarrow \tau\bar{\tau})$ & 1 \% \\\hline
$\text{BR}(H \rightarrow c\bar{c})$       & 4 \% \\\hline
$\text{BR}(H \rightarrow b\bar{b})$       & 4 \% \\\hline
$\text{BR}(H \rightarrow \gamma\gamma)$   & 1 \% \\\hline
$\text{BR}(H \rightarrow Z\gamma)$        & 1 \% \\\hline
$\text{BR}(H \rightarrow gg)$             & 2 \% 
\end{tabular}
\caption{Theory errors used in the analysis. The left table shows the errors on the production cross section and
the right table shows the errors on the branching ratio (taken from~\cite{Lafaye:2009vr}).}
\label{tab:HiggsTheoError}
\end{table}
The production rates for the Higgs sector were calculated in~\cite{Spira:1995mt} and
the branching ratios are taken from 
HDecay~\cite{Djouadi:1997yw,Djouadi:2006bz}.
 All rates and experimental errors are effectively taken from~\cite{duehrssennote}
(with the exception of the modified ttbb channel).
The expected results are listed in Table~\ref{tab:HiggsChannels}.
Experimental systematic errors are shown in Table~\ref{tab:HiggsSystError} and 
the theoretical errors are shown in Table~\ref{tab:HiggsTheoError}

At the ILC the mass of of a 120~\gevcc{} Higgs boson can be measured
with a precision of about 50~\mevcc{}. The expected precision on the 
Higgs coupling measurements is at the percent level.
Additionally precision measurements of the couplings of the Higgs boson are
not restricted to judiciously chosen channels. A model independent analysis 
of the Higgs production and decay is possible. This can 
have consequences on the interpretation 
of the determination of the couplings as will be shown later.

\section{Supersymmetric Observables for a specific Parameter Set}

\begin{figure}[htb]
\begin{minipage}[htb]{0.48\textwidth}
\includegraphics[width=\textwidth]{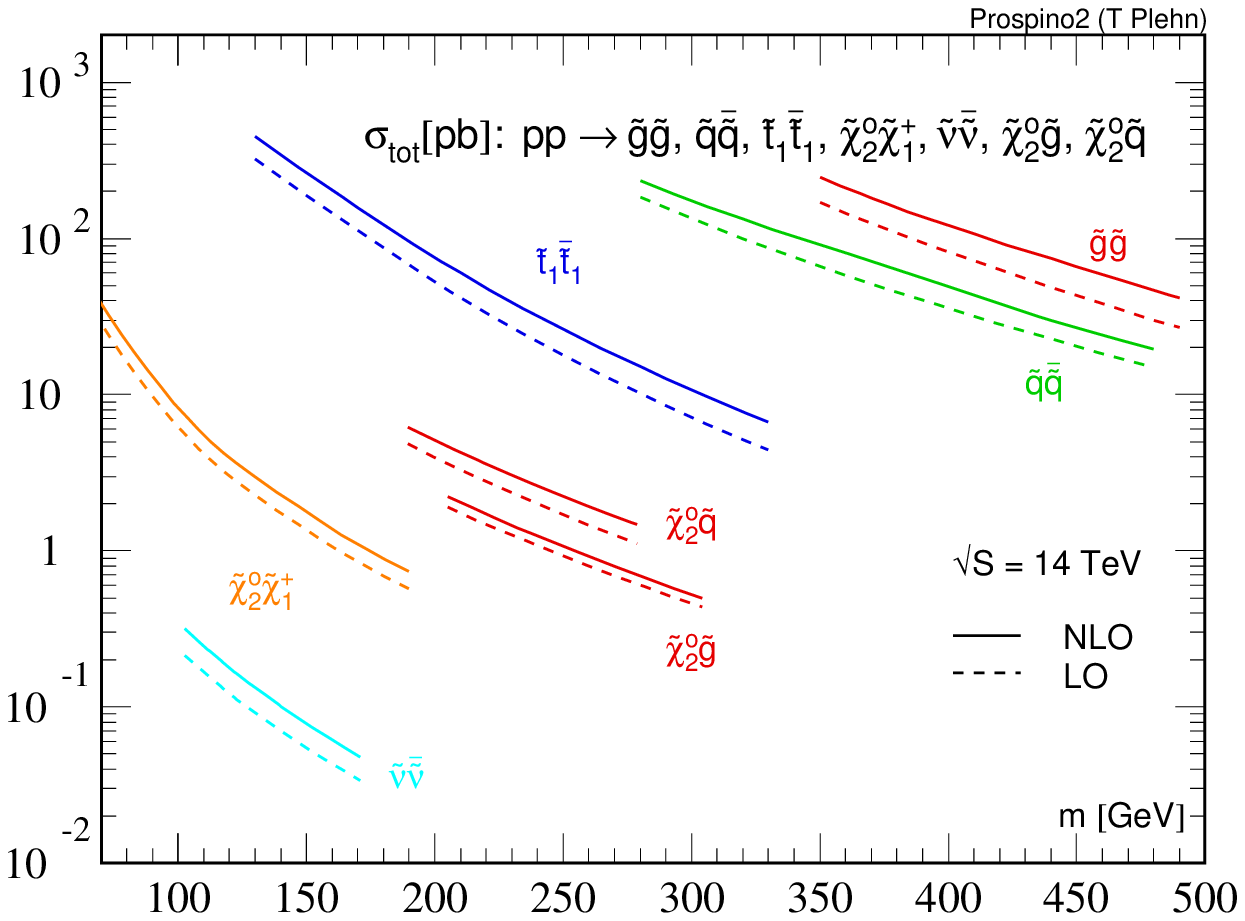}
\caption{The production cross section for supersymmetric particles at the LHC at NLO 
is shown as calculated by Prospino2.0.}
\label{fig:SUSYProd}
\end{minipage}
\hspace{0.3cm}
\begin{minipage}[htb]{0.48\textwidth}
\vspace{0.7cm}
\includegraphics[width=\textwidth]{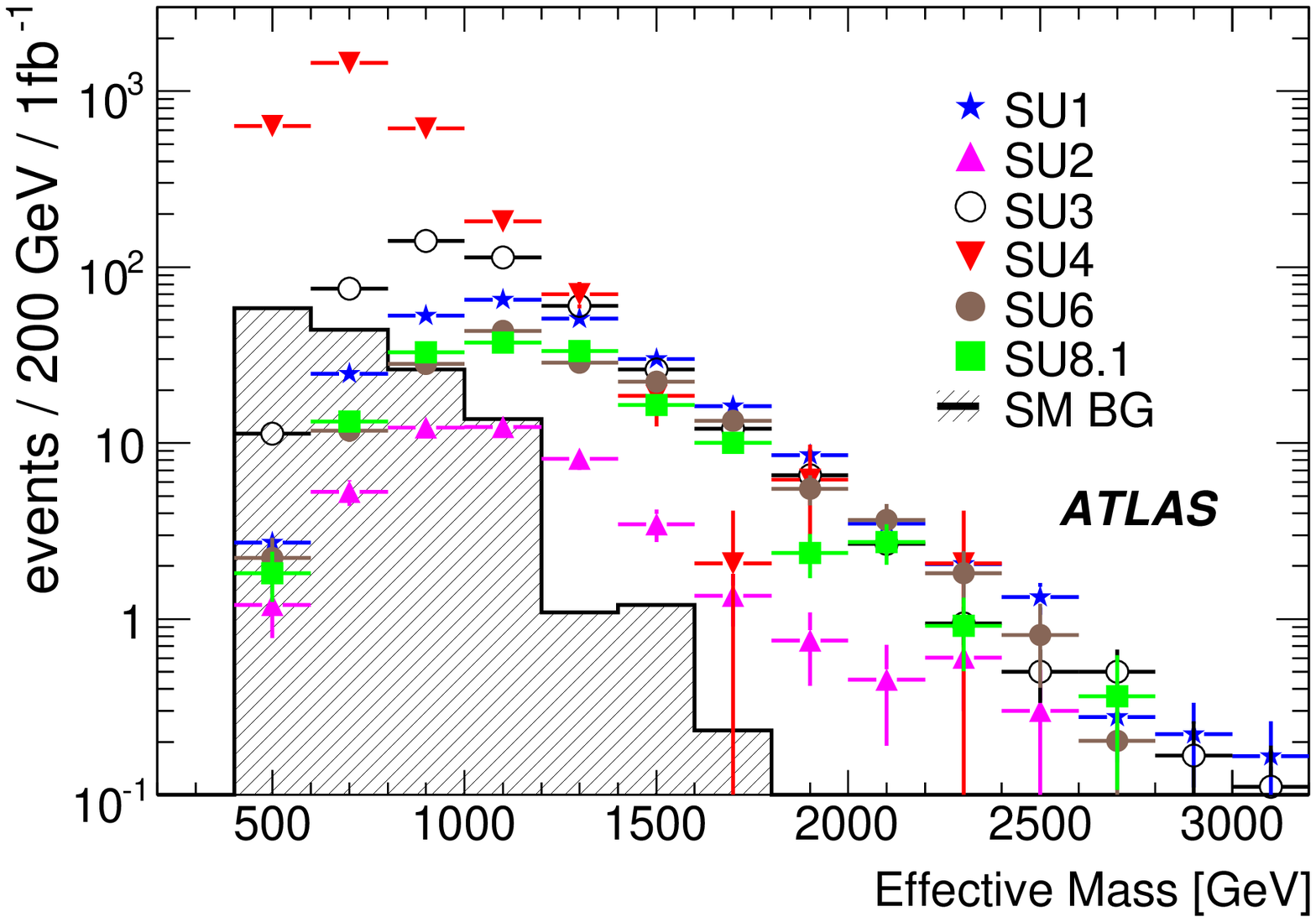}
\vspace{-1.0cm}
\caption{The effective mass is shown for the Standard Model and several supersymmetric parameter sets in ATLAS.}
\label{fig:SUSYdiscover}
\end{minipage}
\end{figure}

The parameter point SPS1a is characterized by moderately heavy squarks
and gluinos, which leads to long cascades including neutralinos
and sleptons. The mass of the
lightest Higgs boson is close to the mass limit determined at LEP.

The production of squarks and gluinos and their decay lead 
to large signals (in contrast to the Higgs sector) above the Standard Model
background. The NLO cross sections calculated by Prospino2.0 are shown in Figure~\ref{fig:SUSYProd}.
The production cross sections for strongly interacting particles are 
several tens of picobarn.
Inclusive analyses such as the search for highly spherical events 
with large missing transverse energy (due to the LSPs) and several 
jets with high transverse momenta, and/or adding the requirement of 
a lepton (or more) will allow the discovery of supersymmetry.
The effective mass, the sum of the jet transverse momenta and the missing transverse
energy, will give a first hint of the scale of supersymmetry. An example for
the effective mass of supersymmetric signals 
and Standard Model background is shown in Figure~\ref{fig:SUSYdiscover}. 
At large effective mass the supersymmetric signal dominates the distribution. It is essential
to have a good understanding of the jet and ETmiss reconstruction to have confidence in
the discovery. In the standard scenarios the LHC will be sensitive to squark and gluino masses
up to 2.5~\tevcc{}.

An important aspect of subsequent measurements of the supersymmetric particles 
is the long decay chain~\cite{Bachacou:1999zb,Allanach:2000kt}
\begin{equation} 
\tilde{q}_L\rightarrow\chi_2^0 q\rightarrow\tilde{\ell}_R\ell q\rightarrow\ell\ell q \chi^0_1 .
\end{equation}
The final state contains at least a hard jet and two opposite sign same 
flavor leptons. 
In this decay chain five edges and thresholds can be calculated
and reconstructed~\cite{Weiglein:2004hn}. One of the invariant mass combinations is the
lepton--lepton edge. As shown in Table~\ref{tab:edges}, with an integrated luminosity of 
300~\fbinv{}, the measurement is not statistics limited, but limited by the knowledge
of the lepton energy scale. Thus attaining the goal of a per mil level energy scale 
is important and justifies the tremendous work performed on this issue 
in test beam and Monte Carlo
simulations.

\begin{table}[htb]
\begin{small} \begin{center}
\begin{tabular}{|ll|r|rrrr|}
\hline
\multicolumn{2}{|c|}{ type of } & 
 \multicolumn{1}{c|}{ nominal } & 
 \multicolumn{1}{c|}{ stat. } & 
 \multicolumn{1}{c|}{ LES } & 
 \multicolumn{1}{c|}{ JES } & 
 \multicolumn{1}{c|}{ theo. } \\
\multicolumn{2}{|c|}{ measurement } & 
 \multicolumn{1}{c|}{ value } & 
 \multicolumn{4}{c|}{ error } \\
\hline
\hline
$m_h$ & 
 & 108.99& 0.01 & 0.25 &      & 2.0 \\
$m_t$ & 
 & 171.40& 0.01 &      & 1.0  &     \\
$m_{\tilde{l}_L}-m_{\chi_1^0}$ & 
 & 102.45& 2.3  & 0.1  &      & 2.2 \\
$m_{\tilde{g}}-m_{\chi_1^0}$ & 
 & 511.57& 2.3  &      & 6.0  & 18.3 \\
$m_{\tilde{q}_R}-m_{\chi_1^0}$ & 
 & 446.62& 10.0 &      & 4.3  & 16.3 \\
$m_{\tilde{g}}-m_{\tilde{b}_1}$ & 
 & 88.94 & 1.5  &      & 1.0  & 24.0 \\
$m_{\tilde{g}}-m_{\tilde{b}_2}$ & 
 & 62.96 & 2.5  &      & 0.7  & 24.5 \\
$m_{ll}^\mathrm{max}$: & three-particle edge($\chi_2^0$,$\tilde{l}_R$,$\chi_1^0$)  
 & 80.94 & 0.042& 0.08 &      & 2.4 \\
$m_{llq}^\mathrm{max}$: & three-particle edge($\tilde{q}_L$,$\chi_2^0$,$\chi_1^0$)  
 & 449.32& 1.4  &      & 4.3  & 15.2 \\
$m_{lq}^\mathrm{low}$: & three-particle edge($\tilde{q}_L$,$\chi_2^0$,$\tilde{l}_R$)
 & 326.72& 1.3  &      & 3.0  & 13.2 \\
$m_{ll}^\mathrm{max}(\chi_4^0)$: & three-particle edge($\chi_4^0$,$\tilde{l}_R$,$\chi_1^0$)
 & 254.29& 3.3  & 0.3  &      & 4.1 \\
$m_{\tau\tau}^\mathrm{max}$: & three-particle edge($\chi_2^0$,$\tilde{\tau}_1$,$\chi_1^0$)
 & 83.27 & 5.0  &      & 0.8  & 2.1 \\
$m_{lq}^\mathrm{high}$: & four-particle edge($\tilde{q}_L$,$\chi_2^0$,$\tilde{l}_R$,$\chi_1^0$)
 & 390.28& 1.4  &      & 3.8  & 13.9 \\
$m_{llq}^\mathrm{thres}$: & threshold($\tilde{q}_L$,$\chi_2^0$,$\tilde{l}_R$,$\chi_1^0$)
 & 216.22& 2.3  &      & 2.0  & 8.7 \\
$m_{llb}^\mathrm{thres}$: & threshold($\tilde{b}_1$,$\chi_2^0$,$\tilde{l}_R$,$\chi_1^0$)
 & 198.63& 5.1  &      & 1.8  & 8.0 \\
\hline
\end{tabular}
\end{center} \end{small} \vspace*{-3mm}
\caption[]{
LHC measurements in SPS1a, taken 
  from~\cite{Weiglein:2004hn}. Shown are the nominal values (from SuSpect) 
  and statistical errors, systematic errors from the lepton (LES)
  and jet energy scale (JES) and theoretical errors. 
  All values are given in GeV. Table taken from~\cite{Lafaye:2007vs}.}
\label{tab:edges}
\end{table}

The edges and thresholds can be expressed as functions of the four intervening masses.
Therefore from the system shown in Table~\ref{tab:edges} the masses of the
supersymmetric particles can be reconstructed,
using either via toy Monte Carlo or a fit. It is important to note that there
is no assumption is made on the underlying
theory. Further signatures, e.g. the squark-R and the sbottoms, 
provide a total of 14~observable and measurable particles at the LHC in SPS1a.
The precision of the mass determination for the LHC is shown in Table~\ref{tab:mass_errors}.
Typically the systematic error on measurements at the LHC coming from 
the jet energy scale is 1\% and 0.1\% for the
lepton energy scale. These energy--scale errors are each taken to be 99\% correlated
separately as discussed in Ref~\cite{Weiglein:2004hn}. 
With integrated luminosities of up to 300~fb$^{-1}$
the statistical error in many cases is smaller than the systematic
error. 

\begin{table}[htb]
\begin{small} \begin{center}
\begin{tabular}{|l|cccc||l|cccc|}
\hline
 & $m_{\rm SPS1a}$ & LHC & ILC & LHC+ILC &
 & $m_{\rm SPS1a}$ & LHC & ILC & LHC+ILC\\
\hline
\hline
$h$  & 108.99& 0.25 & 0.05 & 0.05 &
$H$  & 393.69&      & 1.5  & 1.5  \\
$A$  & 393.26&      & 1.5  & 1.5  &
$H+$ & 401.88&      & 1.5  & 1.5  \\
\hline
$\chi_1^0$ &  97.21& 4.8 & 0.05  & 0.05 &
$\chi_2^0$ & 180.50& 4.7 & 1.2   & 0.08 \\
$\chi_3^0$ & 356.01&     & 4.0   & 4.0  &
$\chi_4^0$ & 375.59& 5.1 & 4.0   & 2.3 \\
$\chi^\pm_1$ & 179.85 & & 0.55 & 0.55 &
$\chi^\pm_2$ & 375.72 & & 3.0  & 3.0 \\
\hline
$\tilde{g}$ &  607.81& 8.0 &  & 6.5 & & & & & \\
\hline
$\tilde{t}_1$ & 399.10&     &  2.0  & 2.0 & & & & & \\
$\tilde{b}_1$ & 518.87& 7.5 &       & 5.7 &
$\tilde{b}_2$ & 544.85& 7.9 &       & 6.2 \\
\hline
$\tilde{q}_L$ &  562.98&  8.7 & &  4.9 &
$\tilde{q}_R$ &  543.82&  9.5 & &  8.0 \\
\hline
$\tilde{e}_L$    & 199.66   & 5.0 & 0.2  & 0.2  &
$\tilde{e}_R$    & 142.65   & 4.8 & 0.05 & 0.05 \\
$\tilde{\mu}_L$  & 199.66   & 5.0 & 0.5  & 0.5  &
$\tilde{\mu}_R$  & 142.65   & 4.8 & 0.2  & 0.2  \\
$\tilde{\tau}_1$ & 133.35   & 6.5 & 0.3  & 0.3  &
$\tilde{\tau}_2$ & 203.69   &     & 1.1  & 1.1  \\
$\tilde{\nu}_e$  & 183.79   &     & 1.2  & 1.2  & & & & & \\
\hline
\end{tabular}
\end{center} \end{small} \vspace*{-3mm}
\caption[]{Errors for the mass determination in SPS1a, taken 
  from~\cite{Weiglein:2004hn}. Shown are the nominal parameter values (from SuSpect),
  the error for the LHC alone, from the ILC alone, and from a combined 
  LHC+ILC analysis (taken from~\cite{Lafaye:2007vs})}
\label{tab:mass_errors}
\end{table}

For the ILC, as a rule of thumb, if particles are light
enough to be produced in pairs given the center-of-mass energy of the
collider, their mass can be determined with an order
of magnitude better precision. Two separate methods provide 
mass measurements. Direct reconstruction, i.e., for the observed
decay products of the sparticles in the detector relies on the precision
of the energy scale of the detector. The second method, a counting 
experiment, is based on the measurement of the cross section as function
of the (variable) center--of--mass energy of the ILC. Here, experimentally,
the precise determination of the beam energy is of the utmost importance,
whereas the challenge for the detector data is the understanding of
the efficiency. 
Both methods have comparable precision in the per mil region, but different
systematical errors.
Additionally discrete quantum numbers like the spin of the
particles can be determined as well by the analysis of the angular 
distributions.

The summary of particle mass measurements is listed in
Table~\ref{tab:mass_errors}, taken from Ref.~\cite{Lafaye:2007vs}. 
The central values are calculated by SuSpect~\cite{Djouadi:2002ze}.  
The coverage of the LHC is better in the strongly interacting sector,
whereas the ILC has the advantage of higher precision in the gaugino/slepton
sector~\cite{Freitas:2003yp,Pierce:1996zz,Fritzsche:2004nf,Oller:2004br}. 
As pointed out in Ref.~\cite{Blair:2002pg,Allanach:2004ud}, the use of the 
ILC measurement of the LSP in the kinematic determination of the squark masses
increases their precision.

\section{Treatment of Errors}
\label{sec:sugra_errors}

A rigorous treatment of the errors is essential to obtain
the most precise and also reliable estimation of the 
fundamental parameters. There are three 
different sources of errors: experimental statistical 
errors, experimental systematic errors and theoretical
errors. Systematic errors can have different sources, such as the
energy scale or the knowledge of the efficiency
and varying degrees of correlation. The picture even depends on the 
measurements used in the analysis: edges and thresholds have independent
statistical errors and correlated energy scale errors, but when translating them 
into masses, the masses are then non-trivially correlated.
Theoretical errors are an estimate 
of unknown higher order corrections.

In SFitter the CKMfitter
prescription for the RFit scheme~\cite{Hocker:2001xe,Charles:2006vd} is followed.  
The experimental errors are treated as Gaussian and the full correlation matrix
is included. The theoretical errors are flat errors. Therefore any
central value within the range is equally probable. This description is particularly well 
adapted as the error covers also missing higher order calculations. In this case
the higher order, once it is calculated, should move the central value of the prediction within the 
error. This behavior would be contrary to the Gaussian expectation.

While in the supersymmetric measurements the errors are essentially Gaussian, for the
study of the Higgs parameters, low counting rates (after background subtraction) are 
expected. A combination scheme of Poisson and Gaussian errors was developed under the
guiding principle that in the Gaussian limit the formula should simplify to the 
quadratic sum of the errors. The detailed procedure is described in the Appendix 
of~\cite{Lafaye:2009vr}.

The contribution to the $\chi^2$ of a given measurement is 
zero within 1$\sigma$ of the theoretical error and outside
of this range the experimental error is used.
Given a set of measurements $\vec d$ and 
a general correlation matrix $C$

\begin{alignat}{7}
\chi^2     &= {\vec{\chi}_d}^T \; C^{-1} \; \vec{\chi}_d  \notag \\ 
|\chi_{d,i}| &=
  \begin{cases}
  0  
          &|d_i-\bar{d}_i | <   \sigma^{\text{(theo)}}_i \\
  \frac{ |d_i-\bar{d}_i | - \sigma^{\text{(theo)}}_i}{ \sigma^{\text{(exp)}}_i}
  \qquad  &|d_i-\bar{d}_i | >   \sigma^{\text{(theo)}}_i \; ,
  \end{cases}
\label{eq:flat_errors}
\end{alignat}

where $\bar{d}_i$ is the $i$-th data point predicted by the model
parameters and $d_i$ the measurement. 

A complication with flat distributions is that in the central
region the log--likelihood can be constant as a function of model
parameters. In those regions these parameters vanish from the counting
of degrees of freedom. 

To determine the errors on the fundamental parameters, two
techniques are used: a direct determination for the best fit using MINUIT 
and using sets of toy measurements. The advantage of
MINUIT is that only one fit is necessary to determine the errors,
but the presence of correlations and the flat theory errors complicate
the error determination. The toy experiments
method is used by SFitter unless stated otherwise. The search for the best--fit point
is performed several (10000) times. The datasets are smeared according
to the experimental and theoretical errors including the correlations
among them.

\section{Measuring mSUGRA}

Determining the parameters from a given set of measurements 
necessitates efficient algorithms. The determination
of the mSUGRA parameters can be used as an illustrative example.
Four parameters have to be determined and one sign. 
The smuon mass depends on \mZero{} and \mOneHalf{}, the neutralino 
mass depends on $\tan\beta$ and \mOneHalf{}, among others.
The measurements are strongly correlated, e.g., via the energy scale
error.

\subsection{Determination of the True Central Value}

The first question to be asked is whether the correct 
parameter set can be found.
Thus the algorithm has to be able to disentangle the different contributions.
The simplest approach is to use a gradient fit, i.e., MINUIT. In a tightly
constrained system such as mSUGRA this is sufficient to find the correct 
parameter set even if the starting values for the parameters are far 
from the true SPS1a values. The sign of the $\mu$ parameter is then 
determined by the quality of the fit: performing the fit with 
the wrong sign of $\mu$ leads to larger $\chi^2$ value than the correct
sign of $\mu$.

However, the result could depend on the starting point when
the fit remains confined to a secondary minima by a ``potential'' 
wall in $\chi^2$. Simulated annealing was developed in Ref.~\cite{Bechtle:2005vt}
to allow to escape the secondary minimum and find the correct solution. 
In SFitter weighted Markov chains were developed. Markov chains 
are linear in the number of parameters, an important property 
for the determination of the MSSM parameters,
and are capable of finding secondary minima. 

First, fully exclusive log-likelihood maps are calculated. From these different
types of projections are possible: Bayesian or frequentist (profile likelihood).
Successive projections over unwanted parameters allow to study the correlations between
parameters. 
In the Bayesian approach a measure, the Bayesian prior,
is introduced and the parameter is integrated 
over. In the frequentist approach the parameter value of the minimum $\chi^2$ 
(maximum likelihood) is retained. The Bayesian approach ensures that after the projection
the distributions have the quality of a probability density function. The advantage of 
the frequentist approach is that the absolute minimum is always retained, however the pdf property
is lost.

\begin{figure}[htb]
\begin{minipage}[htb]{0.48\textwidth}
\centering
\vspace{-1.0cm}
\includegraphics[width=\textwidth]{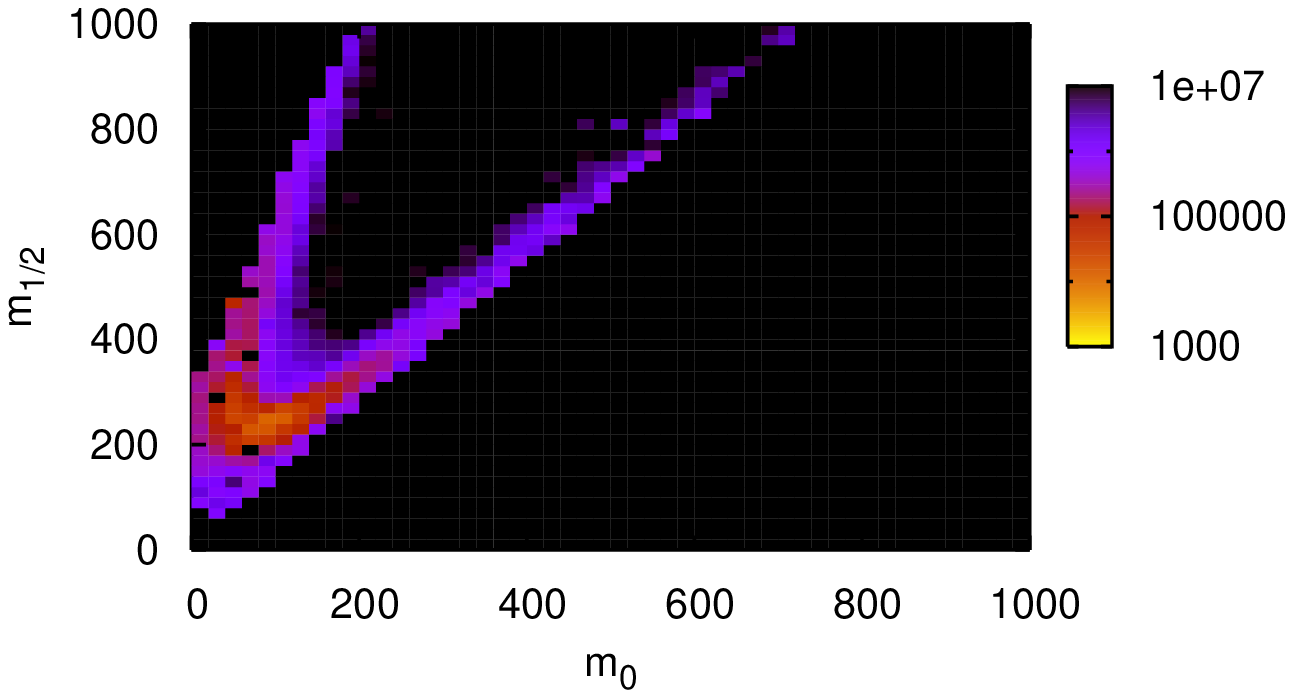}
\vspace{-1.2cm}
\caption{m$_0$--m$_{1/2}$ plane of SPS1a for LHC measurements after the Bayesian projection (from~\cite{Lafaye:2007vs}).} 
\label{fig:mSUGRAbayes}
\end{minipage}
\hspace{0.3cm}
\begin{minipage}[htb]{0.48\textwidth}
\centering
\vspace{-1.0cm}
\includegraphics[width=\textwidth]{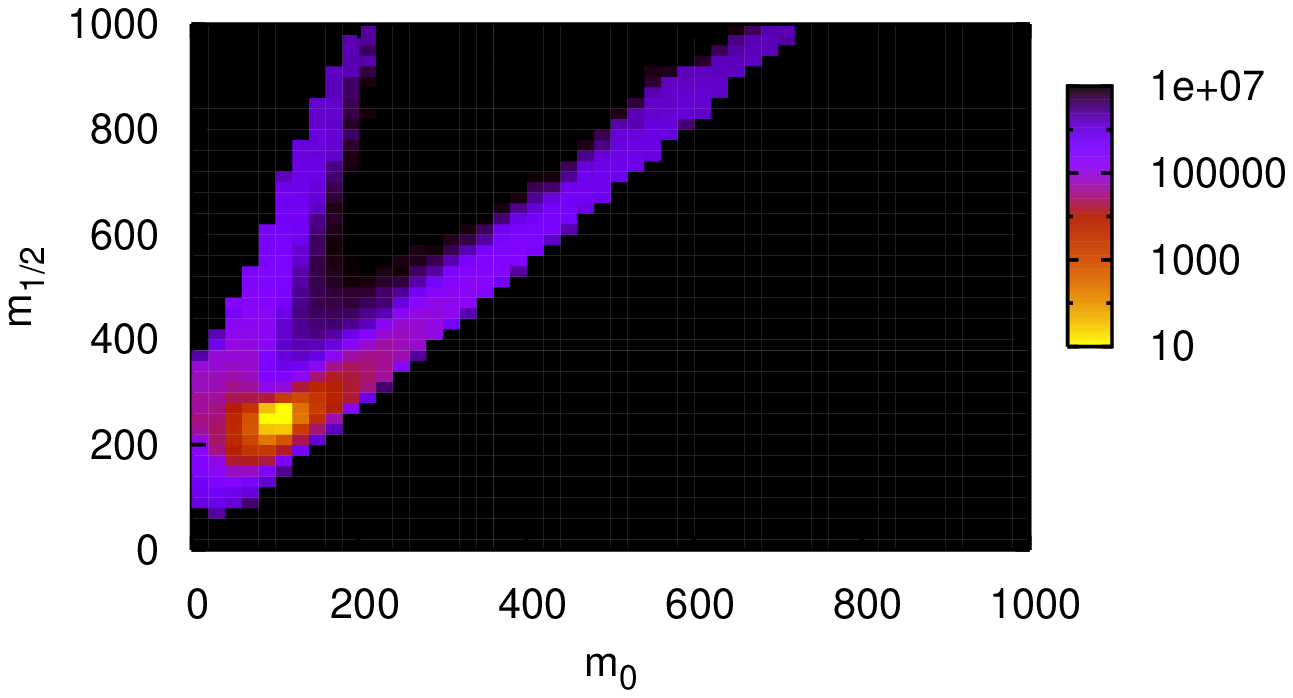}
\vspace{-1.2cm}
\caption{m$_0$--m$_{1/2}$ plane of SPS1a for LHC measurements after the frequentist projection (from~\cite{Lafaye:2007vs}).} 
\label{fig:mSUGRAprofile}
\end{minipage}
\end{figure}

An example is shown in Figure~\ref{fig:mSUGRAbayes} for the Bayesian approach 
with a flat prior and in Figure~\ref{fig:mSUGRAprofile}
for the frequentist approach
in the m$_0$--m$_{1/2}$ plane of SPS1a for LHC measurements.
While the general structure is similar and the correct parameters were determined, 
due to the definition of the frequentist projection the best--fit is brighter. 
In the Bayesian case the sharp maximum/minimum is washed out by the integration 
over large regions with small likelihood (noise effect).
One can also see the parabolic correlation between the parameters as expected from, e.g.,
the sfermion masses.

\begin{table}[htb]
    \begin{tabular}{l|rrrrrr}
     $\chi^2$&m$_0$ & m$_{1/2}$ &$\tan\beta$& A$_0$&$\mu$&$m_t$ \\ \hline
     0.09  &102.0 & 254.0 & 11.5 & -95.2  & $+$ & 172.4 \\
     1.50  &104.8 & 242.1 & 12.9 &-174.4  & $-$ & 172.3 \\
     73.2  &108.1 & 266.4 & 14.6 & 742.4  & $+$ & 173.7 \\
    139.5  &112.1 & 261.0 & 18.0 & 632.6  & $-$ & 173.0 \\
          \dots
     \end{tabular}
\caption{List of minima found in mSUGRA for the LHC measurements (taken from~\cite{Lafaye:2007vs}).}
\label{tab:mSUGRALHC}
\end{table}

The expected error on the top quark mass measurement is 1~\gevcc{} at the LHC and 
an order of magnitude more precise at the ILC (0.12~\gevcc{}). 
The top quark mass therefore has a double role, it is a parameter of the 
model~\cite{Allanach:2005kz,Allanach:2007qj,Allanach:2006jc,Allanach:2007qk}
which can have a large impact on the prediction of masses such as the Higgs boson 
mass~\cite{Haber:1996fp,Degrassi:2002fi,Frank:2006yh}.
The top quark mass is also treated as a measurement with the appropriate error in the
determination of the parameters. In principal all Standard Model parameters would need 
to be treated this way, however, the impact is practically negligible for the 
other parameters, so that they can be fixed to the central value. 
An exception is the b--quark mass for high $\tan\beta$ regions.

The list of secondary minima for the LHC measurements is shown in Table~\ref{tab:mSUGRALHC},
including the top quark mass as a parameter. Even in such a tightly constrained
model with a small number of parameters secondary minima can be observed in addition
to the correct parameter point (first line of Table~\ref{tab:mSUGRALHC}). 
It is particularly interesting to note that the addition of the top quark mass as a parameter
adds a secondary minimum (third line) in its interplay with tri-linear coupling.
However, all secondary minima can easily eliminated by the absolute value of the $\chi^2$
which is higher than for the correct SPS1a parameter set. 
A dataset which was smeared according to the expected errors, including all correlations, 
was used for this determination. Therefore a small shift in the parameters is expected as
the dataset does not correspond to the true value of SPS1a.

\subsection{Determining the Errors on the Parameters}

The second question is to determine the precision with which the parameters 
can be determined. At this point the theoretical errors
have to be specified. The Higgs boson mass prediction is known with a precision
of~3\gevcc{}. Strongly interacting particles such as squarks and gluinos have 
an error of 3\%. Electromagnetically and weakly interacting particles such
as neutralinos and sleptons have an uncertainty of 1\% of their mass 
prediction~\cite{Pierce:1996zz,Fritzsche:2004nf,Oller:2004br}.
The errors defined on the masses were propagated to the edges and thresholds.

\begin{table}[htb]
\begin{small}
\begin{tabular}{|l|r|c|cccc|}
\hline
            & SPS1a  & $\Delta^{\rm theo-exp}_{\rm zero}$
                     & $\Delta^{\rm expNoCorr}_{\rm zero}$ 
                     & $\Delta^{\rm theo-exp}_{\rm zero}$ 
                     & $\Delta^{\rm theo-exp}_{\rm gauss}$ 
                     & $\Delta^{\rm theo-exp}_{\rm flat}$ \\
\hline
            &        & masses 
                     & \multicolumn{4}{c|}{endpoints} \\
\hline
$m_0$       & 100    & 4.11 & 1.08 & 0.50 & 2.97 & 2.17 \\
$m_{1/2}$   & 250    & 1.81 & 0.98 & 0.73 & 2.99 & 2.64 \\
$\tan\beta$ & 10     & 1.69 & 0.87 & 0.65 & 3.36 & 2.45 \\
$A_0$       & -100   & 36.2 & 23.3 & 21.2 & 51.5 & 49.6 \\
$m_t$       & 171.4  & 0.94 & 0.79 & 0.26 & 0.89 & 0.97 \\
\hline
\end{tabular}
\end{small}
\caption[]{Errors on the parameter determination at the LHC in SPS1a. The
  big columns are mass and endpoint measurements.  The
  subscript represents neglected, (probably approximate) Gaussian or
  flat theory errors. The experimental error includes
  correlations unless indicated otherwise in the superscript (taken from~\cite{Lafaye:2007vs}).}
\label{tab:sugra_mass_edge}
\end{table}

At the LHC, one has the choice to either use the edges and thresholds
directly or the masses derived from the edges and 
thresholds~\cite{Bachacou:1999zb,Allanach:2000kt,Gjelsten:2005aw,Gjelsten:2004ki}. A comparison
of the expected precision is shown in Table~\ref{tab:sugra_mass_edge}. 
The precision is improved by the use of edges and thresholds instead of the
masses. In fact, the determination of the mass introduces non-negligible
correlations among them. As these correlations are not
available, there is a loss of information which translates into 
a less precise determination of the fundamental parameters by a factor of 
between two and three for \mZero{} and \mOneHalf{}.
Further evidence for the impact of correlations is seen by the improvement
of the error on \mZero{} when the correlation of the lepton and jet energy scale
errors are taken into account.

Theory errors have an impact on the determination of the parameters at the LHC.
As shown in Table~\ref{tab:sugra_mass_edge}, the proper inclusion of theory errors 
as flat errors reduces the LHC precision significantly. An example
is the Higgs boson mass which is sensitive to $\tan\beta$, but its theoretical
error is an order of magnitude worse than the expected experimental error.

Theoretical errors not only parametrize not-yet calculated high order corrections
but also variations of results between calculations at the same order.
As a check of the consistency of the theory errors, the SPS1a spectrum calculated by 
SoftSUSY was fitted with SuSpect. While the parameter central values were shifted,
they were within 3$\sigma$ of the correct ones adding confidence in the theory
errors.

The table of measurements at the LHC hides a part of the complexity of the determination
of the parameters: three neutralinos are observed, the third lightest, due to Higgsino
nature is not. Therefore one has to be able to infer from the data which 
of the neutralinos were observed. The most difficult example is to wrongly
label the measurement of the fourth neutralino as third. A parameter determination
can be performed and \mZero{} and \mOneHalf{} are off  by up to 1~\gevcc{}. 
The $\chi^2$ value of the fit remains reasonable. While this 
looks like a true difficult secondary minimum, the parameters determined
can be used to predict the spectrum, production cross section and branching ratios. 
Analyzing these results, a larger production of $\chi_4^0$ than $\chi_3^0$ is 
expected in disagreement to the labeling, discarding effectively this secondary 
minimum. It is clear that a lot of thought 
has to go into the definition of the table of observables to make sure that all
underlying/hidden assumptions are tested. An automated algorithm for all does not
seem feasible. 

\begin{table}[htb]
\begin{tabular}{|l|r|ccc|ccc|}
\hline
            & SPS1a  
                     & $\Delta_{\rm endpoints}$ 
                     & $\Delta_{\rm ILC}$ 
                     & $\Delta_{\rm LHC+ILC}$ 
                     & $\Delta_{\rm endpoints}$ 
                     & $\Delta_{\rm ILC}$ 
                     & $\Delta_{\rm LHC+ILC}$ \\
\hline
            &        & \multicolumn{3}{c|}{exp. errors}
                     & \multicolumn{3}{c|}{exp. and theo. errors} \\
\hline
$m_0$       & 100    & 0.50 & 0.18  & 0.13  & 2.17 & 0.71 & 0.58 \\
$m_{1/2}$   & 250    & 0.73 & 0.14  & 0.11  & 2.64 & 0.66 & 0.59 \\
$\tan\beta$ & 10     & 0.65 & 0.14  & 0.14  & 2.45 & 0.35 & 0.34 \\
$A_0$       & -100   & 21.2 & 5.8   & 5.2   & 49.6 & 12.0 & 11.3 \\
$m_t$       & 171.4  & 0.26 & 0.12  & 0.12  & 0.97 & 0.12 & 0.12 \\
\hline
\end{tabular}
\caption[]{Expected errors for mSUGRA at the LHC (endpoints) and 
  ILC. Only absolute errors are given. Flat theory
  errors are used (taken from~\cite{Lafaye:2007vs}).}
\label{tab:sugra_ilc}
\end{table}

The ILC will start taking data after the LHC has taken 
data and discovered SPS1a (if nature is kind enough).
The added value of the ILC is the completion of the particle 
spectrum and the measurement of all particles with an impressive
accuracy, about an order of magnitude improved on the LHC. 
The results are shown in Table~\ref{tab:sugra_ilc}.
The precision improvement by the ILC per se is about a factor 
of five. The improvement of the combination LHC and ILC over
ILC alone is rather marginal. Comparing
the LHC+ILC errors with and without theory errors shows the margin for
the improvement of theory predictions, as shown already for the LHC, justifying the SPA
project~\cite{AguilarSaavedra:2005pw}. The SPA project's goal is 
to improve the theory predictions for the supersymmetry (S) parameter (P) 
analysis (A).

\section{Measuring the MSSM}

mSUGRA has the advantage of being a model with few parameters which
is mostly defined at the GUT scale. The unification of breaking parameters 
at the GUT scale is assumed.
The MSSM, defined at the electroweak scale, allows to 
measure grand unification without imposing it.
The MSSM parameter determination
is far more difficult technically: more parameters have to be 
determined from the same dataset. The reward is to be able to 
extrapolate the MSSM parameters to the GUT scale and thus measure grand unification
or other models such as GMSB.

\subsection{Determination of the True Central Value}

\begin{table}[p]
\begin{tabular}{|l|rrrr|rrrr|}
\hline
                     &\multicolumn{4}{c|}{$\mu<0$}&\multicolumn{4}{c|}{$\mu>0$}\\
\hline
                     &      &      &      &      &SPS1a&      &      &       \\
\hline \hline
$M_1$                &  96.6& 175.1& 103.5& 365.8&  98.3& 176.4& 105.9& 365.3\\
$M_2$                & 181.2&  98.4& 350.0& 130.9& 187.5& 103.9& 348.4& 137.8\\
$\mu$                &-354.1&-357.6&-177.7&-159.9& 347.8& 352.6& 178.0& 161.5\\
$\tan\beta$          &  14.6&  14.5&  29.1&  32.1&  15.0&  14.8&  29.2&  32.1\\
\hline
$M_3$                & 583.2& 583.3& 583.3& 583.5& 583.1& 583.1& 583.3& 583.4\\
$M_{\tilde{\tau}_L}$ & 114.9&2704.3& 128.3&4794.2& 128.0& 229.9&3269.3& 118.6\\
$M_{\tilde{\tau}_R}$ & 348.8& 129.9&1292.7& 130.1&2266.5& 138.5& 129.9& 255.1\\
$M_{\tilde{\mu}_L}$  & 192.7& 192.7& 192.7& 192.9& 192.6& 192.6& 192.7& 192.8\\
$M_{\tilde{\mu}_R}$  & 131.1& 131.1& 131.1& 131.3& 131.0& 131.0& 131.1& 131.2\\
$M_{\tilde{e}_L}$    & 186.3& 186.4& 186.4& 186.5& 186.2& 186.2& 186.4& 186.4\\
$M_{\tilde{e}_R}$    & 131.5& 131.5& 131.6& 131.7& 131.4& 131.4& 131.5& 131.6\\
$M_{\tilde{q}3_L}$   & 497.1& 497.2& 494.1& 494.0& 495.6& 495.6& 495.8& 495.0\\
$M_{\tilde{t}_R}$    &1073.9& 920.3& 547.9& 950.8& 547.9& 460.5& 978.2& 520.0\\
$M_{\tilde{b}_R}$    & 497.3& 497.3& 500.4& 500.9& 498.5& 498.5& 498.7& 499.6\\
$M_{\tilde{q}_L}$    & 525.1& 525.2& 525.3& 525.5& 525.0& 525.0& 525.2& 525.3\\
$M_{\tilde{q}_R}$    & 511.3& 511.3& 511.4& 511.5& 511.2& 511.2& 511.4& 511.5\\
$A_t$ $(-)$          &-252.3&-348.4&-477.1&-259.0&-470.0&-484.3&-243.4&-465.7\\
$A_t$ $(+)$          & 384.9& 481.8& 641.5& 432.5& 739.2& 774.7& 440.5& 656.9\\
$m_A$                & 350.3& 725.8& 263.1&1020.0& 171.6& 156.5& 897.6& 256.1\\
$m_t$                & 171.4& 171.4& 171.4& 171.4& 171.4& 171.4& 171.4& 171.4\\
\hline
\end{tabular}
\caption{Eight best--fitting points in the MSSM with two alternative solutions for $A_t$.
  The $\chi^2$ value for all points 
  is approximately the same, so the ordering of the table is arbitrary. 
  The parameter point closest to the correct point is labeled as SPS1a (taken from~\cite{Lafaye:2007vs}).}
\label{tab:mssm_secondary}
\end{table}

The problem of the higher--dimensional MSSM
parameter space with respect to the mSUGRA case is solved by a combination of 
techniques (for details see~\cite{Lafaye:2007vs}).
In particular a sequence of Markov chains and gradient fits is used.
A second Markov chain step combined with the gradient fit is 
performed in the gaugino--Higgsino subspace and the other parameters
are re-determined subsequently.
This complex procedure leads to the identification of the primary
minimum/maximum and an eightfold ambiguity listed in Table~\ref{tab:mssm_secondary}.
Additionally, alternative
likelihood maxima are triggered by correlations between the rather poorly
measured parameters $A_t$, $\tan\beta$ and the right--handed stop mass.

\begin{figure}[htb] 
\begin{minipage}[htb]{0.48\textwidth}
\centering
\includegraphics[width=\columnwidth]{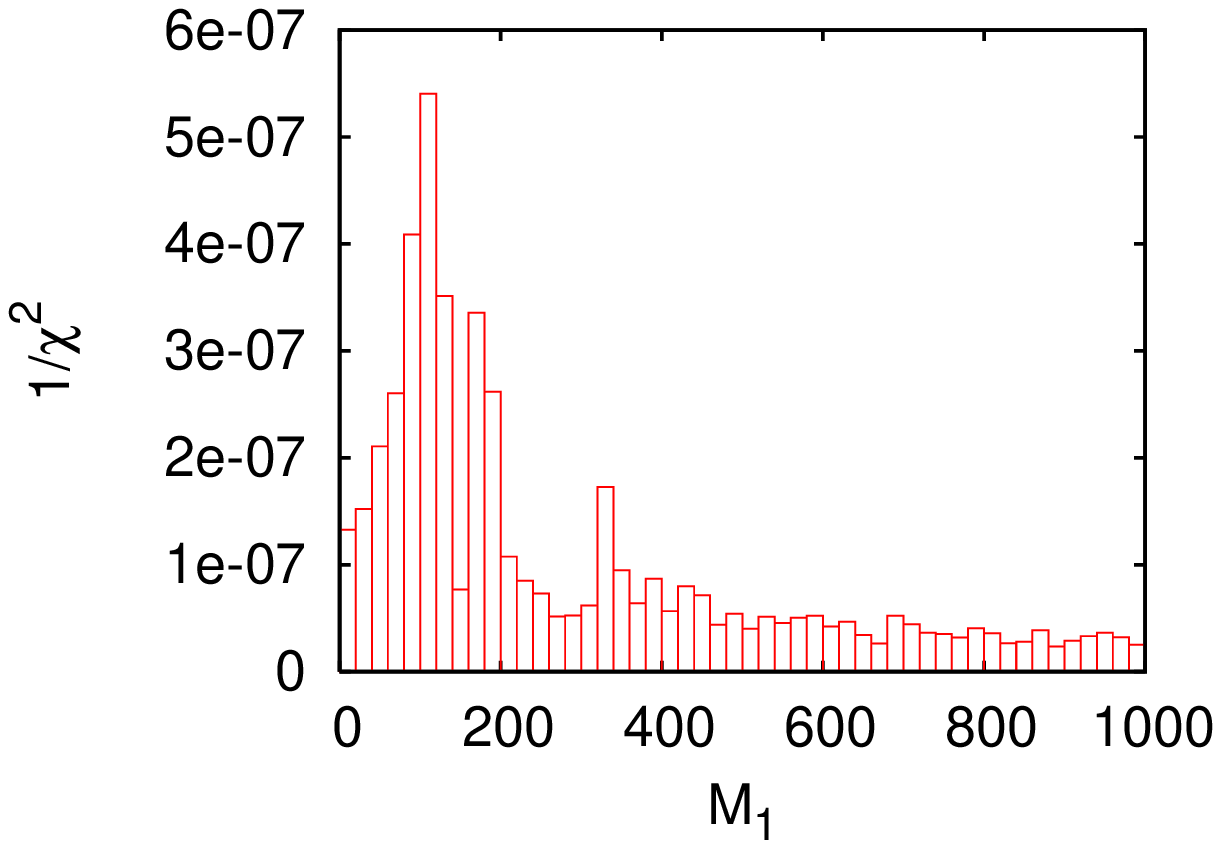} 
\caption[]{The Bayesian projection for \Mone{} is shown for LHC data~\cite{Lafaye:2007vs}).}
\label{fig:m1Bayes}
\end{minipage}
\hspace*{0.2cm}
\begin{minipage}[htb]{0.48\textwidth}
\centering
\includegraphics[width=\columnwidth]{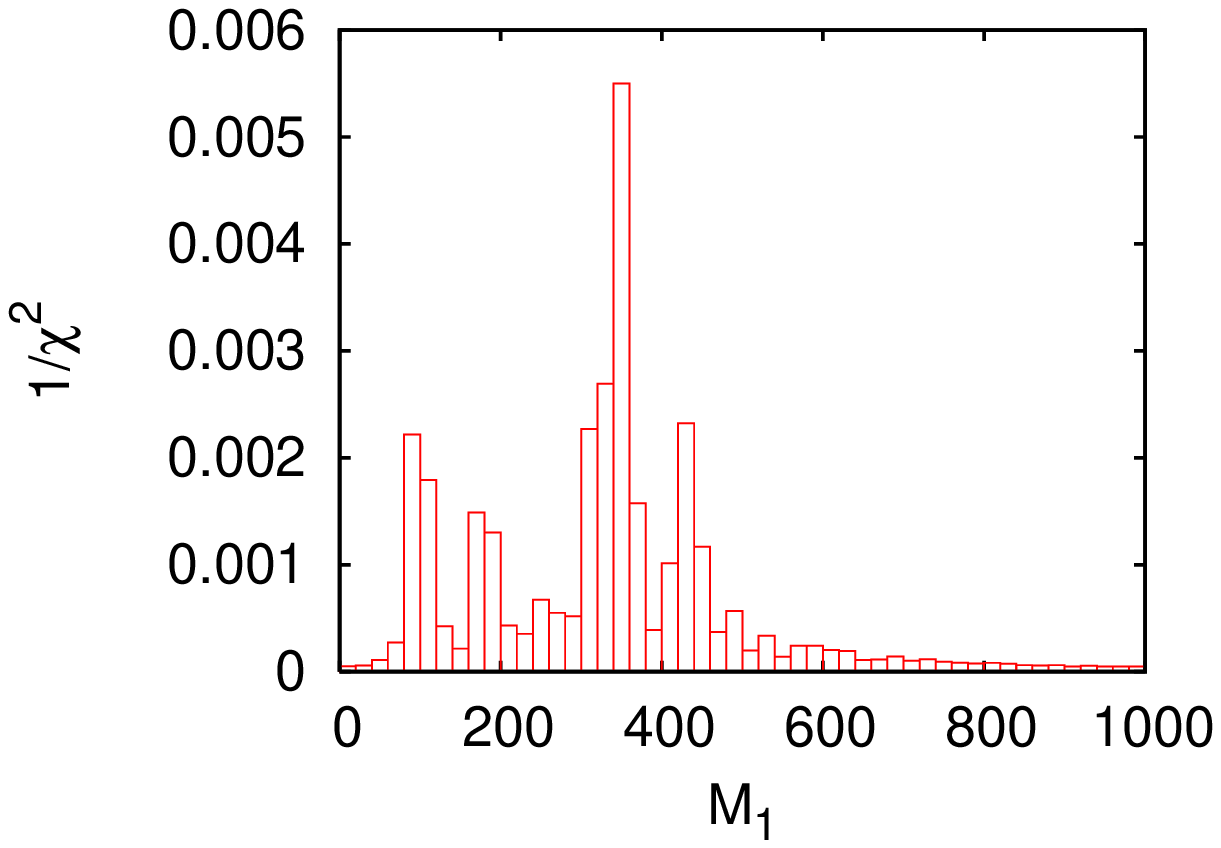} \\
\caption[]{The profile likelihood for \Mone{} is shown for LHC data~\cite{Lafaye:2007vs}).}
\label{fig:m1Prof}
\end{minipage}
\end{figure}

Solutions are observed for the two signs of $\mu$ and for each sign
of $\mu$ four solutions are determined, two of which can be grouped
together by interchanging \Mone{} and \Mtwo{}.
The correct solution differs from the nominal values of SPS1a 
as a smeared dataset was used for this study.
In Figure~\ref{fig:m1Bayes} the Bayesian projection is shown and in Figure~\ref{fig:m1Prof}
the frequentist projection is shown. While in the Bayesian projection the central values
are washed out by noise effects, the profile likelihood allows to identify the location
of the ambiguities.

\subsection{Determining the Errors}

In principle 22 measurements are available at the LHC, thus technically enough
to determine the 19 parameters of the MSSM (including the top quark mass as
a parameter). However, these measurements come from only 15 different masses. 
Therefore, inevitably, some parameters must be fixed in order to determine the errors
with no impact on the determination as these parameters have 
essentially a very weak impact on the observables.
The result is shown in Table~\ref{tab:mssm_ilc} including flat theory errors.

\begin{table}[p]
\begin{tabular}{|l|r@{$\pm$}rr@{$\pm$}rr@{$\pm$}rr|}
\hline
                     & \multicolumn{2}{c}{LHC}    & \multicolumn{2}{c}{ILC}     & \multicolumn{2}{c}{LHC+ILC} & SPS1a \\
\hline
$\tan\beta$          &      10.0 & 4.5             &      12.1 & 7.0             &      12.6 & 6.2             &     10.0 \\
$M_1$                &     102.1 & 7.8             &     103.3 & 1.1             &     103.2 & 0.95            &    103.1 \\
$M_2$                &     193.3 & 7.8             &     194.1 & 3.3             &     193.3 & 2.6             &    192.9 \\
$M_3$                &     577.2 & 14.5            &\multicolumn{2}{c}{fixed 500}&     581.0 & 15.1            &    577.9 \\
$M_{\tilde{\tau}_L}$ &     227.8 & $\mathcal{O}(10^3)$     &     190.7 & 9.1             &     190.3 & 9.8             &    193.6 \\
$M_{\tilde{\tau}_R}$ &     164.1 & $\mathcal{O}(10^3)$     &     136.1 & 10.3            &     136.5 & 11.1            &    133.4 \\
$M_{\tilde{\mu}_L}$  &     193.2 & 8.8             &     194.5 & 1.3             &     194.5 & 1.2             &    194.4 \\
$M_{\tilde{\mu}_R}$  &     135.0 & 8.3             &     135.9 & 0.87            &     136.0 & 0.79            &    135.8 \\
$M_{\tilde{e}_L}$    &     193.3 & 8.8             &     194.4 & 0.91            &     194.4 & 0.84            &    194.4 \\
$M_{\tilde{e}_R}$    &     135.0 & 8.3             &     135.8 & 0.82            &     135.9 & 0.73            &    135.8 \\
$M_{\tilde{q}3_L}$   &     481.4 & 22.0            &     499.4 &$\mathcal{O}(10^2)$      &     493.1 & 23.2            &    480.8 \\
$M_{\tilde{t}_R}$    &     415.8 & $\mathcal{O}(10^2)$     &     434.7 &$\mathcal{O}(4\cdot10^2)$&     412.7 & 63.2            &    408.3 \\
$M_{\tilde{b}_R}$    &     501.7 & 17.9            &\multicolumn{2}{c}{fixed 500}&     502.4 & 23.8            &    502.9 \\
$M_{\tilde{q}_L}$    &     524.6 & 14.5            &\multicolumn{2}{c}{fixed 500}&     526.1 & 7.2             &    526.6 \\
$M_{\tilde{q}_R}$    &     507.3 & 17.5            &\multicolumn{2}{c}{fixed 500}&     509.0 & 19.2            &    508.1 \\
$A_\tau$             &\multicolumn{2}{c}{fixed 0}  &     613.4 & $\mathcal{O}(10^4)$     &     764.7 & $\mathcal{O}(10^4)$     &   -249.4 \\
$A_t$                &    -509.1 & 86.7            &    -524.1 & $\mathcal{O}(10^3)$     &    -493.1 & 262.9           &   -490.9 \\
$A_b$                &\multicolumn{2}{c}{fixed 0}  &\multicolumn{2}{c}{fixed 0}  &     199.6 & $\mathcal{O}(10^4)$     &   -763.4 \\
$A_{l1,2}$           &\multicolumn{2}{c}{fixed 0}  &\multicolumn{2}{c}{fixed 0}  &\multicolumn{2}{c}{fixed 0}  &   -251.1 \\
$A_{u1,2}$           &\multicolumn{2}{c}{fixed 0}  &\multicolumn{2}{c}{fixed 0}  &\multicolumn{2}{c}{fixed 0}  &   -657.2 \\
$A_{d1,2}$           &\multicolumn{2}{c}{fixed 0}  &\multicolumn{2}{c}{fixed 0}  &\multicolumn{2}{c}{fixed 0}  &   -821.8 \\
$m_A$                &     406.3 & $\mathcal{O}(10^3)$     &     393.8 & 1.6             &     393.7 & 1.6             &    394.9 \\
$\mu$                &     350.5 & 14.5            &     354.8 & 3.1             &     354.7 & 3.0             &    353.7 \\
$m_t$                &     171.4 & 1.0             &     171.4 & 0.12            &     171.4 & 0.12            &    171.4 \\
\hline
\end{tabular}
\caption[]{Results for the general MSSM parameter determination in
  SPS1a using flat theory errors. The
  kinematic endpoint measurements are used for the LHC 
  and the mass measurements for the ILC. The LHC+ILC column
  is the combination of the two measurements sets. Shown are the nominal
  parameter values and the result after fits to the different data
  sets (taken from~\cite{Lafaye:2007vs}.}
\label{tab:mssm_ilc}
\end{table}

The stau sector is inevitably poorly determined at the LHC as only one measurement is available 
for the parameters. The stop sector (no measurements in SPS1a) can only be determined
indirectly with a penalty paid in the size of the error. For the ILC several parameters have to 
be fixed as the strongly interacting particles such as squarks and gluinos with the exception of the
lightest stop cannot be produced. The historical motivation for fixing the squark parameters 
in the ILC study was the possibility of concurrent running of LHC and ILC. As this is no longer an
option the more realistic scenario will be that the parameters parameters 
for the ILC-alone study will be fixed to the measurements of the LHC. The fixed parameters in the ILC column
of Table~\ref{tab:mssm_ilc} are within 5\%-15\% of such a scenario.
The main impact of the combination of LHC and ILC 
is observed in the last column of Table~\ref{tab:mssm_ilc}, where all parameters
can be determined, with the exception of the tri-linear couplings of the first two generation, which
have no effect on the observables in the selectron and smuon sector.

Comparing the effect of taking into account the theoretical errors, about a
factor five in precision is lost with today's calculations at the ILC. Its 
precision is decreased
from per mil to 0.5\%. For the LHC roughly a factor two is lost. 
The combination of LHC and
ILC measurements can be particularly useful to determine the link
to dark--matter 
observables~\cite{Buchmueller:2007zk,Ellis:2007fu,Roszkowski:2007va,Trotta:2007pg,Trotta:2006ew,de Austri:2006pe,Ellis:2004bx,Ellis:2003si,Baltz:2006fm}.

\subsection{Extrapolating to the GUT Scale}

\begin{figure}[htb]
\begin{minipage}[htb]{0.48\textwidth}
\centering
\includegraphics[width=\columnwidth]{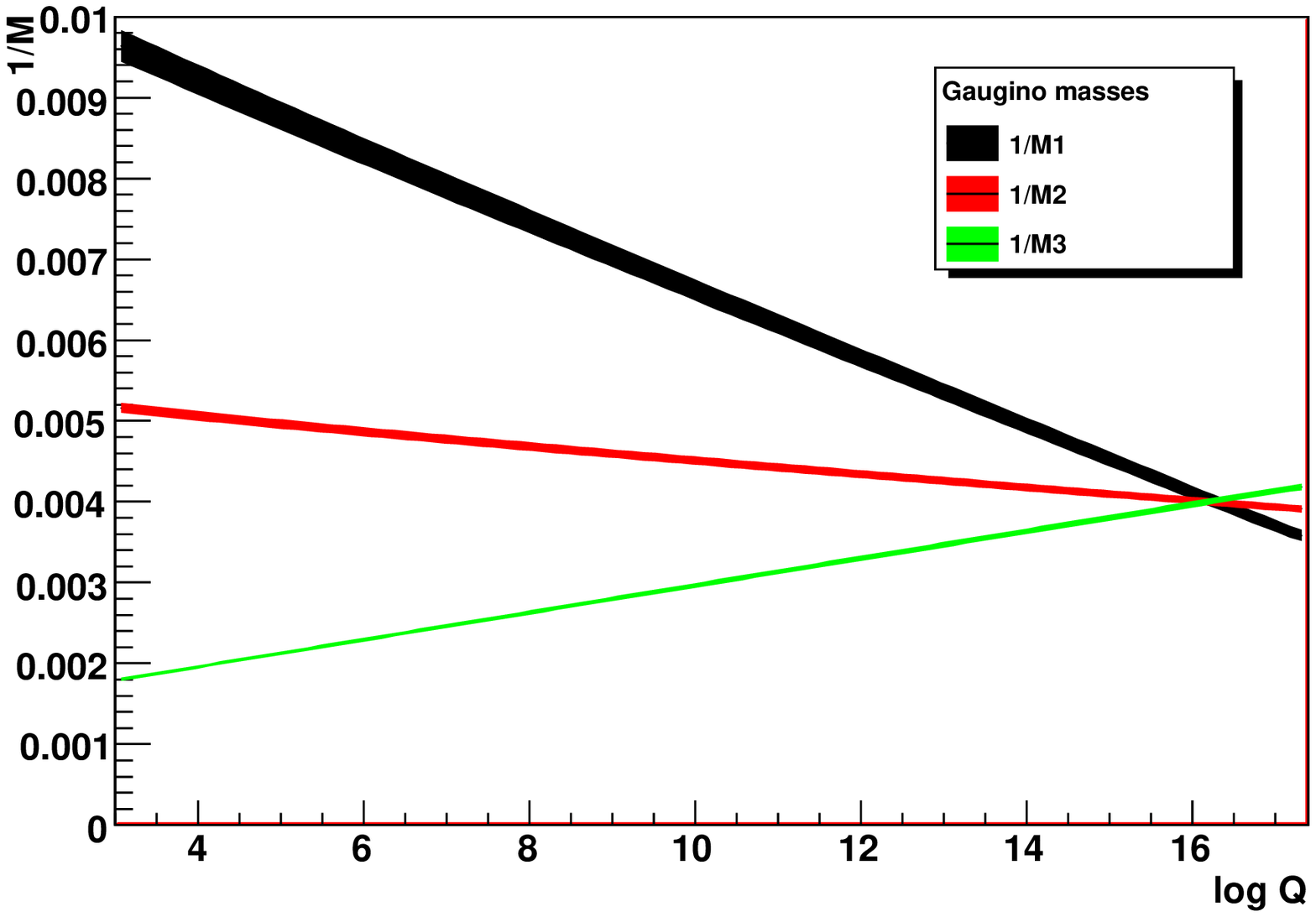}
\vspace{-1.0cm}
\caption{Extrapolation of the inverse of the gaugino mass parameters to the GUT scale for 
the correct solution at the LHC.}
\label{fig:ExtraPoMiCorrect}
\end{minipage}
\hspace{0.2cm}
\begin{minipage}[htb]{0.48\textwidth}
\centering
\includegraphics[width=\columnwidth]{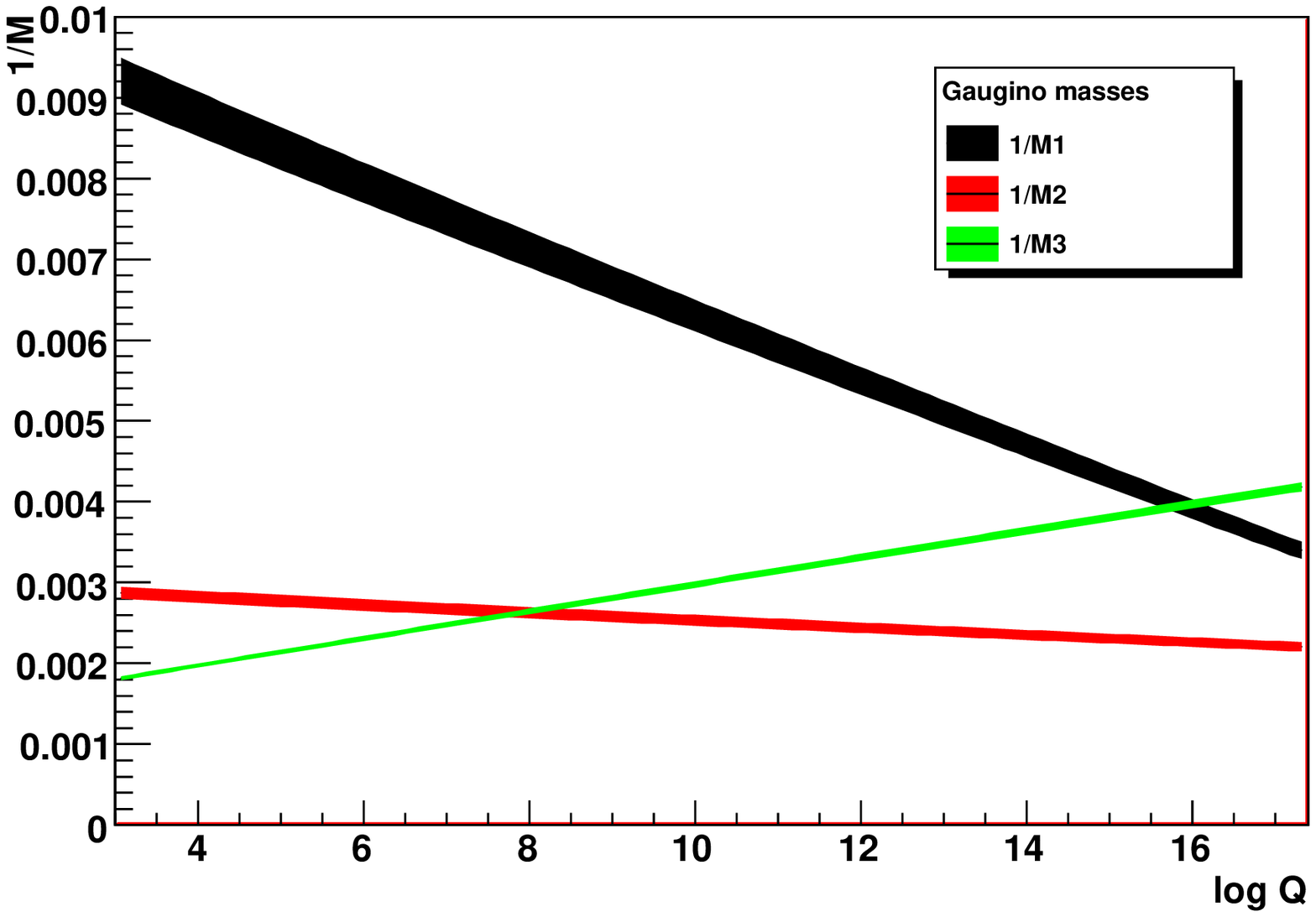}
\vspace{-1.0cm}
\caption{Extrapolation of the inverse of the gaugino mass parameters to the GUT scale for 
a false solution at the LHC.}
\label{fig:ExtraPoMiFalse}
\end{minipage}
\end{figure}

Once the parameters of the weak--scale MSSM--Lagrangian have been
determined, the next step is to extrapolate the parameters all the way
to the Planck scale. Inspired by the apparent unification of the gauge
couplings~\cite{Amaldi:1991cn} in the MSSM the question arises if any
other running parameters unify at a higher scale,
as shown in the pioneering work in~\cite{Blair:2002pg,Allanach:2004ud}.

Technically, upwards running is considerably more
complicated~\cite{Kneur:2008ur} than starting from a
unification--scale and testing the unification hypothesis by comparing
to the weak--scale particle spectrum. For example, it is not
guaranteed that the renormalization group running will converge for
weak--scale input values far away from the top--down prediction. 

\begin{figure}[htb]
\begin{minipage}[htb]{0.48\textwidth}
\centering
\includegraphics[width=\columnwidth]{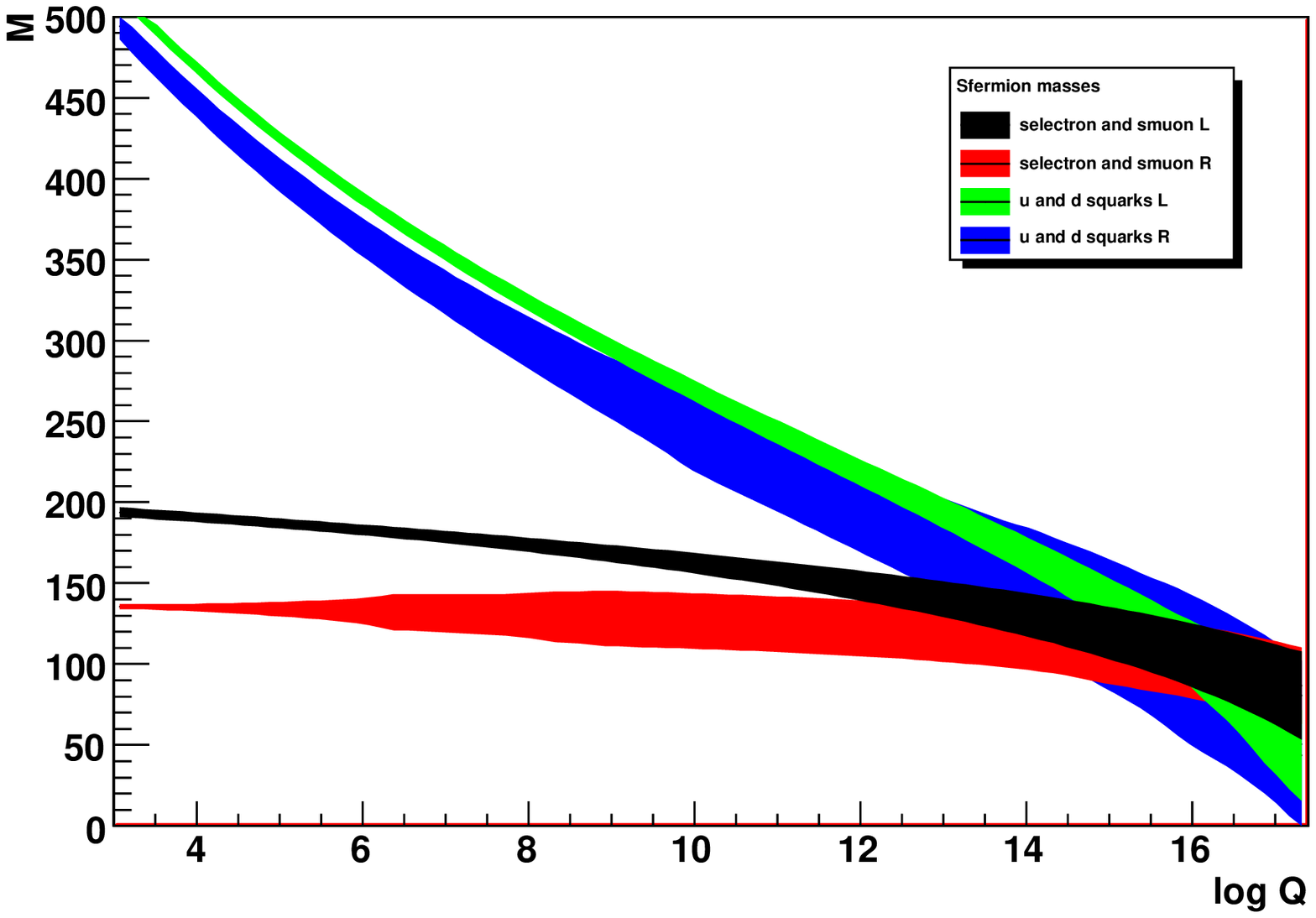}
\vspace{-1.0cm}
\caption{Extrapolation of the first and second generation scalar mass parameters to the GUT scale for 
the correct solution at the LHC.}
\label{fig:ExtraPoM0Correct}
\end{minipage}
\hspace{0.2cm}
\begin{minipage}[htb]{0.48\textwidth}
\centering
\includegraphics[width=\columnwidth]{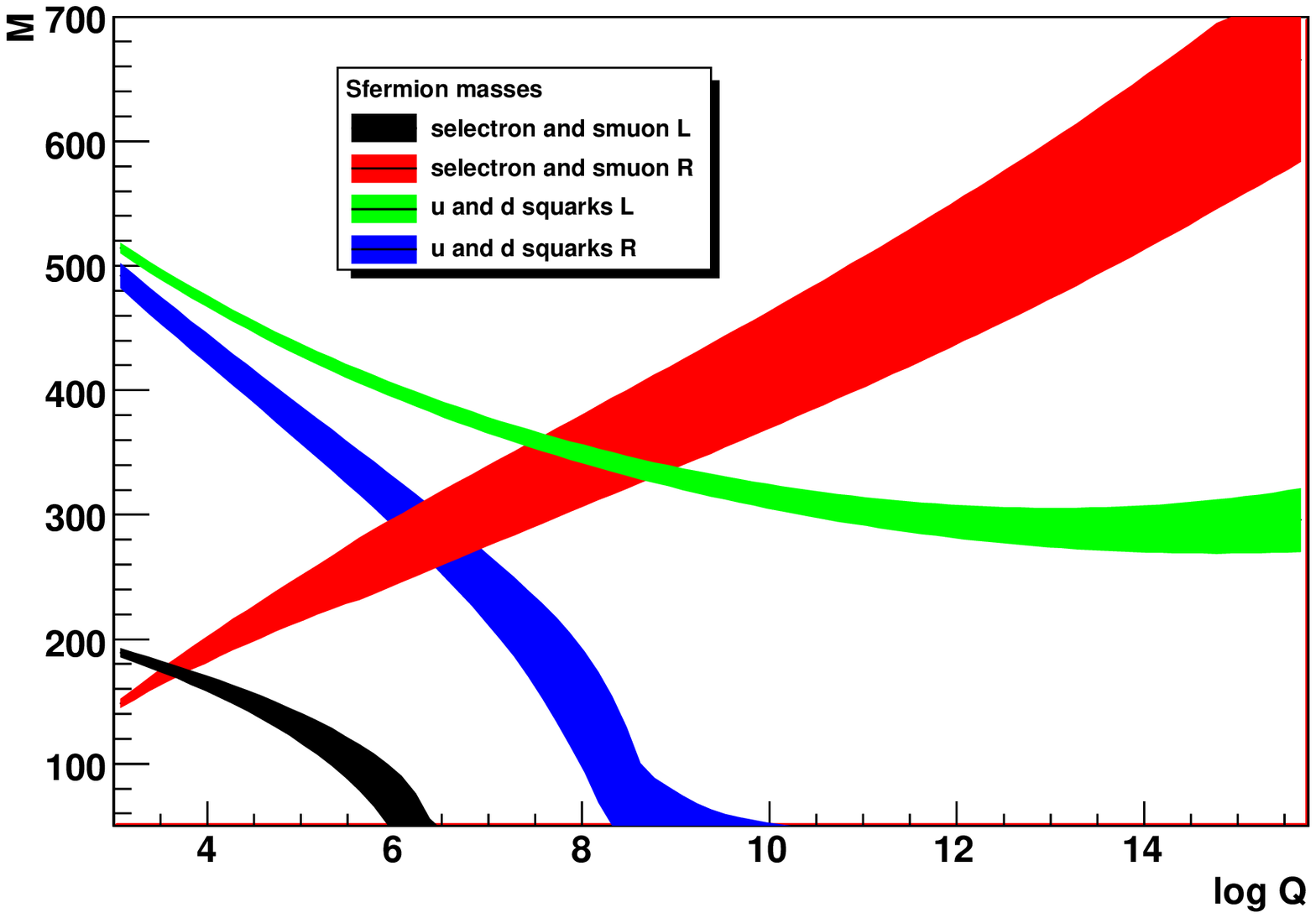}
\vspace{-1.0cm}
\caption{Extrapolation of the first and second generation scalar mass parameters to the GUT scale for 
a false solution at the LHC.}
\label{fig:ExtraPoM0False}
\end{minipage}
\end{figure}

The first step in the extrapolation of the weak scale parameters is to test the
expected measurement of the gaugino mass parameters at the LHC. Figure~\ref{fig:ExtraPoMiCorrect} 
shows the result for the true parameter point. In Figure~\ref{fig:ExtraPoMiFalse} the extrapolation
is performed for 
one of the ambiguous solutions. While grand unification is observed for the correct parameter set 
as expected, in the second case unification is not observed.
The non-unification of six of the solutions is easy to observe. The seventh solution
shows a close miss of unification. It will therefore depend on
the exact precision of the LHC measurement whether it can be distinguished from the correct 
solution which unifies the parameters at the GUT scale. 

For the extrapolation of the scalar mass parameters at the LHC alone the picture 
is even more difficult. In Figure~\ref{fig:ExtraPoM0Correct} the unification of 
the first and second generation scalar mass parameters is observed
for the correct solution, however in Figure~\ref{fig:ExtraPoM0False} the stau parameters
have been moved to different values. This behavior can be understood from the RGEs. In the gaugino
sector the RGEs do not depend on other parameters. In the scalar sector the RGEs are strongly
coupled, so that a wrong value of a stau parameter can wreak havoc in the extrapolation 
of the first and second generation scalar parameters.

If no further measurements are thought of for the SPS1a scenario, the proof and measurement
of grand unification of supersymmetric parameters will have to wait 
until the ILC resolves the ambiguities.
With the ILC a unique solution of the MSSM parameters is determined, 
which can then be extrapolated
to truly measure grand unification of the supersymmetric breaking parameters as shown in 
Chapter~\ref{chap:conclusion}.

\section{Measuring the Higgs Sector}

A more difficult scenario at the LHC would be to discover only a Higgs boson
in two-particle decays.
The question arises naturally if there is more to learn from this 
sector than the Higgs boson mass, which is used also in the supersymmetric 
parameter determination. In analogy to the measurement of the Triple Gauge Couplings
(TGC) at LEP, one can parametrize the Higgs couplings as deviation from the Standard 
Model expectation. 

Thus a tree-level Standard Model Higgs coupling to any particle $j$
is defined as:
\begin{equation}
 g_{jjH} \longrightarrow g_{jjH}^{\text{SM}} \; \left( 1 + \Delta_{jjH}
                                   \right)
\label{eq:Coupl_SM_Change}
\end{equation}
where the $\Delta_{jjH}$ are independent of each other. 
The three loop induced couplings in the
Higgs sector, $g_{ggH}, g_{\gamma\gamma H}$ and $g_{\gamma ZH}$, sensitive
to new physics~\cite{Cacciapaglia:2009ky,Barger:2003rs,Grojean:2004xa,Kanemura:2004mg}, 
are defined in the following way:
\begin{equation}
 g_{jjH} \longrightarrow 
  g_{jjH}^{\text{SM}} \; \left(
   1 + \Delta_{jjH}^{\text{tree}} + \Delta_{jjH} \right)
\label{eq:Coupl_SMeff_Change}
\end{equation}
where $g_{jjH}^{\text{SM}}$ is the loop-induced coupling in the
Standard Model, $\Delta_{jjH}^{\text{tree}}$ is the contribution
from modified tree-level couplings to Standard-Model particles (in the loop), 
$\Delta_{jjH}$ is an additional contribution proper for this effective coupling.

In the scenario treated in this analysis, a Higgs boson mass of 120~\gevcc{},
the total width is defined as the sum of the measured partial widths. 
As the partial width of the decay Higgs to bb is 90\% of the total width,
it is measured reasonably well at the LHC.

The total width can be varied just like the
couplings
\begin{equation} 
\Gamma_\mathrm{tot} = \Gamma^\mathrm{SM}_\mathrm{tot} \; ( 1+\Delta_\Gamma )
\quad .
\label{eq:deltagamma}
\end{equation} 
This parameter can be used of an unobserved channel/anomalous channel.

\subsection{Determination of the True Couplings}
\label{sec:likelihood}

\begin{figure}[htb]
\begin{minipage}[htb]{0.48\textwidth}
\centering
\includegraphics[width=\columnwidth]{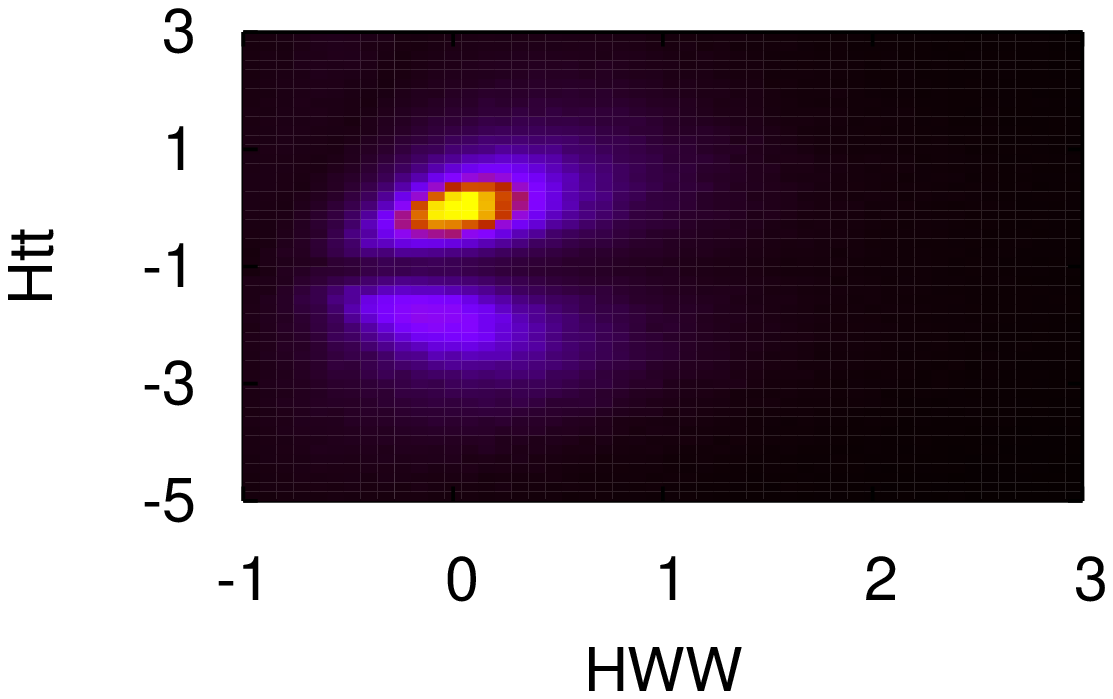}
\vspace{-1.0cm}
\caption{Profile likelihood for the $ttH$ and $WWH$ couplings for 30~\fbinv{} (from~\cite{Lafaye:2009vr}).}
\label{fig:HiggsBayes}
\end{minipage}
\hspace{0.2cm}
\begin{minipage}[htb]{0.48\textwidth}
\centering
\includegraphics[width=\columnwidth]{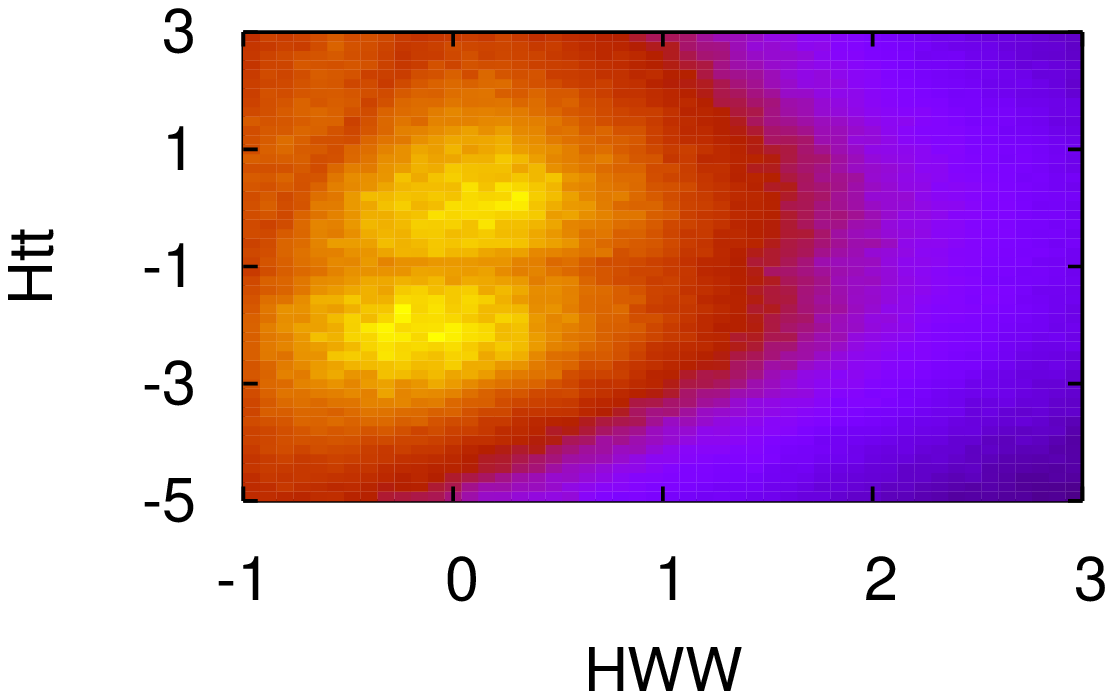}
\vspace{-1.0cm}
\caption{Bayesian projection for the $ttH$ and $WWH$ couplings for 30~\fbinv{} (from~\cite{Lafaye:2009vr}).}
\label{fig:HiggsProfile}
\end{minipage}
\end{figure}

The procedure for the determination of the Higgs boson couplings
follows the strategy applied in the determination of the supersymmetric
couplings. An exclusive likelihood map is calculated for the 
parameters $\Delta jjH$, the deviations from the Standard Model couplings.
From this map different
projections are possible to study correlations.

The analysis is performed for an integrated luminosity of 30~\fbinv{}. 
The errors cannot be scaled simply to 300~\fbinv{} since e.g., 
the systematic error on the mini-jet veto for nominal LHC luminosity is not known.
The first parameter set used for the true dataset (without smearing)
are the tree-level couplings: $WWH$, $ZZH$, $ttH$, $\tau\tau H$ and no genuine
anomalous contribution to the effective couplings is allowed.
The overall phase of the couplings is fixed by requiring $WWH$ to be positive
without loss of generality.

While in the supersymmetric case genuine secondary minima can arise, e.g.,
particularly in the MSSM, in the Higgs couplings no such effect is expected.
In fact, the definition of the parameters leads to quadratic dependence 
of the observables on the parameters.  

The Bayesian projection is shown in
Figure~\ref{fig:HiggsBayes} and the frequentist approach 
of maintaining always the absolute minimum/maximum is shown in Figure~\ref{fig:HiggsProfile}.
From the comparison of the two figures one can conclude that the important integration 
(noise) effect renders the Bayesian approach less performant than the profile likelihood.

The profile likelihood in Figure~\ref{fig:HiggsProfile} shows a well identified correct solution.
It is interesting to note that in spite of the observables being essentially quadratic in 
the parameters, a clear preference for the correct solution is observed for Htt. 
The reason is that the rate of Higgs to two photons is sensitive to the relative sign.

In general, a positive correlation among the couplings is observed among the couplings
induced by the general $(\sigma \cdot \mathrm{BR})$ structure 
of the LHC measurements. A change in the $bbH$ coupling will induce
a coherent compensation on the production side and decay 
side of the observables as the total width in the denominator 
is dominated by the $bbH$ partial width.

\begin{figure}[htb]
\begin{minipage}[htb]{0.48\textwidth}
\centering
 \includegraphics[width=\columnwidth]{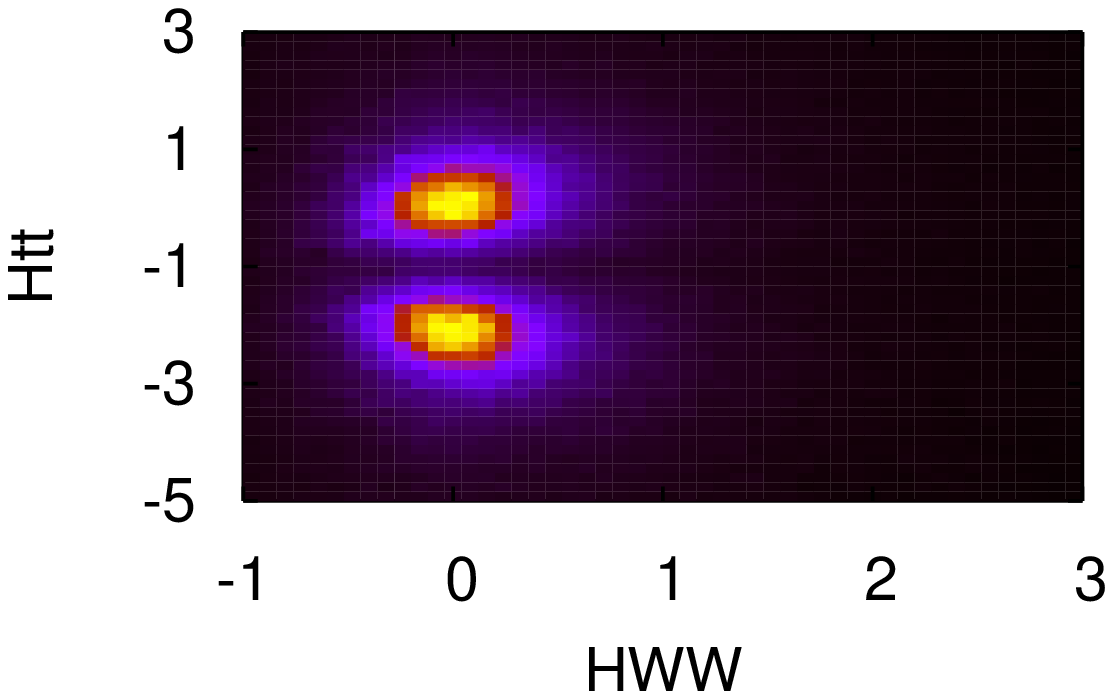}
\vspace{-1.0cm}
\caption[]{Profile likelihoods for $WWH$ versus $ttH$
allowing genuine contributions to the effective couplings (taken from~\cite{Lafaye:2009vr}).}
\label{fig:HiggsHWWHtt}
\end{minipage}
\hspace{0.2cm}
\begin{minipage}[htb]{0.48\textwidth}
 \includegraphics[width=\columnwidth]{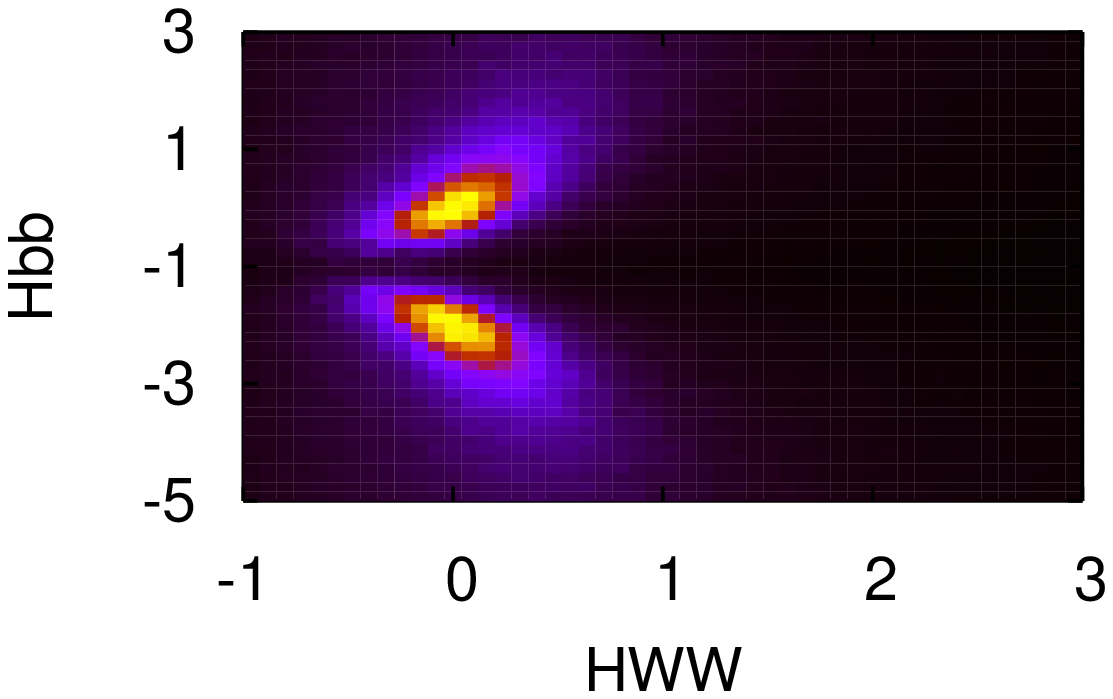}
\vspace{-1.0cm}
\caption[]{Profile likelihoods for $WWH$ versus $bbH$
allowing genuine contributions to the effective couplings (taken from~\cite{Lafaye:2009vr}).}
\label{fig:HiggsHWWHbb}
\end{minipage}
\end{figure}

In Figure~\ref{fig:HiggsHWWHtt} and Figure~\ref{fig:HiggsHWWHbb} the correlations
of the $WWH$ with the $ttH$ and $bbH$ couplings are shown. Here the effective couplings 
are left free and are part of the parameter set to be determined.
The general feature of a positive correlation is still observed, especially clear
for the $bbH$ versus $WWH$ coupling. However, the sign sensitivity of the $\gamma\gamma$ channel
is lost 
due to the additional freedom added by the effective anomalous couplings.

While these results seem to indicate that the absolute couplings of the Higgs boson
can be determined at the LHC, a simple scenario shows that this is not the case.
In the studies discussed above, the $ccH$ coupling was fixed to its Standard Model 
value. What would be the effect of a large anomalous $ccH$ coupling? The observed
rates would be diminished by $\Gamma_H/\Gamma_{H+ccH}$. In the coupling determination
the anomalous couplings would deviate from the Standard Model value and be negative.
This situation is thus indistinguishable from genuine anomalous couplings.

In practice the coherent observation of the reduction of all couplings would 
lead to a study of the $\Delta_\Gamma$ parameter. This parameter is determined in 
a single parameter fit. It cannot be determined in a global fit including the anomalous
Higgs couplings due to the structure of the observables (blind direction).  

Absolute coupling measurements at the LHC therefore depend on the (additional) assumptions
to be made in contrast to coupling ratios. 
A model independent study will have to wait for the ILC.

\subsection{Error determination}

\begin{table}[htb]
\begin{tabular}{l|l|l|ll|l|l|ll}
 &         \multicolumn{4}{c|}{no effective couplings} &
           \multicolumn{4}{c}{with effective couplings} \\
 & 
  RMS & $\sigma_\mathrm{symm}$ & $\sigma_\mathrm{neg}$ & $\sigma_\mathrm{pos}$ & 
  RMS & $\sigma_\mathrm{symm}$ & $\sigma_\mathrm{neg}$ & $\sigma_\mathrm{pos}$ \\\hline 
$m_H$                      
 & $\pm\,0.36$ & $\pm\,0.26$ & $-\,0.26$ & $+\,0.26$ 
 & $\pm\,0.38$ & $\pm\,0.25$ & $-\,0.26$ & $+\,0.25$ \\
$\Delta_{WWH}$             
 & $\pm\,0.31$ & $\pm\,0.23$ & $-\,0.21$ & $+\,0.26$ 
 & $\pm\,0.29$ & $\pm\,0.24$ & $-\,0.21$ & $+\,0.27$ \\
$\Delta_{ZZH}$             
 & $\pm\,0.49$ & $\pm\,0.36$ & $-\,0.40$ & $+\,0.35$ 
 & $\pm\,0.46$ & $\pm\,0.31$ & $-\,0.35$ & $+\,0.29$ \\
$\Delta_{t\bar{t}H}$       
 & $\pm\,0.58$ & $\pm\,0.41$ & $-\,0.37$ & $+\,0.45$ 
 & $\pm\,0.59$ & $\pm\,0.53$ & $-\,0.65$ & $+\,0.43$ \\
$\Delta_{b\bar{b}H}$       
 & $\pm\,0.53$ & $\pm\,0.45$ & $-\,0.33$ & $+\,0.56$ 
 & $\pm\,0.64$ & $\pm\,0.44$ & $-\,0.30$ & $+\,0.59$ \\
$\Delta_{\tau\bar{\tau}H}$ 
 & $\pm\,0.47$ & $\pm\,0.33$ & $-\,0.21$ & $+\,0.46$ 
 & $\pm\,0.57$ & $\pm\,0.31$ & $-\,0.19$ & $+\,0.46$ \\
$\Delta_{\gamma\gamma{}H}$ & 
 \phantom{$\pm$} --- &\phantom{$\pm$} --- &\phantom{$-$} --- &\phantom{$+$} --- 
 & $\pm\,0.55$ & $\pm\,0.31$ & $-\,0.30$ & $+\,0.33$ \\
$\Delta_{ggH}$             & 
 \phantom{$\pm$} --- &\phantom{$\pm$} --- &\phantom{$-$} --- &\phantom{$+$} --- 
 & $\pm\,0.80$ & $\pm\,0.61$ & $-\,0.59$ & $+\,0.62$ \\
$m_b$                      
 & $\pm\,0.073$& $\pm\,0.071$& $-\,0.071$& $+\,0.071$
 & $\pm\,0.070$& $\pm\,0.071$& $-\,0.071$& $+\,0.072$\\
$m_t$                      
 & $\pm\,1.99$ & $\pm\,1.00$ & $-\,1.03$ & $+\,0.98$ 
 & $\pm\,1.99$ & $\pm\,0.99$ & $-\,1.00$ & $+\,0.98$ 
\end{tabular}
\caption{Errors on the measurements from 10000 toy experiments
for an integrated luminosity of $30\ \fbinv{}$ (taken from~\cite{Lafaye:2009vr}).}
\label{tab:toyerrors}
\end{table}

The errors on the Higgs boson couplings are determined with 
10000 toy datasets. The datasets were smeared according to the experimental
and theoretical errors, including the correlations among them. 
The results are shown in Table~\ref{tab:toyerrors} without and 
with anomalous effective couplings.

The RMS is shown, as well as the fit of a symmetric Gaussian 
and the fit of two separate Gaussian for the 
positive and negative side of the coupling.
The expected precision for the coupling measurement is typically between
25\% and 45\%. As statistics increase (higher luminosity) 
the difference between RMS and the Gaussian fits should decrease.

Adding the effective couplings to the parameter sets, one would expect
an increase of the errors. While this is true in most cases, e.g., the $ZZH$
coupling is decreased. One can understand this effect by the impact of correlations.
The rate of the ZZ final state is the upper limit. With a positive correlation between
the effective coupling (production side) and the $HZZ$ (decay side), the effective
values that the $ZZH$ coupling can take on are actually reduced, leading to a smaller error.

\begin{table}[htb]
\begin{tabular}{r|l|ll|l|ll}
         & \multicolumn{3}{c|}{no effective couplings} &
           \multicolumn{3}{c}{with effective couplings} \\
 & 
  $\sigma_\mathrm{symm}$ & $\sigma_\mathrm{neg}$ & $\sigma_\mathrm{pos}$ & 
  $\sigma_\mathrm{symm}$ & $\sigma_\mathrm{neg}$ & $\sigma_\mathrm{pos}$ \\\hline 
$\Delta_{ZZH}/\Delta_{WWH}$             & $\pm\,0.46$ & $-\,0.36$ & $+\,0.53$ 
 & $\pm\,0.41$ & $-\,0.40$ & $+\,0.41$ \\
$\Delta_{t\bar{t}H}/\Delta_{WWH}$       & $\pm\,0.30$ & $-\,0.27$ & $+\,0.32$ 
 & $\pm\,0.51$ & $-\,0.54$ & $+\,0.48$ \\
$\Delta_{b\bar{b}H}/\Delta_{WWH}$       & $\pm\,0.28$ & $-\,0.24$ & $+\,0.32$ 
 & $\pm\,0.31$ & $-\,0.24$ & $+\,0.38$ \\
$\Delta_{\tau\bar{\tau}H}/\Delta_{WWH}$ & $\pm\,0.25$ & $-\,0.18$ & $+\,0.33$ 
 & $\pm\,0.28$ & $-\,0.16$ & $+\,0.40$ \\
$\Delta_{\gamma\gamma{}H}/\Delta_{WWH}$ & 
  \phantom{$\pm$} --- & \phantom{$-$} --- & \phantom{$+$} --- 
 & $\pm\,0.30$ & $-\,0.27$ & $+\,0.33$ \\
$\Delta_{ggH}/\Delta_{WWH}$             & 
  \phantom{$\pm$} --- & \phantom{$-$} --- & \phantom{$+$} --- 
 & $\pm\,0.61$ & $-\,0.71$ & $+\,0.46$ 
\end{tabular}
\caption{Errors on the ratio of couplings, corresponding to
  Table~\ref{tab:toyerrors} (taken from~\cite{Lafaye:2009vr}).}
\label{tab:toyerrors_ratio}
\end{table}

The errors on the ratios are shown in Table~\ref{tab:toyerrors_ratio}. As many 
experimental and theoretical errors are correlated, one would expect that the
ratios are measured more precisely than the absolute couplings. However the effect
is relatively small. This is essentially due to the statistics of the signal
for 30~\fbinv{}. For the experimental error the statistical error is much
larger than the systematic error which is canceled in the ratio. 
The definition of the flat errors implies that outside the theoretical 
errors, the $\chi^2$ varies only with the experimental error. 
As the experimental error is much larger, the theoretical errors do not have a large impact.
This has been checked by removing the theoretical errors entirely, which has only 
a marginal effect.

\begin{figure}[htb]
\includegraphics[width=0.32\textwidth]{figures/sfitter_markovbins3.HWW_Hbb.phaseW.truesubjet.eps}
\includegraphics[width=0.32\textwidth]{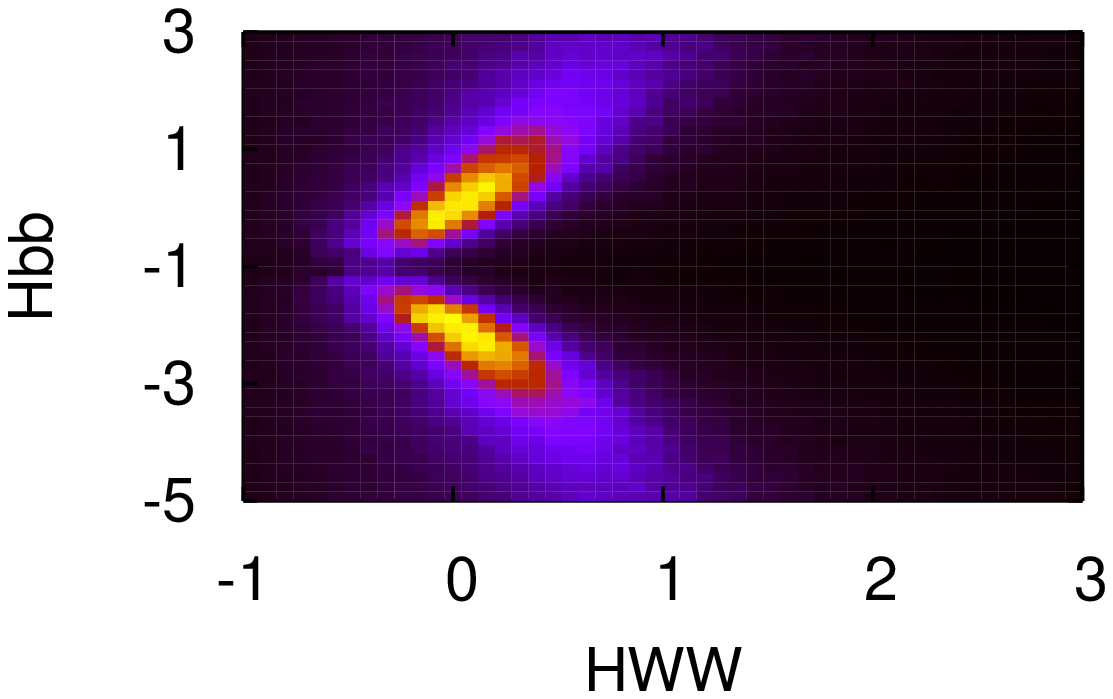}
\includegraphics[width=0.32\textwidth]{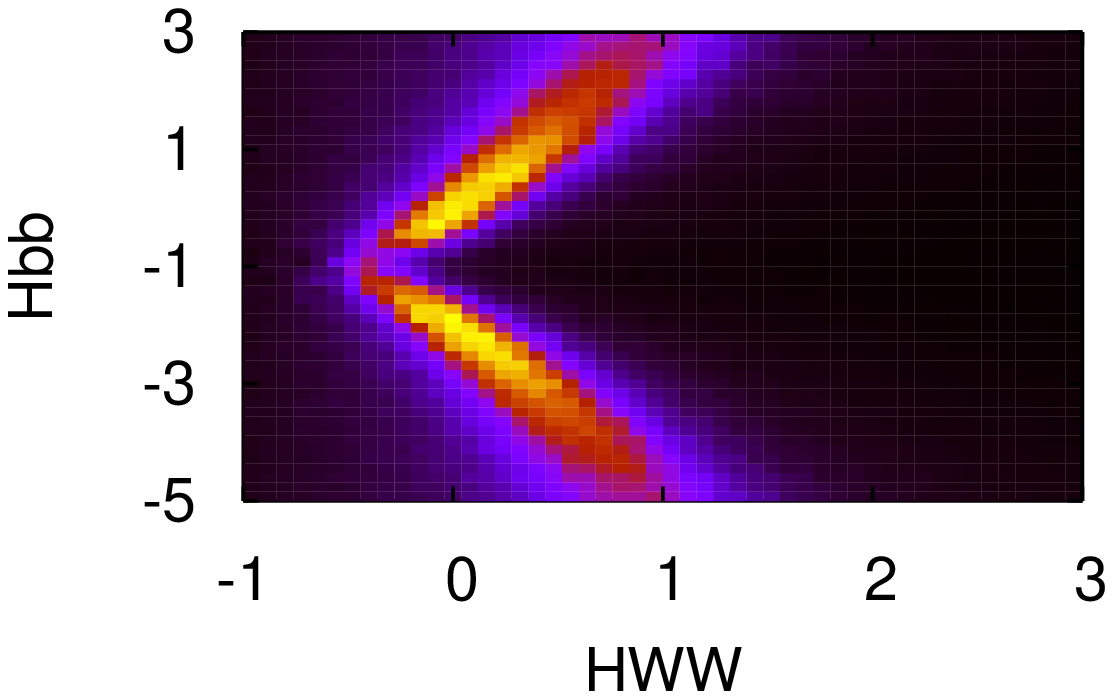}
\caption[]{Impact of the subjet analysis. The sensitivity is 100\% (left),
50\% (middle) and entirely removed (right) (taken from~\cite{Lafaye:2009vr}).}
\label{fig:Hbb}
\end{figure}

To illustrate the importance of the measurement of the $bbH$ coupling, in 
Figure~\ref{fig:Hbb} the effect of the subjet analysis is 
shown with full sensitivity,  
reducing its sensitivity by 50\% and removing it entirely. 
ATLAS has confirmed the theoretical study within recently 10\%, i.e., 3.7$\sigma$ significance
versus 4.2$\sigma$ which is closest to the full sensitivity in Figure~\ref{fig:Hbb}.

\subsection{Supersymmetric Higgs Scenarios}

In the favorable case of low mass supersymmetry, supersymmetric particles will 
be discovered well before the lightest Higgs boson. However, there are scenarios
where the supersymmetric particles are out of reach or need very high integrated
luminosity to be discovered and the Standard Model Higgs boson is the first 
``supersymmetric'' particle discovered. In such a case it is interesting to
ask whether the measurement of the Higgs boson couplings could be used 
to probe new physics in the absence of/or ignoring direct evidence.

Two scenarios were studied in~\cite{Lafaye:2009vr} of which only one 
will be discussed here as an illustration. The parameter set is a modified
version of SPS1a.
The study is based on the 
assumption that only a Higgs boson has been measured with a mass of 120~\gevcc{}
and no supersymmetric measurements have been made.

\begin{table}[htb]
\begin{tabular}{l|r|r|r|r|r|r|r|r}
 &
$\Delta_{WWH}$           &
$\Delta_{ZZH}$           &
$\Delta_{ttH}$           &
$\Delta_{bbH}$           &
$\Delta_{\tau\tau H}$    &
$\Delta_{\gamma\gamma H}$&
$\Delta_{ggH}$           &
$m_H$                    \\\hline
true &
$-0.13$ & $-0.13$ & $-0.19$ & $3.27$ & $3.29$ & $0.19$ & $-0.28$ &
$120.0$ \\
errors &
$\pm 0.45$ &
$\pm 0.61$ &
$\pm 0.63$ &
$\pm 2.34$ &
$\pm 3.35$ &
$\pm 0.99$ &
$\pm 1.12$ &
$\pm 0.29$ \\
 & 
$- 0.43$   &
$- 0.99$   &
$- 0.60$   &
$- 3.68$   &
$- 3.23$   &
$- 0.70$   &
$- 0.69$   &
$- 0.29$   \\
 & 
$+ 0.48$   & 
$+ 0.52$   & 
$+ 0.65$   & 
$+ 1.52$   & 
$+ 3.58$   & 
$+ 1.30$   & 
$+ 1.46$   & 
$+ 0.30$    
\end{tabular}
\caption{True couplings for the SPS1a-inspired scenario. The errors
were calculated for 30~\fbinv{} with 10000 toy experiments (taken from~\cite{Lafaye:2009vr}).}
\label{tab:sps1alow}
\end{table}

The datasets were modified to take into account the modification of the couplings.
With respect to SPS1a, $\tan\beta$ was modified to 7, $A_t$ to
$-1100~\gevcc{}$ and $m_A$ to $151~\gevcc{}$. The
true values for all Higgs couplings are listed in the first line of
Table~\ref{tab:sps1alow}. 
Comparing these deviations with the Standard
Model error bands in Table~\ref{tab:toyerrors}, it seems daunting to draw
conclusions from the individual measurements, but in addition 
to the individual precision of the coupling measurements, the full correlation
matrix (added information) is used.

The first question to be asked is whether the Standard Model is excluded.
The exclusion limit for the Standard Model is calculated by determining
the $\chi^2$ distribution for smeared Standard Model datasets, calculated with the
fixed Standard Model couplings. The integral over $\chi^2$  to 90\% of the datasets 
gives the 90\% confidence level. Then the new physics (SPS1a-modified) datasets
are used and the $\chi^2$ is calculated with the Standard Model couplings. 
The two dimensional results are that for 
$77\%$ of the toy experiments the Standard Model is excluded at a
confidence level of $90\%$. 
A discovery is much more difficult, for 4\% of the toy experiments the new physics scenario
is a better description of the datasets than the Standard Model. 
Using the Higgs coupling measurements, one may be able to exclude the Standard Model, but a discovery 
will be difficult.

\cleardoublepage

\chapter{Conclusions}
\label{chap:conclusion}

After many years of preparation, the LHC has started operations in 2008.
The first collisions are expected in the run 2009/2010 which will
start in November 2009. 
The increase of the center--of--mass energy beyond
the TeVatron will open up a new energy regime
to search for new physics.

The first year(s) of LHC and ATLAS/CMS operations will
concentrate on the understanding of the detector using Standard Model 
processes such as the production and decay of the Z~boson and 
lower mass resonances such as the J/$\psi$. An essential
step for the discovery of new physics will be to prove and improve the 
understanding of the Standard Model at the LHC.

Once the detectors and the backgrounds are understood, the search for 
new physics, the Higgs boson, supersymmetry and others (extra dimensions)
will commence. 
If supersymmetry is at low mass as hinted by electroweak
precision data, large signals are expected 
and many observables will be measured.

Disentangling the fundamental parameters from the measurements 
necessitates powerful tools such as SFitter and Fittino
as observables depend on different combinations of them.
Additionally, the parameter space has large dimensions, 
thus efficient ways of sampling are needed. 

The project SFitter has developed a framework for such an endeavor. 
Tools such as weighted Markov chains have been developed to 
sample the parameter space. Interfaces to the most precise 
theoretical predictions of observables have been built.
A rigorous treatment of statistical, systematic and theoretical 
errors, correlated or not, is provided.

Using SFitter, two projects have been described here. The 
Higgs sector was studied in the Standard Model as well as in supersymmetry,
using only the expected measurements for a low mass
Higgs. The supersymmetric parameters have been studied
at the electroweak scale in the MSSM. The parameters have also
been studied at the GUT scale in a top-down approach (mSUGRA).

In addition to these studies, SFitter has also analyzed the impact of
$(g-2)_\mu$ on the precision of the determination of supersymmetric 
parameters in a Les Houches project~\cite{:2008gva}.
In~\cite{Turlay:2009} the LHC potential for the parameter determination
of Split-Supersymmetry was studied.
New SFitter projects are already on the way. 
From one, excerpts were already shown in this paper: 
a study of the extrapolation of the supersymmetric parameters from 
the electroweak scale to the high scale. 
Another study will be the implication of the supersymmetric parameter determination
at the LHC on the prediction of the relic density. Here we also 
plan to include direct/indirect detection results. 

\begin{figure}[htb]
\begin{minipage}[htb]{0.48\textwidth}
\centering
\includegraphics[width=\textwidth]{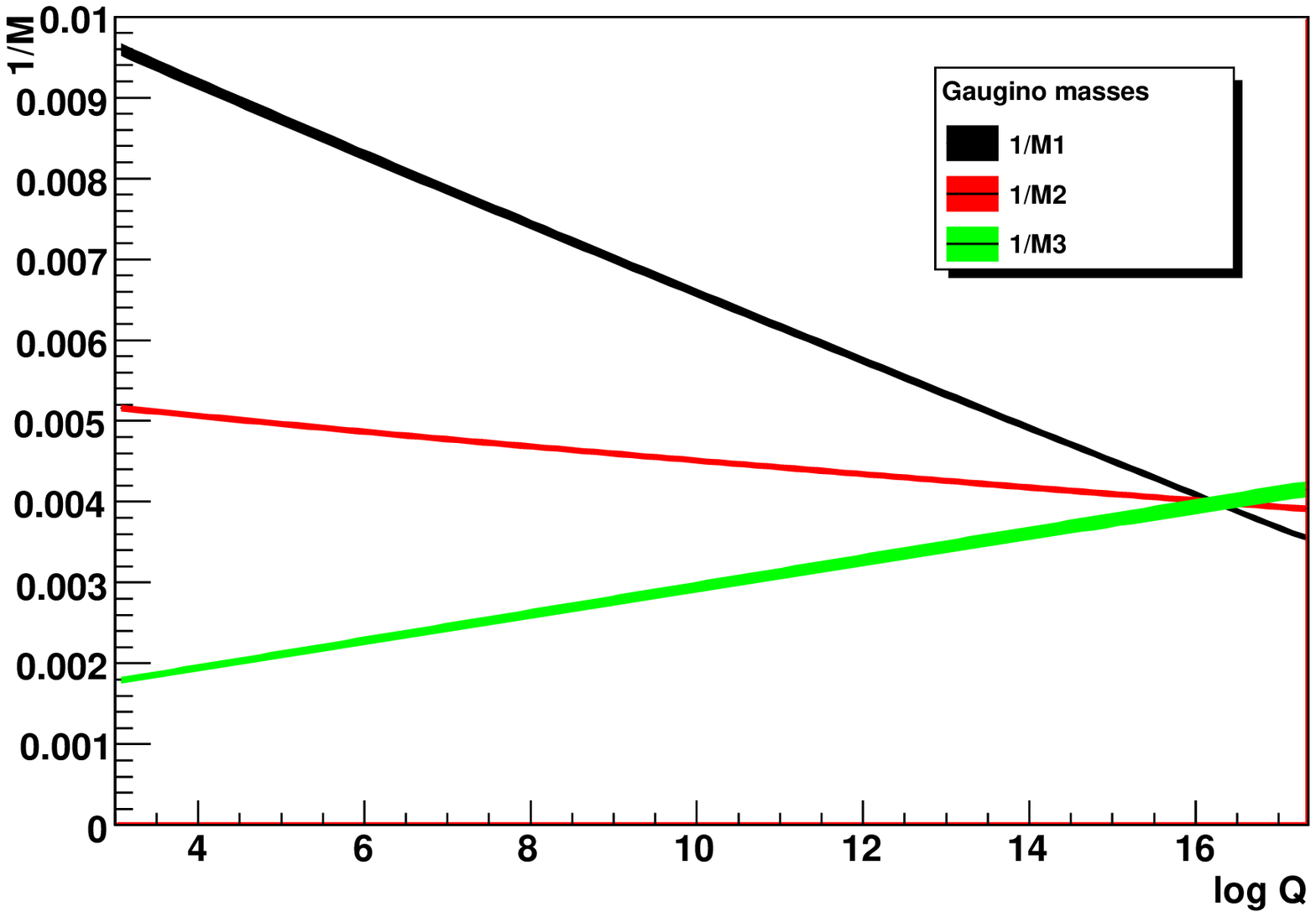}
\caption{Extrapolation of the gaugino mass parameters from the electroweak
scale to the GUT scale using LHC and ILC measurements.}
\label{fig:LHCILCgauginos}
\end{minipage}
\hspace{0.2cm}
\begin{minipage}[htb]{0.48\textwidth}
\centering
\includegraphics[width=\textwidth]{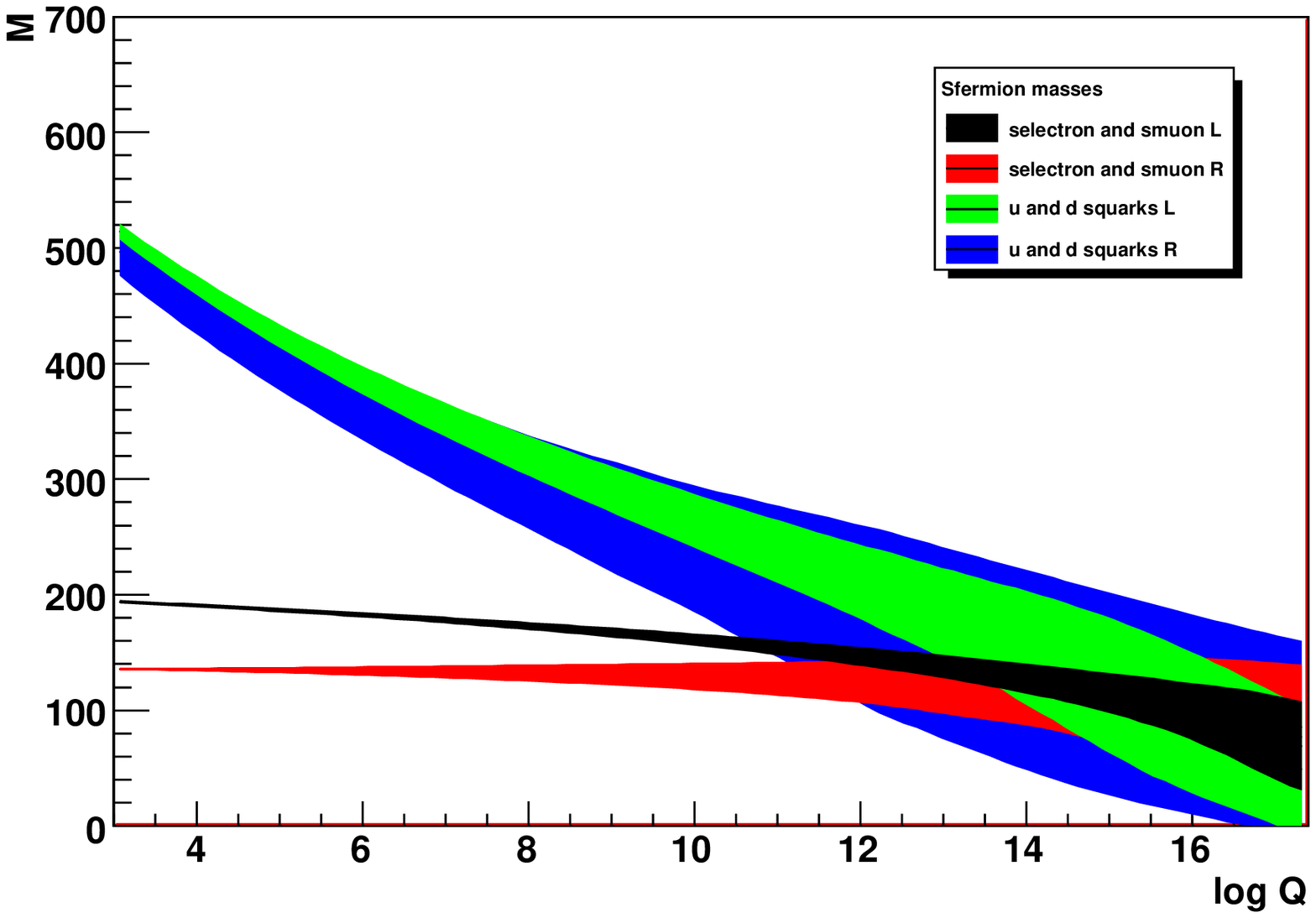}
\caption{Extrapolation of the scalar mass parameters from the electroweak
scale to the GUT scale using LHC and ILC measurements.}
\label{fig:LHCILCmScalars}
\end{minipage}
\end{figure}

The tools developed by SFitter can also be used for other models where it
is difficult to disentangle for the measurements the fundamental parameters.
Nature be supersymmetric or not, with the LHC/ATLAS now and an ILC 
in the future, particle physics will continue to provide exciting
measurements and insights into nature. Maybe SFitter will take us,
supersymmetry existing, with the LHC and ILC to the GUT scale, close to the Planck scale, 
as shown in Figures~\ref{fig:LHCILCgauginos} for the gaugino mass parameters 
and in Figure~\ref{fig:LHCILCmScalars} for the scalar mass parameters 
of the first and second generation.

\cleardoublepage

\cleardoublepage

\section*{Acknowledgements}

Pierre Bin\'etruy m'a appel\'e en decembre 2002 pour me demander 
si je voulais bien m'occuper d'un groupe dans le GDR
Supersym\'etrie. Cet appel a in fine donn\'e naissance \`a SFitter. 
Je voudrais donc remercier les SFitter: R\'emi Lafaye, Tilman Plehn
et Michael Rauch pour le travail en commun d\'ecrit partiellement
dans ce manuscrit. J'ai beaucoup appr\'eci\'e de travailler avec
Jean-Loic Kneur, Michael D\"uhrssen et Claire Adam sur des projets SFitter.

Je suis encore surpris qu'on m'ait laiss\'e toucher l'\'electronique
d'ATLAS. Une amie de longue date a remarqu\'e: ``Tu fais
des \'etudes de physique, mais tu es incapable de r\'eparer une prise 
\'electrique.'' Le travail sur des FEBs, stressant et int\'eressant,
a \'et\'e l'effort de toute une \'equipe. Merci \`a
Dominique Breton le magicien de l'\'electronique, 
Isabelle Falleau et Antoine P\'erus qui ont fait des miracles pour \'ecrire et 
faire fonctionner le soft.
Sans le travail efficace (qui n'a pas \'et\'e facile) de Patrick Favre, Alain Bozone, 
St\'ephane Trochet, Olivier Bohner, Mich\`ele Quentin, Bernard de Bennerot, 
Matthieu Lechowski ainsi que plusieurs \'etudiants/stagiaires nous n'aurions jamais pu tenir
les delais. Pierre Imbert s'est d\'evou\'e pour faire la navette entre le r\'eparateur
et l'\'equipe afin d'acc\'el\'erer. L'\'equipe de transport du LAL a d\'emontr\'e qu'il faut eviter
de sous-traiter...(vous connaissez GEFCO?)

Cot\'e software et egamma, je remercie David Rousseau et Srini Rajagopalan pour avoir
guid\'e mes pas et leur support quand je coordonnais ATLAS-egamma. 
Merci \`a Fred Derue pour sa contribution permanente et stable \`a egamma.
J'ai appris beaucoup de Daniel Froidevaux qui a du souffrir avec une personne aussi bavarde
que moi.

Un grand merci \`a Genevi\`eve Gilbert et aux membres du service mission et 
des services administratives du LAL pour leur aide quotidien et leur
support dans la gestion du GDR Terascale.

Je remercie les rapporteurs, Sally Dawson, (Fittino) Klaus Desch et Horst Oberlack, 
mon chef quand j'\'etais coordinateur egamma, de m'avoir fait l'honneur de venir.
Je remercie chaleureusement Jean Orloff, mon chef en tant que directeur 
du GDR Supersym\'etrie pendant quatre ans, Guy Wormser, directeur du LAL et Daniel 
Fournier, mon chef du groupe ATLAS pendant 11~ans et p\`ere du calorim\`etre
\'electromagn\'etique d'ATLAS, d'avoir accept\'e de faire partie du jury.

Des remerciements ne seraient pas complets sans remercier les membres du groupe ATLAS,
Laurent, mon chef lors du test beam pendant des ann\'ees pour de nombreux conseils, 
Louis, Lydia, 
David, Marumi, RD, Jean-Baptiste, Caroline et beaucoup d'autres.
J'ai particuli\`erement appr\'eci\'e de travailler avec Emmanuel Turlay, Andrea Thamm, 
Adrien Renaud.

Je voudrais remercier le conseil scientifique de l'Universit\'e Paris-Sud
pour sa contribution \`a mon repertoire d'anecdotes. 

Finalement un grand merci \`a mon p\`ere, Christiane, Sophie et Rebecca pour tout!
\cleardoublepage

\setlength{\oddsidemargin}{0.cm}

\end{document}